\documentclass[11pt,a4paper]{article}

\usepackage[a4paper, hmargin=2.5cm, vmargin=3cm]{geometry}

\usepackage{setspace}
\usepackage{fontspec}
\usepackage{xeCJK} %

\newfontfamily\CJKFont{NotoSansCJKsc-Regular}[Path = fonts/, Extension = .otf]
\newfontfamily\EmojiFont{NotoEmoji-Regular}[Path = fonts/, Extension = .ttf] 

\usepackage{authblk} %
\usepackage{graphicx} %
\RequirePackage{xcolor}
\usepackage[hidelinks]{hyperref}
\usepackage{xurl}
\usepackage{tikz}
\usepackage{tikz-network}
\usepackage[most]{tcolorbox}
\usepackage{makecell}
\usepackage{bm}
\usepackage{multirow}
\usepackage{standalone}
\usepackage{amssymb}
\usepackage{array}
\usepackage{ragged2e}
\usepackage{longtable}
\usepackage{booktabs}
\usepackage[commandnameprefix=ifneeded]{changes}
\usepackage{placeins}
\usepackage{pdflscape}

\usepackage{tabularray}
\usepackage{tabularx}
\usepackage{subcaption}
\usepackage{float}
\usepackage{graphicx}
\usepackage{codehigh}
\usepackage[normalem]{ulem}
\UseTblrLibrary{booktabs}
\UseTblrLibrary{siunitx}

\NewTableCommand{\tinytableDefineColor}[3]{\definecolor{#1}{#2}{#3}}

\usetikzlibrary{shapes.misc}
\tikzset{cross/.style={cross out, draw=black, minimum size=2*(#1-\pgflinewidth), inner sep=0pt, outer sep=0pt},
	cross/.default={1pt}}

\def\ShiftsProposalN/{\num{1202}}
\def\ShiftsProposalPct/{\num{14.81}}
\def\ShiftsProposalMax/{\num{30}}
\def\ShiftsProposalMedian/{\num{1.0}}
\def\ShiftsProposalStd/{\num{2.64}}
\def\ShiftsSpaceN/{\num{229}}
\def\ShiftsSpacePct/{\num{64.15}}
\def\ShiftsSpaceNHighTvlM/{\num{30}}
\def\ShiftsSpaceNHighTvlB/{\num{2}}
\def\ShiftsDAOContributorN/{\num{1457}}
\def\ShiftsContributorN/{\num{1362}}
\def\ShiftsContributorPct/{\num{3.37}}
\def\ShiftsContributorsProposalN/{\num{728}}
\def\ShiftsContributorsProposalPct/{\num{60.57}}
\def\ContributorsN/{\num{17561}}
\def\SpaceHasContributors/{\num{12272}}
\def\SpaceContributorsAvg/{\num{1.43}}
\def\ContributorComponents/{\num{10651}}
\def\ProjGDCSpaceAvgDegree/{\num{0.64}}
\def\SpaceAssociateVotes/{\num{794}}
\def\SpaceSelfAssociates/{\num{529}}
\def\ProposalChoiceScoreNotMatch/{\num{17}}
\def\ProposalNotFinal/{\num{1806}}
\def\ProposalNotFinalPerc/{\num{3.84}}
\def\ProposalSingleCoicePct/{\num{86.97}}
\def\SpacesSelfAssociatesDecided/{\num{178}}
\def\SpacesSelfAssociatesDecidedPct/{\num{20.41}}
\def\ProposalsSelfAssociatesDecided/{\num{2100}}
\def\ProposalsSelfAssociatesDecidedPct/{\num{5.98}}
\def\ProposalsNoneSelfAssociatesDecided/{\num{1847}}
\def\SpacesToValidate/{\num{367}}
\def\VotesToValidate/{\num{473306}}
\def\VotesConsistency/{\num{461402}}
\def\VotesConsistencyPct/{\num{97.48}}
\def\ProposalsTimeOut/{\num{0}}
\def\SpacesRaw/{\num{12294}}
\def\ProposalsRaw/{\num{76851}}
\def\VotersRaw/{\num{1603994}}
\def\VotesRaw/{\num{8365707}}
\def\ContributionRaw/{\num{11949}}
\def\ContrVotesRaw/{\num{316900}}
\def\SpacesClean/{\num{872}}
\def\ProposalsClean/{\num{35124}}
\def\VotersClean/{\num{986557}}
\def\VotesClean/{\num{5240622}}
\def\ContributionClean/{\num{7478}}
\def\ContrVotesClean/{\num{191507}}
\def\SpacesValidated/{\num{357}}
\def\ProposalsValidated/{\num{8116}}
\def\VotersValidated/{\num{119413}}
\def\VotesValidated/{\num{438668}}
\def\ContributionValidated/{\num{3927}}
\def\ContrVotesValidated/{\num{22878}}
\def\SpacesValidatedPct/{\num{40.94}}
\def\ProposalsValidatedPct/{\num{23.11}}
\def\VotersValidatedPct/{\num{12.1}}
\def\VotesValidatedPct/{\num{8.37}}
\def\ContributionValidatedPct/{\num{52.51}}
\def\ContrVotesValidatedPct/{\num{11.95}}
\def\SpacesNVPmoreOne/{\num{580}}
\def\SpacesNVPmoreFifty/{\num{66}}
\def\SpacesNVPmoreFiftyPct/{\num{7.54}}
\def\SpacesNVPmoreTen/{\num{297}}
\def\SpacesNVPmoreTenPct/{\num{33.94}}
\def\SpacesNVPmoreHundred/{\num{9}}
\def\VpMedian/{\num{4.26}}
\def\VpStd/{\num{21.22}}
\def\StrategiesN/{\num{208}}
\def\StrategiesErcPct/{\num{53.85}}
\def\SpacesOrigN/{\num{9505}}
\def\ProposalsOrigN/{\num{76851}}
\def\VotersOrigN/{\num{1603994}}
\def\SpacesSingleProposalPct/{\num{42.67}}
\def\SpacesFollowersLessFivePct/{\num{54.05}}
\def\SpacesLessTwoVotesPct/{\num{50.85}}
\def\SpacesSelectedN/{\num{967}}
\def\ProposalsSelectedN/{\num{47037}}
\def\VotesSelectedN/{\num{7193060}}
\def\ProposalsSelectedPct/{\num{61.21}}
\def\VotesSelectedPct/{\num{87.61}}

\newcommand{\DaoBiasRawProposalN}{\num{223760}}
\newcommand{\DaoBiasRawSnapshotFlaggedProposalN}{\num{143}}
\newcommand{\DaoBiasRawSpaceN}{\num{32995}}
\newcommand{\DaoBiasRawChoiceN}{\num{653913}}
\newcommand{\DaoBiasRawVoteN}{\num{57626386}}

\newcommand{\DaoBiasFinalProposalN}{\num{127281}}
\newcommand{\DaoBiasFinalSpaceN}{\num{19201}}
\newcommand{\DaoBiasFinalChoiceN}{\num{398218}}
\newcommand{\DaoBiasFinalVoteMassN}{\num{50759796}}
\newcommand{\DaoBiasDatasetSpaceRemovedN}{\num{13794}}
\newcommand{\DaoBiasDatasetSpaceReductionPct}{\num{41.81}}
\newcommand{\DaoBiasDatasetProposalRemovedN}{\num{96479}}
\newcommand{\DaoBiasDatasetProposalReductionPct}{\num{43.12}}
\newcommand{\DaoBiasDatasetChoiceRemovedN}{\num{255695}}
\newcommand{\DaoBiasDatasetChoiceReductionPct}{\num{39.1}}
\newcommand{\DaoBiasDatasetVoteRemovedN}{\num{6866590}}
\newcommand{\DaoBiasDatasetVoteReductionPct}{\num{11.92}}

\newcommand{\DaoBiasExcludedSnapshotFlaggedProposalN}{\num{83}}
\newcommand{\DaoBiasExcludedSnapshotFlaggedProposalPct}{\num{58.04}}

\newcommand{\DaoBiasSinglePreferenceProposalN}{\num{118373}}
\newcommand{\DaoBiasSinglePreferenceProposalPct}{\num{93}}
\newcommand{\DaoBiasSinglePreferenceChoiceN}{\num{293818}}
\newcommand{\DaoBiasSinglePreferenceChoicePct}{\num{73.78}}
\newcommand{\DaoBiasMultiPreferenceProposalN}{\num{8908}}
\newcommand{\DaoBiasMultiPreferenceProposalPct}{\num{7}}
\newcommand{\DaoBiasMultiPreferenceChoiceN}{\num{104400}}
\newcommand{\DaoBiasMultiPreferenceChoicePct}{\num{26.22}}
\newcommand{\DaoBiasVotingTypeBasicProposalN}{\num{8975}}
\newcommand{\DaoBiasVotingTypeBasicProposalPct}{\num{7.05}}
\newcommand{\DaoBiasVotingTypeBasicChoiceN}{\num{26925}}
\newcommand{\DaoBiasVotingTypeBasicChoicePct}{\num{6.76}}
\newcommand{\DaoBiasVotingTypeSingleChoiceProposalN}{\num{109398}}
\newcommand{\DaoBiasVotingTypeSingleChoiceProposalPct}{\num{85.95}}
\newcommand{\DaoBiasVotingTypeSingleChoiceChoiceN}{\num{266893}}
\newcommand{\DaoBiasVotingTypeSingleChoiceChoicePct}{\num{67.02}}
\newcommand{\DaoBiasVotingTypeWeightedProposalN}{\num{4671}}
\newcommand{\DaoBiasVotingTypeWeightedProposalPct}{\num{3.67}}
\newcommand{\DaoBiasVotingTypeWeightedChoiceN}{\num{79043}}
\newcommand{\DaoBiasVotingTypeWeightedChoicePct}{\num{19.85}}
\newcommand{\DaoBiasVotingTypeQuadraticProposalN}{\num{1793}}
\newcommand{\DaoBiasVotingTypeQuadraticProposalPct}{\num{1.41}}
\newcommand{\DaoBiasVotingTypeQuadraticChoiceN}{\num{10591}}
\newcommand{\DaoBiasVotingTypeQuadraticChoicePct}{\num{2.66}}
\newcommand{\DaoBiasVotingTypeRankedChoiceProposalN}{\num{907}}
\newcommand{\DaoBiasVotingTypeRankedChoiceProposalPct}{\num{0.71}}
\newcommand{\DaoBiasVotingTypeRankedChoiceChoiceN}{\num{5031}}
\newcommand{\DaoBiasVotingTypeRankedChoiceChoicePct}{\num{1.26}}
\newcommand{\DaoBiasVotingTypeApprovalProposalN}{\num{1537}}
\newcommand{\DaoBiasVotingTypeApprovalProposalPct}{\num{1.21}}
\newcommand{\DaoBiasVotingTypeApprovalChoiceN}{\num{9735}}
\newcommand{\DaoBiasVotingTypeApprovalChoicePct}{\num{2.44}}
\newcommand{\DaoBiasFinalChoicePerProposalMean}{\num{3.13}}
\newcommand{\DaoBiasFinalChoicePerProposalStd}{\num{8.29}}
\newcommand{\DaoBiasFinalChoicePerProposalMedian}{\num{2}}
\newcommand{\DaoBiasFinalProposalPerSpaceMean}{\num{6.63}}
\newcommand{\DaoBiasFinalProposalPerSpaceStd}{\num{38.71}}
\newcommand{\DaoBiasFinalProposalPerSpaceMedian}{\num{1}}

\newcommand{\DaoBiasStageThreeLoadedFinalProposalN}{\num{169689}}

\newcommand{\DaoBiasVoteExpansionFinishedProposalN}{\num{148056}}
\newcommand{\DaoBiasVoteExpansionFavoriteRowN}{\num{267437}}
\newcommand{\DaoBiasVoteExpansionFavoriteErrorRowN}{\num{13554}}

\newcommand{\DaoBiasLlmKappaLlama}{\num{0.69}}
\newcommand{\DaoBiasLlmKappaLlamaAgreement}{substantial}
\newcommand{\DaoBiasLlmKappaLlamaThreePointThree}{\num{0.68}}
\newcommand{\DaoBiasLlmKappaLlamaThreePointThreeAgreement}{substantial}
\newcommand{\DaoBiasLlmKappaOai}{\num{0.36}}
\newcommand{\DaoBiasLlmKappaOaiAgreement}{fair}
\newcommand{\DaoBiasLlmKappaChatgpt}{\num{0.61}}
\newcommand{\DaoBiasLlmKappaChatgptAgreement}{substantial}
\newcommand{\DaoBiasLlmKappaHighestSource}{Llama 3.1}
\newcommand{\DaoBiasHumanEvaluatedProposalN}{\num{590}}

\newcommand{\DaoBiasHumanEnglishProposalN}{\num{560}}
\newcommand{\DaoBiasHumanChineseProposalN}{\num{20}}
\newcommand{\DaoBiasHumanJapaneseProposalN}{\num{5}}
\newcommand{\DaoBiasHumanSpanishProposalN}{\num{5}}
\newcommand{\DaoBiasHumanMaxChoiceN}{\num{10}}
\newcommand{\DaoBiasHumanProposalReviewMinN}{\num{3}}

\newcommand{\DaoBiasHumanEvaluatorN}{\num{375}}

\newcommand{\DaoBiasHumanLowEffortEvaluatorN}{\num{6}}
\newcommand{\DaoBiasHumanAverageTaskMinutes}{\num{14.09}}
\newcommand{\DaoBiasHumanLabeledChoiceN}{\num{1603}}

\newcommand{\DaoBiasHumanKappa}{\num{0.71}}
\newcommand{\DaoBiasHumanKappaAgreement}{substantial}
\newcommand{\DaoBiasSuggestedNoSuggestionPctLlama}{\num{57.9}}
\newcommand{\DaoBiasSuggestedNoSuggestionPctLlamaThreePointThree}{\num{72.5}}
\newcommand{\DaoBiasSuggestedNoSuggestionPctOai}{\num{64}}
\newcommand{\DaoBiasSuggestedNoSuggestionPctChatgpt}{\num{53.3}}
\newcommand{\DaoBiasSuggestedNoSuggestionPctMin}{\num{53.3}}
\newcommand{\DaoBiasSuggestedNoSuggestionPctMinSource}{ChatGPT}
\newcommand{\DaoBiasSuggestedNoSuggestionPctMax}{\num{72.5}}
\newcommand{\DaoBiasSuggestedNoSuggestionPctMaxSource}{Llama 3.3}
\newcommand{\DaoBiasSuggestedLlamaThreePointThreeReviewedProposalN}{\num{89479}}
\newcommand{\DaoBiasSuggestedLlamaThreePointThreeNotReviewedProposalN}{\num{37859}}
\newcommand{\DaoBiasSuggestedLlamaThreePointThreeBasicExclusionProposalN}{\num{8998}}
\newcommand{\DaoBiasSuggestedLlamaThreePointThreeConflictProposalN}{\num{67}}
\newcommand{\DaoBiasSuggestedLlamaThreePointThreeNoSuggestionProposalN}{\num{64831}}
\newcommand{\DaoBiasSuggestedLlamaThreePointThreePositiveProposalN}{\num{24581}}
\newcommand{\DaoBiasSuggestedApprovePositivePctMin}{\num{99.8}}

\newcommand{\DaoBiasSuggestedApprovePrecisionMin}{\num{0.64}}
\newcommand{\DaoBiasSuggestedApprovePrecisionMax}{\num{0.77}}
\newcommand{\DaoBiasSuggestedApproveRecallMin}{\num{0.46}}
\newcommand{\DaoBiasSuggestedApproveRecallMax}{\num{0.58}}
\newcommand{\DaoBiasSuggestedApproveFOneMin}{\num{0.54}}
\newcommand{\DaoBiasSuggestedApproveFOneMax}{\num{0.65}}

\newtcbtheorem{Summary}{\bfseries Summary}{enhanced,drop shadow={black!90!white},
	coltitle=black,
	top=0.3in,
	attach boxed title to top left=
	{xshift=1.5em,yshift=-\tcboxedtitleheight/2},
	boxed title style={size=small,colback=lightgray}
}{summary}

\newcommand\identity{1\kern-0.25em\text{l}}

\usepackage[authoryear,round]{natbib} %
\title{Voting Biases in Decentralized Autonomous \\ Organization (DAO) Governance}

\author[1,2]{Stefano Balietti}  %
\author[3,4]{Pietro Saggese\thanks{Corresponding author. E-mail: pietro.saggese@imtlucca.it, saggese@csh.ac.at}}  %
\author[1,4,5]{Markus Strohmaier} %

\affil[1]{University of Mannheim}
\affil[2]{Akku Labs, Private Sector}
\affil[3]{IMT School for Advanced Studies Lucca}
\affil[4]{Complexity Science Hub Vienna}
\affil[5]{GESIS - Leibniz Institute for the Social Sciences}

\begin{document}

\maketitle

\begin{abstract}

Decentralized Autonomous Organizations (DAOs) replace formal managerial authority with collective decision-making, by handing both proposal rights and token-weighted votes directly to community members. 
Yet, as in corporate votes, where management sets the agenda and its proposals rarely fail, DAO votes are strikingly one-sided.
Using 127 281 proposals and about 51 million votes cast on Snapshot, the leading off-chain DAO voting platform, between 2020 and 2023, we ask which features of the ballot are associated with this support. We relate each choice's voting-power share to three observable features: its position in the list, whether its stance expresses approval, and whether the proposal author selected it. Each identifies one voting bias: position bias, approval bias, and author-selected choice bias. %
Author-selected choices show the strongest and most robust association, receiving a voting-power share 58.8 percentage points higher than non-author-selected ones, compared with 27.1 percentage points for approving choices and 7.7 percentage points for first-listed ones.  About one third of the author bias comes from the author's own voting power; that advantage is smaller in the largest protocols and in proposals with higher participation. The results are robust across alternative specifications and outcomes.  Our findings suggest that the agenda power that management holds in corporate voting persists even without a formal hierarchy, and that ballot design and the visibility of proposers' preferences are governance choices rather than neutral implementation details.

\vspace{0.3cm}

\vspace{0.2cm}

\end{abstract}

\clearpage

\section{Introduction}

Organizations that decide by vote must first decide what to vote on, and whoever writes the proposal has a form of agenda power~\citep{romer1978political,baron_bargaining_1989}. In traditional corporations, this power rests with management, and its effects are visible. Shareholder votes overwhelmingly end in approval: management-sponsored proposals and director elections typically receive very high support, and contested outcomes are the exception rather than the rule~\citep{yermack2010shareholder,cunat2012vote}. When votes are close, management prevails far more often than chance would predict~\citep{listokin2008management,bach2019close}, %
and shareholders respond to visible signals from management and proxy advisors~\citep{malenko2016role}. Decentralized Autonomous Organizations (DAOs), a novel governance model that relies on blockchain technology and promises to deliver efficient and democratic coordination mechanisms, remove the board and the manager and hand both proposal rights and token-weighted votes directly to community members~\citep{hassan2021decentralized,han2025review}. They therefore offer a natural setting to ask whether the same regularities persist once the formal hierarchy is ruled out: to what extent do DAO votes also skew toward approval, and if so, what are the determinants that explain how that advantage operates? In particular, does the author of a proposal enjoy an advantage similar to management's?

In a DAO, governance rules are encoded in smart contracts, and voting rights are distributed through governance tokens, so that each member's voting power is proportional to the tokens they hold~\citep{wang2019decentralized}.
DAOs govern decentralized finance (DeFi) protocols~\citep{Auer2023}, online communities and funding initiatives~\citep{weidener2025decentralization}, and virtual economies~\citep{Goldberg2022,balietti2025crypto}.
The stakes are substantial: the DAOs controlling DeFi protocols such as MakerDAO, %
Uniswap, %
and Compound %
manage treasuries worth hundreds of millions of dollars~\citep{Sun2022}.%

Decisions are taken by voting on \emph{governance proposals} that can affect any aspect of the organization: its technical infrastructure, %
its economic incentives, %
and the allocation of its funds. %
A growing literature documents frictions in this process.
Token ownership is highly concentrated~\citep{barbereau2022defi,appel2023control}, %
participation is low~\citep{Barbereau2023}, and holders with the power to change outcomes rarely use it~\citep{Fritsch2022a,Feichtinger2023}.
Large holders and insiders vote strategically~\citep{kitzler2024governance}, voting rights are valued differently across voters~\citep{rossello2026voters}, and proposal authors---or simply, proposers---trade on the information in their proposals~\citep{cong2025centralized}.
These frictions coexist with another puzzle: when DAO proposals reach formal voting, support is often highly skewed toward a single winning side.
In protocol governance, \citet{messias2024understanding} find that proposals on the DeFi lending protocol Compound receive on average 88.6\% of votes in favor, with a median of 97.7\%.
On the early on-chain platforms DAOstack, Aragon, and DAOhaus, \citet{faqir2021comparative} report proposal acceptance rates between about 75\% and 90\%, and argue that off-chain discussion and the cost of failed proposals may screen out proposals before a vote.
\cite{cong2025centralized} find that, across a subsample of selected proposals in 216 DAOs, the winning option receives on average 84.4\% of votes cast.

Prior work thus establishes that DAO voting records strong support, but does not investigate systematically the entire voting process, whether the winning choices approve or reject the proposals, and leaves open which features of the ballot are associated with that support.
Established literature from related fields suggests three characteristics of a ballot that can shape how voting power is allocated: %
how options are presented, the processes through which proposals reach a vote, and the social signals attached to them.
First, earlier-listed choices may benefit from interface and order effects, consistent with survey response-order and ballot-order research \citep{krosnick1987evaluation,koppell2004effects,ho2008estimating,meredith2013causes}; we call this the \textit{position bias}.
Second, approval-oriented choices may be favored because proposals are screened, discussed, and revised before reaching a formal vote~\citep{maug2009shareholders,faqir2021comparative,messias2024understanding}; we call this the \textit{approval bias}.
Third, choices selected by proposal authors may attract support because less-informed voters defer to a better-informed participant~\citep{feddersen1996swing}, because the author votes with their own tokens, or because their visible vote seeds a bandwagon~\citep{callander2007bandwagons,knight2010momentum,muchnik2013social}; we call this the \textit{author bias}.\footnote{We note that we use \textit{bias} to denote a systematic association with voting-power share, not as proof that any feature causally distorts preferences.}

To our knowledge, these forms of bias have not yet been conceptualized and empirically examined in DAO voting systems.
In this paper, we thus investigate the extent to which the high approval rates and one-sided outcomes %
can be explained by the three biases identified above, and how much these overlap with one another.  %
Accordingly, we ask:

\vspace{-0.15cm}

\begin{description}
	\item[\textbf{RQ1 (Position).}] Do earlier-listed choices receive a disproportionate share of voting power, compared to subsequent options?
	\vspace{-0.15cm}
	\item[\textbf{RQ2 (Approval).}] Are approval-oriented choices systematically favored, and is this advantage distinct from list position?
	\vspace{-0.15cm} 
	\item[\textbf{RQ3 (Author).}] Do author-selected choices receive a voting-power advantage beyond what position and stance already explain? And how much of this advantage remains once the author's own voting power is removed?
\end{description} 

\vspace{-0.15cm}

We study \DaoBiasFinalProposalN{} proposals and around 51 million votes cast on Snapshot, the leading off-chain voting platform for DAOs, between October 2020 and November 2023.
We relate each choice's share of voting power to these three features in fractional-logit models with DAO fixed effects.
The author advantage is by far the largest: an author-selected choice receives a voting-power share 58.8 percentage points higher than that of an otherwise similar choice, compared with 27.1 percentage points for an approval-oriented choice and 7.7 percentage points for a first-listed choice.
Because we observe the author's own vote, we can decompose this advantage.
In proposals where the author voted and other voters contributed positive voting power, that advantage falls from 45.6 to 29.7 percentage points when the author's own voting power is removed: roughly one third is mechanical, and two thirds comes from other voters.
The approval advantage, in turn, falls to 7.7 percentage points once we condition on the proposer having voted, suggesting that much of what looks like a general preference for approval moves together with the author's own support.
Finally, the author advantage is smaller in the largest DeFi protocols (27--33 percentage points) and once proposals with very low participation are excluded (39.7 percentage points), that is, where stakes and participation are higher.
These results hold across alternative outcomes, samples, voting systems, and fixed-effect structures.

We contribute to three strands of literature.
First, we extend the economics of DAO governance~\citep{han2025review,rossello2024blockholders,cong2025centralized} from the distribution of voting power to the mechanics of the ballot itself. To our knowledge, this paper provides the first large-scale evidence on Snapshot governance at the level of individual ballot choices. Moreover, to measure choice stance at scale, we also develop a stance classification procedure validated against language-model and human evaluations, which we make available for future work.
Second, we add to the literature on shareholder voting and agenda power~\citep{listokin2008management,bach2019close,maug2009shareholders} by documenting a large proposer advantage in organizations without managers, and by decomposing it into the author's own votes and the support of other voters, a decomposition that corporate data rarely allow.
Third, we contribute to the economics of voting under limited information~\citep{feddersen1996swing,battaglini2010swing,knight2010momentum} and to ballot-order research~\citep{ho2008estimating} with evidence from a setting in which %
tallies are public and the ballot is designed by an interested party.

These results matter for the design of decentralized organizations and potentially for other settings where proposal authors' choices and running tallies are visible to voters.
First, if a large share of the support for an option is associated with who proposed it rather than with its content, then token-weighted voting aggregates preferences less independently than its transparency suggests, and proposal authors---often founders, core developers, or large holders---have more influence than their token holdings alone imply.
Second, systematic advantages for particular kinds of choices may penalize dissenting positions and mask limited deliberation, suggesting that seemingly technical features of the voting process can have institutional consequences.
Third, ballot features such as the order of choices, the visibility of the author's vote, and the display of live tallies are governance design choices with distributional consequences, rather than neutral interface details.
Finally, and more broadly, understanding these associations is crucial to shedding light on the quality and robustness of collective decision-making in DAOs. Identifying systematic biases is essential to evaluate the legitimacy of DAO governance and, ultimately, to guide improvements in the design of decentralized decision-making systems.

The paper is structured as follows. Section \ref{sec:background} provides background information and reviews the related literature. Section \ref{sec:datacleaning} describes the data and outlines the data processing and cleaning procedures. Sections \ref{sec:desc_analysis} and \ref{sec:reg_analyses} respectively present the descriptive results and the main empirical analyses. %
Section~\ref{sec:discussion} interprets the results and discusses implications for governance design and limitations. Section~\ref{sec:Conclusion} concludes.

\section{Background and Literature Review}%
\label{sec:background}

A growing literature studies Decentralized Autonomous Organizations and token-based governance. Theory analyzes how token voting interacts with platform growth and control by dominant participants \citep{abadi2024token,ferreira2023corporate} and conflicts between large and small token holders \citep{han2023dao}; comprehensive reviews are provided by \citet{biais2023advances} and \citet{han2025review}. Empirically, voting power is highly concentrated \citep{barbereau2022defi,Fritsch2022a,appel2023control,fracassi2024decentralized,Sun2022}, large holders vote strategically \citep{rossello2024blockholders,han2023dao}, voting rights are priced heterogeneously across voters \citep{rossello2026voters}, proposal managers trade on the information in their proposals \citep{cong2025centralized}, and delegation shapes who effectively decides \citep{hall2024liquid}; see also \citet{Laturnus2023}, \citet{Goldberg2022} and \citet{kitzler2024governance}. These works focus on who holds and exercises voting power. We instead study what happens inside the ballot, and show that the author's choice, the stance of an option and its position are systematically associated with how that power is allocated.

\subsection{Decentralized Autonomous Organizations (DAOs)}

Decentralized Autonomous Organizations are a novel organizational structure designed to offer a decentralized alternative to traditional corporate governance~\citep{hassan2021decentralized,rossello2024blockholders}.
Although they primarily emerged as a means of governing blockchain-based decentralized applications (dApps), such as DeFi protocols, they can also be used to coordinate online communities or participants in a virtual economy without any underlying dApp~\citep{constitution2021documentation}.
Their key components are deployed on distributed ledger technology: smart contracts encode the governance rules that determine who can make decisions and how; governance tokens distribute this decision-making power across community members; and treasuries, also managed via smart contracts, hold the DAO's funds. 
By relying on blockchain technologies, DAOs aim to remove central authorities and support more democratic and participatory governance~\citep{balietti2025slaying}.

\paragraph{Decision making in DAOs.}
In DAO governance decision-making, community members collectively propose, evaluate, and implement changes.

\textit{Proposals and choices.}
Decisions are taken through \textit{governance proposals} that submit for collective consideration a proposed change, action, or decision on the DAO structure or the underlying project.\footnote{Examples of governance proposals are available here:~\cite{Uniswap2022deploy, curve2020inflation, Compound2023migration, Uniswap2023donation, Uniswap2023fees}.}
A proposal consists of a title and a body, which define the question and the suggested change, and a set of choices, which define the alternatives to decide on.
The proposal \textit{author}, or \textit{proposer}, drafts the choices, sets their order, and can vote on the proposal like any other token holder.
Proposals range from decisions with a direct effect on the organization, such as smart contract upgrades, stipends to members, board elections, or changes to protocol parameters (e.g., trading fees, interest rates, or which tokens to list), to broader consultations, such as opinion polls or selections among pre-selected alternatives (e.g., the official logo of the organization).
Choices can be expressed as simple binary responses, numbers, strings of text, or other arbitrary content.
We define the intended meaning or position expressed by a choice as its \textit{stance}. In most proposals, choices simply indicate whether to approve or reject the proposal content, but they can also encode more complex stances.

\textit{Voting and execution.}
Once a proposal is created, token holders cast their vote, with voting power usually proportional to the governance tokens they control.
Votes can be recorded directly on the blockchain (on-chain voting), which allows automatic execution through smart contracts, or on external platforms (off-chain voting) such as Snapshot~(\citeyear{snapshot2023documentation}), where votes are authenticated via on-chain addresses and voting power is computed from on-chain balances, but votes are recorded off-chain.
While on-chain voting is more secure and transparent, it incurs high transaction costs, and most DAOs today rely on off-chain voting.
Platforms like Snapshot support several voting systems, which differ in how voters express preferences and how votes are aggregated (see Table~\ref{tab_voting_systems_catalog}).
The most widely adopted is single-choice voting~\citep{Wang2022b}, where users indicate one single preference; other systems allow voters to split their voting power across several choices or to rank them.
In its default configuration, a Snapshot ballot displays the author's identity, the individual votes, and the running tally throughout the voting period.
Because Snapshot votes are recorded off-chain, accepted proposals are typically executed on-chain by trusted members, but the substantive decision is taken on the Snapshot ballot itself.

\subsection{Voting, Agenda Power, and Information}
\label{sec:econ_lit}

Token-weighted DAO voting is a close analogue to shareholder voting in corporations, where control rights are also allocated in proportion to holdings \citep{yermack2010shareholder}. A robust finding in this literature is that the party that sets the agenda tends to win: management-sponsored proposals pass at very high rates, and management wins close votes disproportionately often \citep{listokin2008management,bach2019close}.
Theories of agenda control show that the right to formulate the alternatives confers power even when the final decision is taken by majority vote \citep{romer1978political,baron_bargaining_1989}.
Observed support also depends on which proposals reach a vote at all \citep{maug2009shareholders} and on the signals of other participants, such as peer funds and proxy advisors \citep{matvos2010heterogeneity,malenko2016role}.
In a DAO, the proposal author combines these roles: like management, they set the agenda, choosing the options, their wording, and their order; like a proxy advisor, they can publicly signal a preferred option; and like a blockholder, they can vote with their own tokens.
Position, stance, and author selection are therefore all, in part, outcomes of agenda design.

\subsection{Biases in Voting and Online Collaborative Platforms}

Voting and online collaborative platforms translate heterogeneous individual judgments into aggregate rankings, ratings, or voting outcomes, and are therefore exposed to well-documented behavioral and interface effects.
Three channels are most relevant to our setting, and correspond to the three biases we study: presentation effects (position bias), selection before the vote and social desirability (approval bias), and information, social influence, and reputation signals (author bias).

\paragraph{Presentation effects.}
Survey response-order and ballot-order studies show that the first-listed response or candidate can receive an advantage~\citep{krosnick1987evaluation}.
The voting literature provides estimates coming from settings where order is assigned at random, such as the California alphabet lottery \citep{ho2008estimating,miller1998impact,meredith2013causes}, and shows effects concentrated in low-information races \citep{koppell2004effects}; in online political primaries, too, candidates listed at the top of the screen received a measurable advantage~\citep{marolla2024voting}.

The literature on search and recommendation settings reaches similar findings. Users are more likely to inspect and select items displayed higher in search results~\citep{joachims2005accurately,craswell2008experimental}, and analogous effects have been documented on Reddit and Hacker News~\citep{stoddard2015popularity}, StackExchange~\citep{burghardt2017myopia,burghardt2018quantifying,liu2025counterfactual}, recommender and crowdsourcing systems~\citep{collins2018study,burghardt2020origins}, and cultural markets~\citep{hogg2014disentangling,abeliuk2017taming}.
Other interface effects, such as attention, trust and anchoring biases~\citep{agarwal2019addressing,yang2012anchoring} have been studied in online settings, but are less relevant in Snapshot, which presents voters with a simple list of choices.

\paragraph{Pre-vote selection and positive-outcome skew.}
High approval rates may reflect a broader positive-outcome skew rather than voter preferences alone.
One channel is social desirability: voters may be inclined to support approval-oriented choices when approval is perceived as cooperative, constructive, or aligned with community goals \citep{grimm2010social}.
Another channel operates before the vote: DAO proposals are often discussed, revised, or informally tested in forums, so that, as in corporate settings \citep{maug2009shareholders}, those reaching a formal ballot are mostly those expected to pass.
Studies of DAOstack, DAOhaus, and Aragon~\citep{faqir2021comparative}, as well as Snapshot~\citep{Wang2022b,messias2024understanding} report high approval rates and link them to off-chain discussion and the cost of submitting proposals likely to fail.

\paragraph{Information, social influence, and reputation signals.}
When voters differ in how well informed they are, deferring to better-informed participants can be individually rational: in the swing voter's curse, uninformed voters abstain or vote so as to let the informed decide \citep{feddersen1996swing,battaglini2010swing}.
When earlier choices are observable, later voters may also rationally follow their predecessors, producing herding and cascades \citep{banerjee1992simple,bikhchandani1992theory}, bandwagons \citep{callander2007bandwagons}, and momentum \citep{knight2010momentum,morton2015exit}.

Similar dynamics can be observed in online, digital environments: social influence increases inequality and unpredictability in cultural markets~\citep{salganik_cultural_market_2006}, prior ratings causally affect later ones~\citep{muchnik2013social}, sequential voting creates herding~\citep{celis2016sequential}, and even limited social influence can undermine the wisdom of crowds~\citep{lorenz_2011}.
Recent DAO studies are consistent with this view: late voters are more likely to support leading alternatives~\citep{yaish2024strategic} and emerging work explicitly studies information cascades in DAO voting~\citep{buddensiek2025information}.

Reputation and authority provide another channel: online platforms often attach visible status signals to users, and reputation systems can affect trust and perceived quality~\citep{resnick2000reputation}.
Prior work has studied authority in social media and Q\&A platforms~\citep{pal2011identifying,paul2012authoritative}, linked user reputation to perceived contribution quality~\citep{tausczik2011predicting,liang2017knowledge}, and shown that reputation can affect persuasion~\citep{manzoor2024influence}.
The closest study to ours, \citet{dev2019quantifying}, jointly estimates reputation, social influence, and position biases on the StackExchange platform.
Because Snapshot ballots display the author's identity, their vote, and the running tally, the author's choice may thus act as an informative, persuasive, or focal signal for the community members. %
  
\vspace{0.2cm}

Taken together, these strands of literature establish that collective evaluations are shaped by choice order, visible prior behavior, and user-level authority signals.
Our contribution is to bring these channels into the DAO governance setting and quantify their associations at the proposal-choice level.
Specifically, we ask how much of the high support observed in Snapshot governance is associated with the stance of a choice, its position in the list, and whether it is selected by the proposal author.

\section{Data}
\label{sec:datacleaning}

The original dataset, spanning from October 2020 to November 2023, was obtained from Snapshot.org and contains all votes conducted through the platform, with information on each proposal like author, title, body, voting system, and ordered list of choices, and on each vote, like voter address, selected choice, and voting power~\citep{snapshot2023documentation}.
To validate it and ensure consistency, we perform several sanity checks, as discussed in Appendix~\ref{sec:app_data}. %
The cleaned dataset contains $\sim$51 million votes cast on $\sim$127,000 proposals. Table~\ref{tab:dataset_summary} reports summary statistics.

\subsection{Data Processing}

Before proceeding with the analysis, we homogenize two sources of heterogeneity across Snapshot proposals: the voting systems used to cast votes, and the formats used to express choices. Further details for each of the following subsections are reported in Appendix~\ref{sec:app_data}.

\subsubsection{Multi-preference vote expansion}

As Table~\ref{tab_voting_systems_catalog} shows, most proposals (93\%) use single-preference voting, where one vote maps to one choice; the remainder use multi-preference systems (e.g., weighted or quadratic voting), where a single vote can be split across several choices. To make these different voting systems comparable, we map multi-preference ballots to choice-level rows using two alternative schemes: \emph{favorite-choice}, which assigns each ballot's unit mass (i.e., the vote) to its top-weighted choice, and \emph{support-fraction}, which divides mass proportionally across all supported choices. Both approaches conserve each ballot's total mass and results are highly comparable. In the remainder of the paper, we rely primarily on the favorite-choice scheme (see Appendix \ref{appendix_processing_votes} for details).

\begin{table}[t]
	\caption{Raw and cleaned Snapshot dataset, October 2020--November 2023.}
	\label{tab:dataset_summary}
	\centering
	\begin{tabular*}{\textwidth}{@{\extracolsep{\fill}}l r r r r}
		\toprule
		& \textbf{Raw} & \textbf{Cleaned} & \textbf{Removed} & \textbf{Reduction} \\
		\midrule
		Spaces     & \DaoBiasRawSpaceN{}    & \DaoBiasFinalSpaceN{}     & \DaoBiasDatasetSpaceRemovedN{}    & \DaoBiasDatasetSpaceReductionPct{}\%    \\
		Proposals  & \DaoBiasRawProposalN{} & \DaoBiasFinalProposalN{}  & \DaoBiasDatasetProposalRemovedN{} & \DaoBiasDatasetProposalReductionPct{}\% \\
		Choices    & \DaoBiasRawChoiceN{}   & \DaoBiasFinalChoiceN{}    & \DaoBiasDatasetChoiceRemovedN{}   & \DaoBiasDatasetChoiceReductionPct{}\%   \\
		Votes      & \DaoBiasRawVoteN{}     & \DaoBiasFinalVoteMassN{}  & \DaoBiasDatasetVoteRemovedN{}     & \DaoBiasDatasetVoteReductionPct{}\%     \\
		\bottomrule
	\end{tabular*}
\end{table}

\begin{table}[tbp]
	\centering
	\footnotesize
	\caption{Snapshot voting systems and their prevalence in the analytic sample}
	\label{tab_voting_systems_catalog}
    \begin{tabularx}{\textwidth}{lXcc}
		\toprule
		\makecell{\textbf{Voting}\\\textbf{system}} &
		\makecell{\textbf{Voting system description}} &
		\makecell{\textbf{Proposals}\\\textbf{N (\%)}} &
		\makecell{\textbf{Choices}\\\textbf{N (\%)}} \\
		\midrule
		\multicolumn{4}{l}{\textit{Single-preference voting systems}} \\
		\specialrule{0.001pt}{2pt}{4pt}
		\quad Basic &
		Pick one from three choices: Yes, No, Abstain. &
		\DaoBiasVotingTypeBasicProposalN{} (\DaoBiasVotingTypeBasicProposalPct\%) &
		\DaoBiasVotingTypeBasicChoiceN{} (\DaoBiasVotingTypeBasicChoicePct\%) \\
		\quad Single-choice &
		Pick one from any number of choices. &
		\DaoBiasVotingTypeSingleChoiceProposalN{} (\DaoBiasVotingTypeSingleChoiceProposalPct\%) &
		\DaoBiasVotingTypeSingleChoiceChoiceN{} (\DaoBiasVotingTypeSingleChoiceChoicePct\%) \\
		\midrule
		\textbf{Subtotal} & &
		\textbf{\DaoBiasSinglePreferenceProposalN{} (\DaoBiasSinglePreferenceProposalPct\%)} &
		\textbf{\DaoBiasSinglePreferenceChoiceN{} (\DaoBiasSinglePreferenceChoicePct\%)} \\
		\midrule
		\multicolumn{4}{l}{\textit{Multi-preference voting systems}} \\
		\specialrule{0.001pt}{2pt}{4pt}
		\quad Weighted &
		Express both preference and intensity (summing to 100) for any number of choices. &
		\DaoBiasVotingTypeWeightedProposalN{} (\DaoBiasVotingTypeWeightedProposalPct\%) &
		\DaoBiasVotingTypeWeightedChoiceN{} (\DaoBiasVotingTypeWeightedChoicePct\%) \\
		\quad Quadratic &
		Express both preference and intensity for any number of choices. Voting power grows linearly, while intensity grows quadratically. &
		\DaoBiasVotingTypeQuadraticProposalN{} (\DaoBiasVotingTypeQuadraticProposalPct\%) &
		\DaoBiasVotingTypeQuadraticChoiceN{} (\DaoBiasVotingTypeQuadraticChoicePct\%) \\
		\quad Ranked-choice &
		Rank choices in order of preference. &
		\DaoBiasVotingTypeRankedChoiceProposalN{} (\DaoBiasVotingTypeRankedChoiceProposalPct\%) &
		\DaoBiasVotingTypeRankedChoiceChoiceN{} (\DaoBiasVotingTypeRankedChoiceChoicePct\%) \\
		\quad Approval &
		Approve any number of choices with equal weight. &
		\DaoBiasVotingTypeApprovalProposalN{} (\DaoBiasVotingTypeApprovalProposalPct\%) &
		\DaoBiasVotingTypeApprovalChoiceN{} (\DaoBiasVotingTypeApprovalChoicePct\%) \\
		\midrule
		\textbf{Subtotal} & &
		\textbf{\DaoBiasMultiPreferenceProposalN{} (\DaoBiasMultiPreferenceProposalPct\%)} &
		\textbf{\DaoBiasMultiPreferenceChoiceN{} (\DaoBiasMultiPreferenceChoicePct\%)} \\
		\bottomrule
	\end{tabularx}
    {\parbox{\linewidth}{\footnotesize \textit{Notes}: Percentages are computed over the \DaoBiasFinalProposalN{} proposals and \DaoBiasFinalChoiceN{} choices retained in the cleaned dataset. Voting systems are grouped into single-preference (one vote per ballot) and multi-preference (one ballot spread over several choices).}}
\end{table}

\vspace{-0.25cm} 

\subsubsection{Choice stance detection}

As discussed in Section~\ref{sec:background}, choices may express similar stances using different formats across proposals, ranging from simple binary responses to numbers, strings of text, and other arbitrary content. For instance, an approving choice can be formulated as a simple `Yes' text string, but also as a `thumbs up' emoji, or a longer text answer. We thus devise an approach to classify each choice into a simplified and comparable set of stances: \textit{Approve}, \textit{Reject}, \textit{Abstain}, or \textit{Other}. %
We implement this categorization approach using three different methods: (i) heuristics, our main tool in the analysis, (ii) large language models, and (iii) crowdworker evaluations. We use the latter two as validation tools. 

\paragraph{Heuristics.}
The pipeline has three stages. First, a keyword classifier assigns an initial stance to each choice by matching its text against a curated multi-language keyword list. Second, we manually review cases the classifier cannot reliably resolve (such as negated phrasing, abstention language, and choices represented through emojis) and adjust their labels accordingly. Third, we apply a small set of documented corrections at the level of the whole DAO space or the whole proposal to fix systematic mislabeling that the first two stages miss.
Finally, we enforce a rule that a governance proposal must contain at least one Approve and one Reject choice for either label to apply: proposals with only Approve (1,945) or only Reject (4,274) choices are relabeled as Other (see Appendix \ref{appendix_heuristics} for details).

\paragraph{Large language models (LLMs) evaluations.}
To assess the robustness of the heuristic classification, we validate it with external LLM evaluations of both open source and commercial models (Llama 3.1/3.3, GPT-4o, ChatGPT-4o). 
We instruct each LLM to categorize all proposals and related stances following the prompts reported in Appendix~\ref{sec_llm_prompts}. 
Notably, this allows us to also examine whether the body of a proposal implicitly or explicitly contains a cue pointing readers toward a voting direction.
We treat this \textit{suggested stance} as an LLM-coded textual cue in the proposal text, not as verified author intent or evidence that voters were persuaded.
We thus obtain two variables for each LLM that respectively capture whether and in which direction the text suggests a vote, and classify each choice as `Abstain', `Approve', `Other', or `Reject'.

To measure the agreement across the two classification methods, we compute Cohen's Kappa between our heuristics and each LLM (using model-specific denominators so Llama 3.3's sparser coverage is handled consistently). The values are \DaoBiasLlmKappaLlama{} for Llama 3.1, \DaoBiasLlmKappaLlamaThreePointThree{} for Llama 3.3, \DaoBiasLlmKappaOai{} for OAI, and \DaoBiasLlmKappaChatgpt{} for ChatGPT, indicating \DaoBiasLlmKappaLlamaAgreement{}, \DaoBiasLlmKappaLlamaThreePointThreeAgreement{}, \DaoBiasLlmKappaOaiAgreement{}, and \DaoBiasLlmKappaChatgptAgreement{} agreement, respectively. \DaoBiasLlmKappaHighestSource{} has the highest agreement with the heuristic-based approach (see Appendix \ref{appendix_llm_eval} for details).

\paragraph{Human evaluations.}
We further validate the heuristic approach by measuring the extent to which it agrees with external human evaluations; this procedure is implemented with the platform  NodeGame~\citep{balietti_nodegame_2017}. 

First, we create a stratified multilingual sample of \DaoBiasHumanEvaluatedProposalN{} representative random proposals by computing the embeddings of each proposal, reducing their dimensionality using UMAP, and clustering them with the K-means algorithm. 
Next, using the online labor market for research Prolific, we assign a group of evaluators (N = \DaoBiasHumanEvaluatorN{}) the task of manually labeling five proposals, so that each proposal is reviewed at least \DaoBiasHumanProposalReviewMinN{} times.
Finally, we adopt a majority rule approach to aggregate the evaluations provided by different crowdworkers and assign each stance a unique label (`Abstain', `Approve', `Other', or `Reject'). In case of a tie, we leave the label unassigned, and we fill in unassigned stances when the label can be inferred from the remaining choices of the same proposal.

We compare the heuristic-based classification and the crowdworkers' evaluations: the Cohen's Kappa is \DaoBiasHumanKappa{}, which indicates \DaoBiasHumanKappaAgreement{} agreement. %
Further information on crowdworkers' demographics and the aggregation of their evaluations is reported in Appendix \ref{appendix_human_eval}. 

The code to replicate all analyses is available at \url{https://github.com/rand176/dao-voting-biases}. Task instructions and the anonymized crowdworker labels are available at \url{https://osf.io/sbmkd}.

\section{Biases in Decentralized Governance}
\label{sec:desc_analysis}

This section documents how voting power is distributed across the choices of a proposal, focusing on three features of a Snapshot ballot that correspond to the candidate biases motivating this study: the order in which choices are listed, the stance each choice expresses, and whether the proposal author backs it. Consistent with the introduction, we use \textit{bias} to denote a systematic association with voting-power share, not a causal effect. We recall each bias and define the corresponding variables, each operationalized as a binary indicator at the choice level:

\begin{itemize}
	
	\item \textit{Position bias}: choices listed earlier may attract more support because of order and interface effects, as documented in survey response-order and ballot-order research. We define a choice as \textit{first-listed} if it occupies position one in the ordered list of choices shown to voters. %
	
	\item \textit{Approval bias}: choices that approve the proposal may be favored, either because proposals are discussed and filtered before reaching a formal vote, or because approving is perceived as cooperative. We define a choice as \textit{approval-oriented} if its stance is classified as Approve by the procedure described in Section~\ref{sec:datacleaning}.
	
	\item \textit{Author bias}: the choice backed by the proposal author may attract support through the author's own tokens, through deference of less-informed voters to a better-informed proposer, or through a bandwagon triggered by the author's visible vote. We define a choice as \textit{author-selected} if the address that created the proposal cast a vote for that choice.
	
\end{itemize}

The outcome of a proposal is determined by voting power: the choice with the highest voting power is implemented. 
Operationally, it is a weighted sum of individual votes per choice; DAOs are free to choose how the weights are computed, but most assign them in proportion to the governance tokens held by each voter.\footnote{That means that, in principle, a single contrarian vote from somebody with a large weight could significantly shift the distribution of voting power across the choices of a proposal. However, as shown in Table \ref{table_stats_vp_n}, this is rarely the case: voting power and individual votes are highly correlated (95\%); hence, in the following, we only focus on voting power, but the results apply to both.}
To compare proposal outcomes regardless of how voting power is computed, in all our analyses we focus on two normalized measures computed at the choice level:

\begin{itemize}
	
	\item \textit{Share}: the voting power of a choice divided by the total amount of voting power of all choices in a proposal; e.g., in a proposal with two choices with voting power 10 and 20, the total voting power is 30, and the share of the two choices is respectively 1/3 and 2/3.
	
	\item \textit{Excess / deficit}: the difference between the actual share and a uniformly random share; e.g., in a proposal with four choices, the random share is 0.25; choice $A$, with a share of 0.6, would have an excess voting power of $0.6-0.25=0.35$, while choice $B$ with a share of 0.1 would have a deficit of $-0.15$.
	
\end{itemize}

The uniform share serves as the no-bias benchmark: absent any systematic advantage, a choice would on average receive zero excess.

\begin{figure}[t]
	\centering
	\includegraphics[width=1\linewidth]{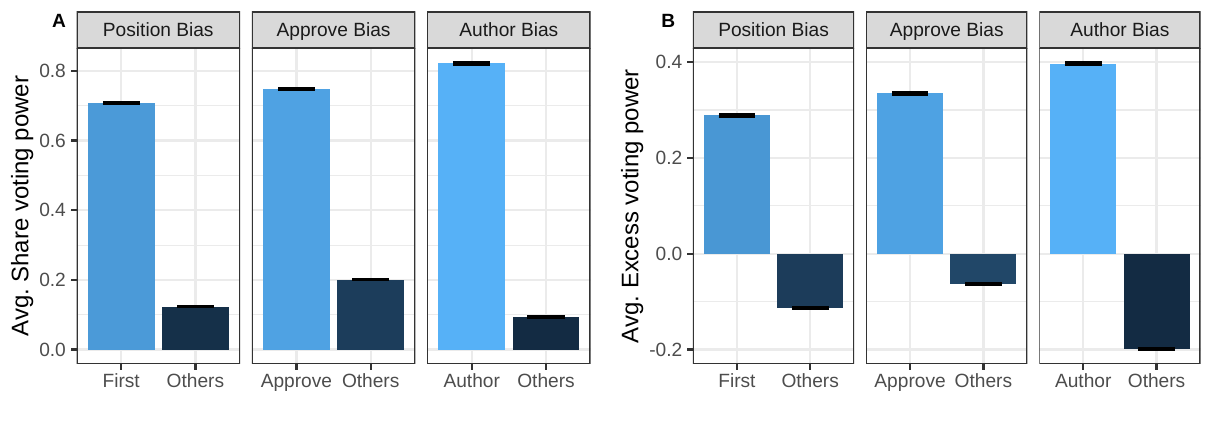}
	\caption{\textbf{Share and excess of choice-level voting power for the three indicators under study}. \textbf{A.} Outcome: mean voting-power share for choices that are first-listed, approval-oriented, or selected by the proposal author. \textbf{B.} Outcome: mean excess voting power relative to a uniform-within-proposal benchmark. Error bars show 95\% confidence intervals for mean estimates. Across both measures, the choice associated with each bias captures far more voting power than the alternative ones, indicating that each channel is associated with a substantial, non-random concentration of support.}
	\label{fig:three_biases_teaser}
\end{figure}

Fig.~\ref{fig:three_biases_teaser} documents the relationship between voting-power share and excess/deficit for the three biases under study and provides a first overview across all governance proposals. 
The three indicators are not mutually exclusive: the same choice can be first-listed, approval-oriented, and author-selected at once, and in practice these features frequently coincide.\footnote{
	Note that in this figure, as well as in the following, error bars represent 95\% confidence intervals of the mean estimates. 
	Share does not sum to one and excess does not sum to zero in this figure because proposals with different numbers of choices are pooled together; both quantities sum to their respective totals (one and zero) only within proposals sharing the same number of choices, a comparison we hold fixed in the sections that follow.
}
The descriptive evidence below therefore documents each association separately; %
then, Section~\ref{sec:reg_analyses} disentangles them within a regression framework.

\subsection{Position Bias}

We find a highly imbalanced distribution of voting power across choices, with the first choice seizing a disproportionate share of voting power: 70\% across all proposals (see Table \ref{table_stats_vp_n}) with a peak of 79\% for proposals of type Basic (see Table \ref{table_stats_3biases_bytype_firstchoice}). This pattern is consistent with strong order effects.

\begin{figure}[h]
    \centering
    \begin{subfigure}[t]{0.41\linewidth}
        \centering
        \includegraphics[width=\linewidth]{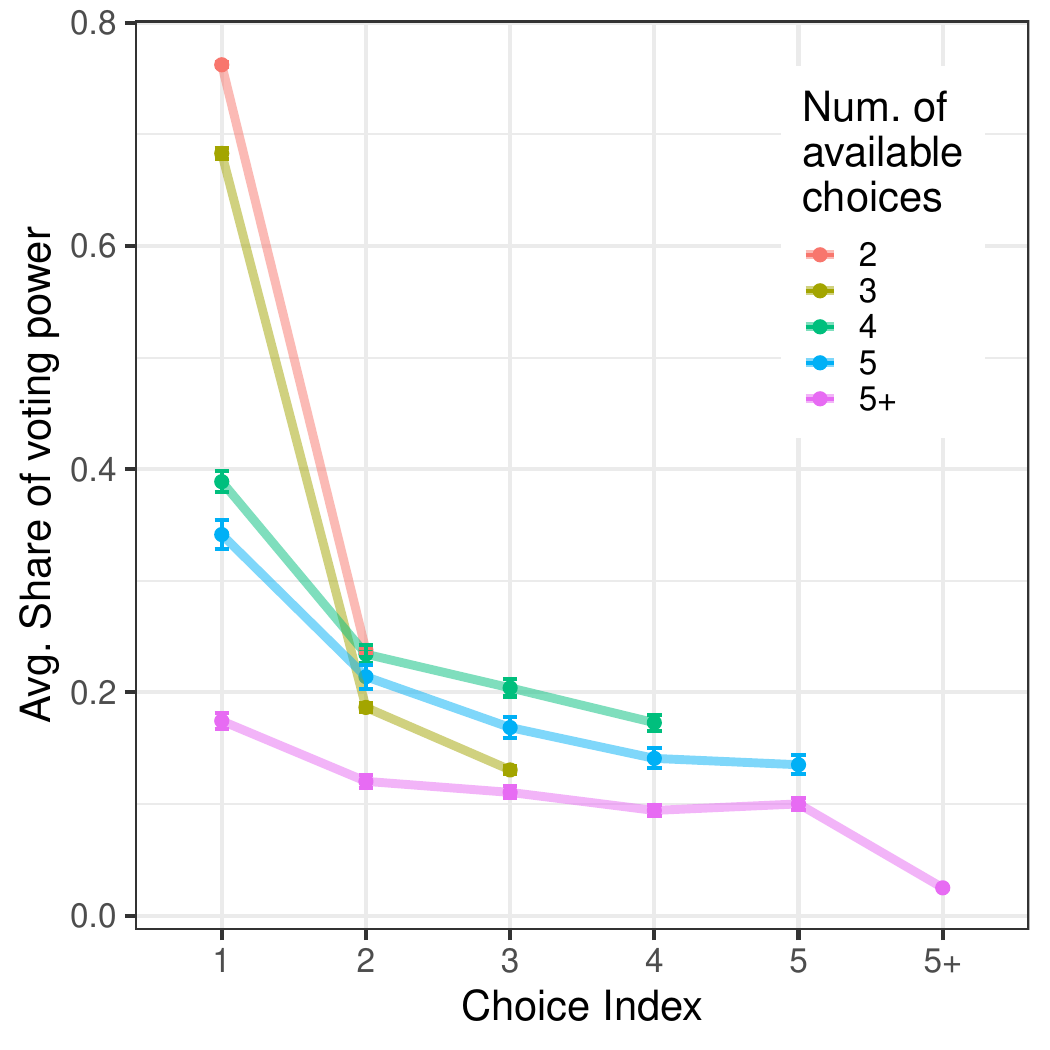}
        \caption{Voting-power share.}
    \end{subfigure}
    \hfill
    \begin{subfigure}[t]{0.57\linewidth}
        \centering
        \includegraphics[width=\linewidth]{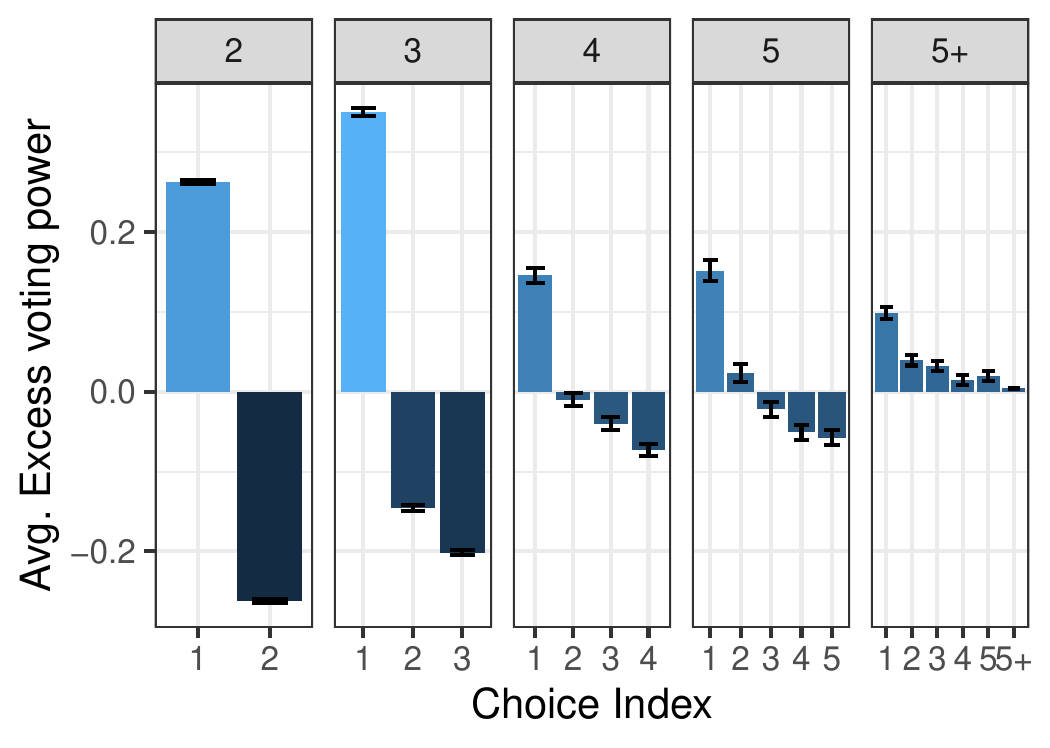}
        \caption{Excess voting power.}
    \end{subfigure}
    \caption{\textbf{Choice position and voting power by number of choices in a proposal.} Mean voting-power share (\textbf{A}) and mean excess voting power (\textbf{B}) by choice position and number of choices in the proposal. Excess is  relative to a uniform-within-proposal benchmark by choice position and number of choices. Error bars show 95\% confidence intervals for mean estimates.
    The positional advantage is higher in two- and three-choice proposals, but it can be observed across all ballot lengths.} 
    \label{fig:pos_bias}
\end{figure}

Fig. \ref{fig:pos_bias}A shows how the share of voting power changes across proposals with different numbers of choices. The association is strongest on proposals with two and three choices; in the latter, the first choice captures around 80\% of total voting power. These proposals often follow an approve/reject pattern, as in the Basic voting system, which always has exactly three choices. Fig. \ref{fig:pos_bias}B shows the excess voting power for proposals with different numbers of choices: the extra voting share captured by the first choice is drawn proportionally from choices with higher indexes. To quantify this pattern, however, we also need to account for stance and author selection, as we do in the next sections.

\subsection{Approval Bias}

\begin{figure}[H]
    \centering
    \includegraphics[width=0.8\linewidth]{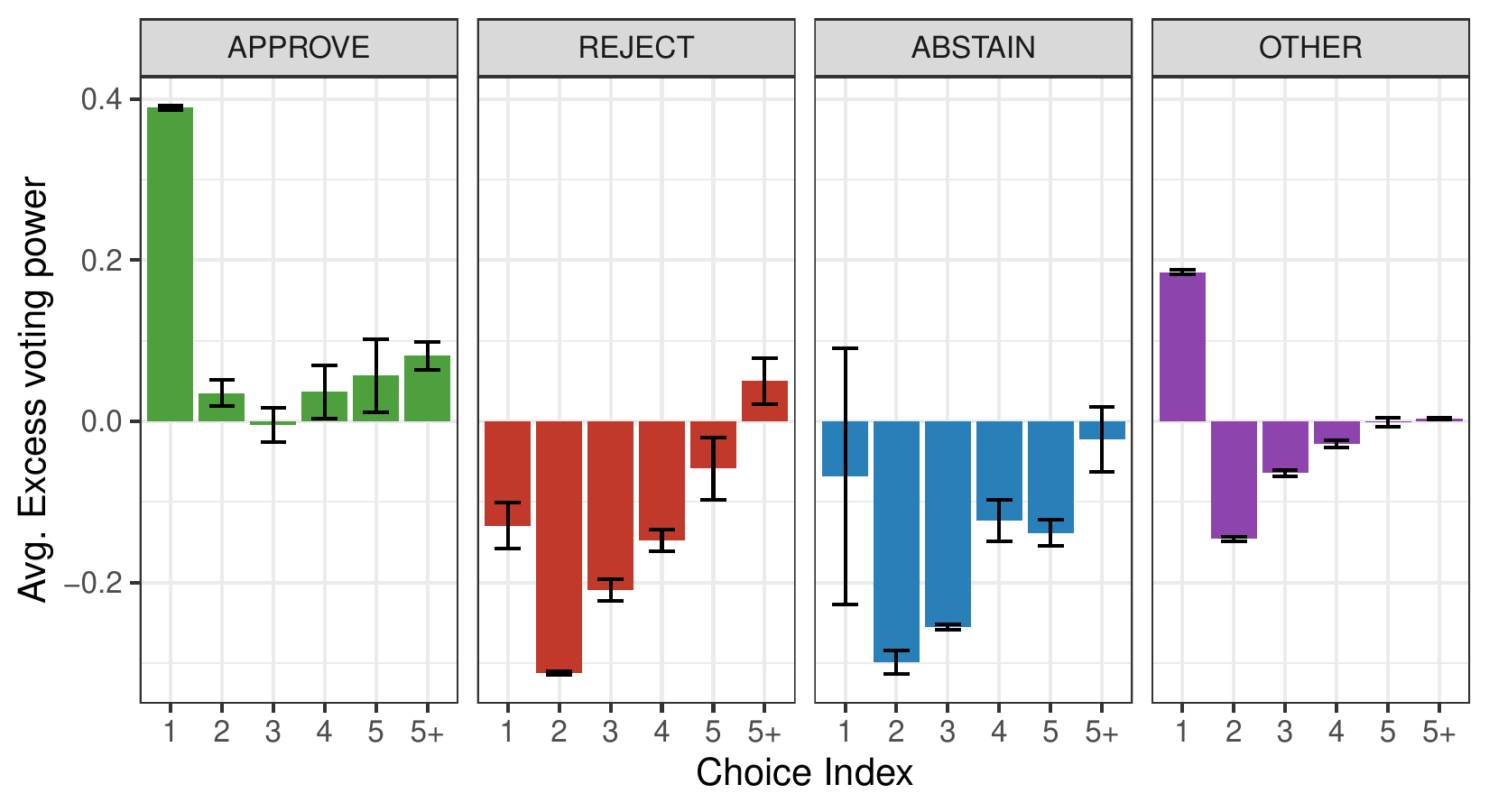}
    \caption{\textbf{Choice stance, list position, and excess voting power.} Excess voting power is plotted separately for each stance and by choice position. Excess is relative to a uniform-within-proposal benchmark. Error bars show 95\% confidence intervals for mean estimates. All stances obtain a boost in voting power when placed in position 1, but the Approve stance obtains the largest gain and has limited deficit at other positions. }
    \label{fig:avg_excess_vp_choice_stance}
\end{figure}

The stance of a choice---i.e., whether it is approving or rejecting the proposal---may influence its voting share. In fact, choices with an Approve stance overwhelmingly capture the largest share of voting power across all proposals (about 80\%, see Fig. \ref{fig:approve_bias_app}A). However, this dominance is not evenly distributed across all choice positions, but predominantly driven by a large excess of voting power in the first position (see Fig. \ref{fig:avg_excess_vp_choice_stance}). All stances get boosted for being placed in the first position, but their excess voting power is much more modest. Overall, at every position the Approve stance has a positive delta compared to other stances, and the same pattern can be observed disaggregating proposals by number of choices (see Fig. \ref{fig:avg_excess_vp_choice_stance_by_num_choices}).

To study the interaction between position and approval-oriented stance, we need to look at how often the Approve stance is placed in position one versus other positions. Fig. \ref{fig:approve_bias_app}B and Table \ref{table_stats_stance_choice_idx} show that Approve choices are much more likely to occupy the first choice for proposals with two or three choices, which are more often associated with approve/reject proposals. There are at least two explanations for this observation. First, by design, the popular Basic voting system places Approve choices in position one. Second, proposers may place approval-oriented choices first. To better understand this point, we next examine author selection.

\subsection{Author Bias}

\begin{figure}[H]
	\centering
	\includegraphics[width=0.7\linewidth]{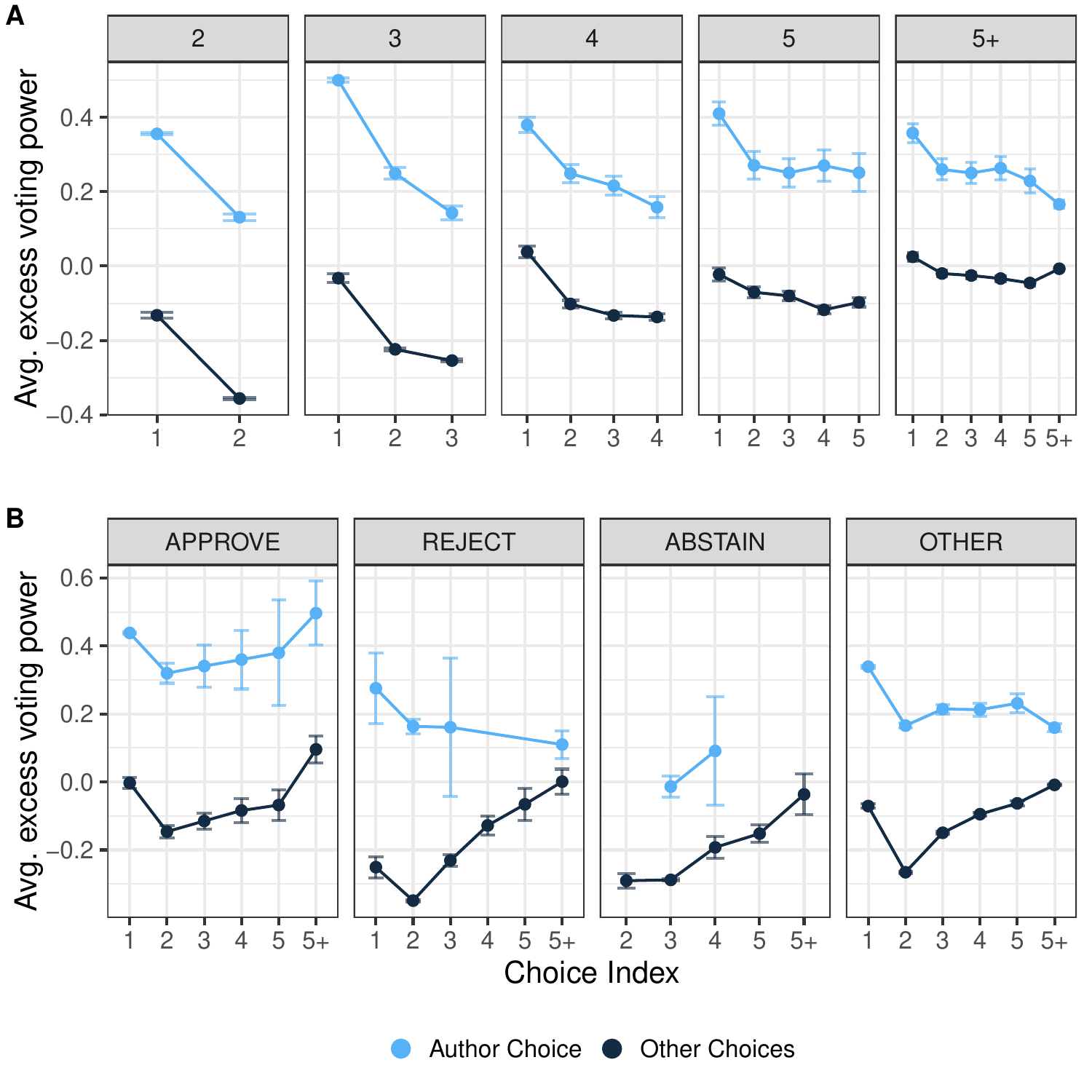}
	\caption{\textbf{Author-selected choices and excess voting power.} \textbf{A.} Excess voting power by number of choices, choice position, and whether the proposal author selected the choice. \textbf{B.} Excess voting power by stance, choice position, and whether the proposer selected the choice; data points averaging 10 observations or fewer are excluded. Excess is relative to a uniform-within-proposal benchmark. Error bars are 95\% confidence intervals of the means. In all cases, author-selected choices receive larger voting power compared to each alternative outcome.}
	\label{fig:author_bias}
\end{figure}

Proposal authors might influence outcomes with their own choices. Fig. \ref{fig:author_bias}A shows that the choice selected by the author receives the largest amount of voting power for any choice position and for any number of choices. This boost is larger for the first choice than for other choices, suggesting that the author-choice association is larger than the position association, which still plays an important role.
Fig. \ref{fig:author_bias}B compares approval stance and author selection, highlighting two descriptive patterns. First, even when disaggregated by stance, author-selected choices show positive excess voting power across positions, although the size varies. Second, the author boost is stable across all choice positions only for the Approve stance, while it is decreasing for Reject and Other stances (see also Fig. \ref{fig:author_boost_by_stance_and_pos}); interestingly, the highest boost overall is for the Reject choices at positions two and three. 

Overall, the results in this section suggest a nested descriptive pattern: author-selected choices have the largest voting-power advantage, being at the first position also generates gains, and the effect of approval-oriented choices is partly intertwined with author selection and first placement. In the next section, we evaluate this pattern with an econometric framework.

\section{Disentangling Biases: A Regression Approach}
\label{sec:reg_analyses}

The descriptive results show that the three candidate biases are tightly entangled.
Table~\ref{table_stats_3biases} makes this overlap explicit: first-position choices receive most voting power, but first-position choices are also much more likely to be approving choices and author-selected choices.
Raw averages therefore cannot tell whether an author-selected choice attracts voting power because it is first, because it approves the proposal, because it was selected by the author, or because these indicators co-occur.
We use regression models to separate these conditional associations while preserving the observational interpretation of the analysis.

\begin{table}[!tbp]
\centering
\caption{Overlap among position, approval, and author-choice indicators}
\label{table_stats_3biases}
\small
\setlength{\tabcolsep}{3pt}
\renewcommand{\arraystretch}{1.08}
\begin{tabular*}{\linewidth}{@{\extracolsep{\fill}}lrrrrrrrr@{}}
\toprule
& \multicolumn{2}{c}{Voting power} & \multicolumn{2}{c}{Approve} & \multicolumn{2}{c}{Author choice} & \multicolumn{2}{c}{All choices} \\
\cmidrule(lr){2-3}\cmidrule(lr){4-5}\cmidrule(lr){6-7}\cmidrule(l){8-9}
Choice rank & Mean & SD & Mean & SD & Mean & SD & Share (\%) & N \\
\midrule
\multicolumn{9}{@{}l}{\textit{Panel A: all proposals}} \\
1 & \num{0.70} & \num{0.39} & \num{0.40} & \num{0.49} & \num{0.73} & \num{0.44} & \num{31.96} & 127,281 \\
2 & \num{0.22} & \num{0.35} & \num{0.01} & \num{0.12} & \num{0.21} & \num{0.41} & \num{31.96} & 127,281 \\
3 & \num{0.14} & \num{0.26} & \num{0.02} & \num{0.15} & \num{0.16} & \num{0.37} & \num{9.87} & 39,303 \\
4 & \num{0.14} & \num{0.24} & \num{0.03} & \num{0.16} & \num{0.20} & \num{0.40} & \num{3.48} & 13,839 \\
5 & \num{0.11} & \num{0.22} & \num{0.02} & \num{0.14} & \num{0.18} & \num{0.38} & \num{2.04} & 8,130 \\
6 & \num{0.09} & \num{0.18} & \num{0.01} & \num{0.11} & \num{0.16} & \num{0.37} & \num{1.28} & 5,110 \\
7 & \num{0.07} & \num{0.15} & \num{0.02} & \num{0.13} & \num{0.18} & \num{0.39} & \num{0.97} & 3,853 \\
8 & \num{0.06} & \num{0.14} & \num{0.01} & \num{0.10} & \num{0.18} & \num{0.39} & \num{0.81} & 3,209 \\
9 & \num{0.05} & \num{0.13} & \num{0.01} & \num{0.11} & \num{0.17} & \num{0.38} & \num{0.69} & 2,766 \\
10 & \num{0.05} & \num{0.12} & \num{0.01} & \num{0.10} & \num{0.17} & \num{0.37} & \num{0.62} & 2,482 \\
10+ & \num{0.01} & \num{0.06} & \num{0.01} & \num{0.09} & \num{0.09} & \num{0.29} & \num{16.31} & 64,964 \\
\addlinespace[0.35em]
\multicolumn{9}{@{}l}{\textit{Panel B: proposals with up to three choices}} \\
1 & \num{0.74} & \num{0.37} & \num{0.44} & \num{0.50} & \num{0.77} & \num{0.42} & \num{44.95} & 113,442 \\
2 & \num{0.23} & \num{0.35} & \num{0.01} & \num{0.09} & \num{0.21} & \num{0.41} & \num{44.95} & 113,442 \\
3 & \num{0.13} & \num{0.27} & \num{0.01} & \num{0.08} & \num{0.13} & \num{0.33} & \num{10.09} & 25,464 \\
\bottomrule
\end{tabular*}
{\parbox{\linewidth}{\footnotesize \textit{Notes}:  Mean voting-power share, approval indicator, and author-choice indicator by choice rank. Panel A uses all proposals in the final cleaned dataset; %
Panel B restricts the same statistics to proposals with up to three choices. Author-choice means omit rows without reconstructed author-choice information.}}
\end{table}

\subsection{Model Design and Samples}

\paragraph{Model}The unit of analysis are the proposal's choices, and the outcome is the voting-power share of each choice (bounded between zero and one). We estimate fractional-logit models with a quasibinomial logit link, which models fractional outcomes without converting the voting-power share into a binary outcome. The reported coefficient estimates are on the logit scale; substantive interpretation therefore relies on the response-scale average marginal effects (AMEs) reported in the robustness tables.

\paragraph{Specification} We control for the three bias indicators (author selection, approving stance, choice position) plus a set of covariates: relative proposal date, number of choices, author-choice visibility among the first six interface positions,\footnote{Six is a constraint of the user interface of Snapshot.} relative choice-text length, and voting system in use. In addition, because the proposer may or may not have voted, we include an indicator that is equal to one if the proposal author voted, and zero otherwise. The main fixed effect is the DAO space, which absorbs persistent differences across governance communities (robustness checks also test proposal fixed effects to compare choices within the same proposal). Standard errors are clustered by space throughout because choices and proposals from the same DAO are not independent observations.

The primary regression sample uses choices from all single-preference proposals, where the mapping from a ballot to a choice is most direct. 
We also use alternative sample specifications and model variations to verify the robustness of our findings. Table~\ref{tab:robustness-variation-key} defines our experiments, the short names used to identify them in the robustness tables, and reports the fitted row count for each of them.

\begin{table}[!tbp]
\centering
\caption{Robustness specification key}
\label{tab:robustness-variation-key}
\small
\setlength{\tabcolsep}{3pt}
\renewcommand{\arraystretch}{1.10}
\begin{tabularx}{\linewidth}{@{}r>{\RaggedRight\arraybackslash}p{0.12\linewidth}>{\RaggedRight\arraybackslash}p{0.14\linewidth}>{\RaggedRight\arraybackslash}Xr@{}}
\toprule
\# & Short name & Dimension & Definition & N \\
\midrule
1 & Main & Baseline & Single-preference proposals: single-choice and basic voting systems. & 293,816 \\
\midrule
2 & Vote-count & Outcome & Vote-count share (number of votes divided by total proposal votes) as outcome. & 293,816 \\
\midrule
3 & Auth-voted & Participation & Restricted to proposals where the author voted. & 172,171 \\
4 & Low-part. & Participation & Excludes proposals with fewer than two votes. & 221,774 \\
5 & Vote-control & Participation & Adds log total proposal votes. & 293,816 \\
\midrule
6 & TVL-25 & TVL & Restricted to Top 25 Ethereum-TVL DAOs matched against DefiLlama ranking. & 11,410 \\
7 & TVL-50 & TVL & Restricted to Top 50 Ethereum-TVL DAOs matched against DefiLlama ranking. & 14,393 \\
8 & TVL-100 & TVL & Restricted to Top 100 Ethereum-TVL DAOs matched against DefiLlama ranking. & 15,468 \\
\midrule
9 & Proposal FE & Fixed effects & Uses proposal fixed effects. & 293,816 \\
\midrule
10 & No top-5 & Space influence & Excludes the five largest contributing spaces. & 272,743 \\
\midrule
\midrule
11 & Expanded & Proposal type & Includes approval and ranked-choice voting systems. & 308,582 \\
12 & Full & Proposal type & Includes all voting systems, including weighted and quadratic. & 398,216 \\

\bottomrule
\end{tabularx}
\parbox{\linewidth}{\footnotesize \textit{Notes}: Variations 2-10 are variations of the main single-preferences sample; variations 11-12 expand the sample to multi-preference proposals. The TVL samples contain the DAOs matched against the Top 100 Decentralized Finance (DeFi) protocols ranked by DefiLlama (TVL refers to the median daily Ethereum total value locked by DeFi protocols over the study period); accordingly, TVL-100 contains less than 100 DAOs because not every protocol can be matched (see Appendix \ref{sec:app:defillama} for details).} %
\end{table}

\subsection{Results}

{\small
\begin{table}[!tbp]
\centering
\caption{Nested fractional-logit specifications for the main single-preference sample}
\label{tab:robustness-nested-single-preference}
\small
\setlength{\tabcolsep}{4pt}
\renewcommand{\arraystretch}{1.10}
\begin{tabularx}{\linewidth}{@{}>{\RaggedRight\arraybackslash}Xrrrrrr@{}}
\toprule
& (1) & (2) & (3) & (4) & (5) & (6) \\
\midrule
Author choice & \makecell[r]{3.347***\\(0.086)} & \makecell[r]{3.347***\\(0.086)} & \makecell[r]{3.348***\\(0.087)} & \makecell[r]{3.348***\\(0.087)} & \makecell[r]{3.352***\\(0.086)} & \makecell[r]{2.073***\\(0.088)} \\
Approving choice & \makecell[r]{1.750***\\(0.141)} & \makecell[r]{1.750***\\(0.141)} & \makecell[r]{1.754***\\(0.140)} & \makecell[r]{1.754***\\(0.140)} & \makecell[r]{1.751***\\(0.139)} & \makecell[r]{1.768***\\(0.139)} \\
Choice rank & \makecell[r]{-0.529***\\(0.038)} & \makecell[r]{-0.528***\\(0.038)} & \makecell[r]{-0.517***\\(0.039)} & \makecell[r]{-0.517***\\(0.039)} & \makecell[r]{-0.511***\\(0.039)} & \makecell[r]{-0.510***\\(0.039)} \\
Choice rank squared & \makecell[r]{0.003***\\(0.000)} & \makecell[r]{0.003***\\(0.000)} & \makecell[r]{0.003***\\(0.000)} & \makecell[r]{0.003***\\(0.000)} & \makecell[r]{0.003***\\(0.000)} & \makecell[r]{0.003***\\(0.000)} \\
\hline
Author voted & \makecell[r]{-1.546***\\(0.039)} & \makecell[r]{-1.546***\\(0.039)} & \makecell[r]{-1.546***\\(0.039)} & \makecell[r]{-1.524***\\(0.037)} & \makecell[r]{-1.526***\\(0.037)} & \makecell[r]{-0.949***\\(0.039)} \\
Relative proposal date &  & \makecell[r]{-0.000\\(0.000)} & \makecell[r]{-0.000\\(0.000)} & \makecell[r]{-0.000\\(0.000)} & \makecell[r]{-0.000\\(0.000)} & \makecell[r]{-0.000\\(0.000)} \\
Number of choices &  &  & \makecell[r]{-0.029**\\(0.009)} & \makecell[r]{-0.029**\\(0.009)} & \makecell[r]{-0.030**\\(0.009)} & \makecell[r]{-0.030**\\(0.009)} \\
Author choice in first six &  &  &  & \makecell[r]{-0.026+\\(0.013)} & \makecell[r]{-0.026+\\(0.013)} & \makecell[r]{-0.667***\\(0.045)} \\
Author choice x first-six visibility &  &  &  &  &  & \makecell[r]{1.412***\\(0.091)} \\
Relative choice text length &  &  &  &  & \makecell[r]{-0.118*\\(0.059)} & \makecell[r]{-0.122*\\(0.059)} \\
\hline
Voting-type controls & Yes & Yes & Yes & Yes & Yes & Yes \\
N & 293816 & 293816 & 293816 & 293816 & 293816 & 293816 \\
\bottomrule
\end{tabularx}
\parbox{\linewidth}{\footnotesize \textit{Notes}: Coefficient estimates are on the logit scale with DAO-space clustered standard errors in parentheses. Significance markers: + $p<0.1$, * $p<0.05$, ** $p<0.01$, *** $p<0.001$.}
\end{table}

}

\begin{figure}[tbp]
	\centering
	\includegraphics[width=\linewidth]{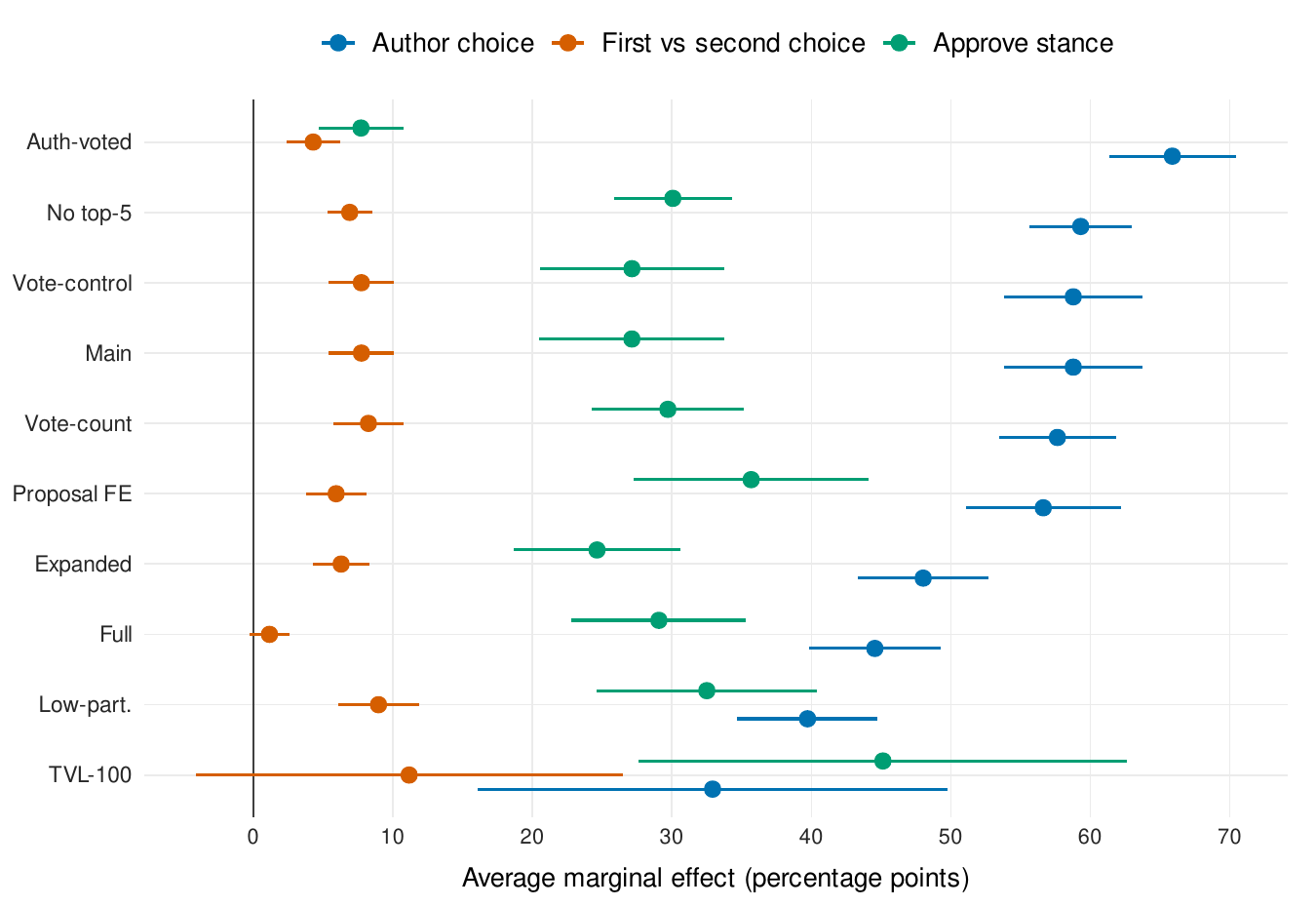}
	\caption{\textbf{Three-bias average marginal effects (AMEs) across main sample and robustness specifications}. Points show response-scale AMEs in percentage points; horizontal intervals are Bonferroni FWER-adjusted across the 36 shared AMEs. \textit{Main} corresponds to the primary specification; others are robustness checks. Across all robustness specifications except the Top-TVL subsample, the author-choice AME exceeds both the position effect (first vs. second choice) and the approval-stance effect; in the Top-TVL case, the difference is not statistically significant. Author-choice point estimates remain positive across specifications, while approval-stance point estimates are larger in the Top-TVL subsample. \textit{TVL-25} and \textit{TVL-50} rows omitted for readability; estimates are comparable to \textit{TVL-100}, see Table \ref{tab:robustness-three-bias-summary} in the Appendix for details.
	}
	\label{fig:three-bias-forest}
\end{figure}

Table~\ref{tab:robustness-nested-single-preference} reports nested fractional-logit specifications for the main single-preference sample. Across all six specifications, the author-choice coefficient remains stable and is the strongest predictor of voting-power share among the three biases, followed by approving stance and choice rank.
Because these coefficients are estimated on the logit scale, we turn to average marginal effects (AMEs) for their substantive interpretation: AMEs express how many percentage points a choice's voting-power share changes, on average, when the corresponding indicator varies, holding other covariates fixed.
Figure~\ref{fig:three-bias-forest} reports these AMEs for our Main specification alongside all robustness checks. In the Main specification, an author-selected choice gains 58.8 percentage points of voting-power share relative to an otherwise similar choice, compared with 27.1 points for having an approving stance, and 7.7 points for being listed first rather than second. This ordering holds across the shared robustness specifications except the Top-TVL subsample, where the approval-stance AME exceeds the author-choice AME.

To better understand the role of author bias, we conducted an additional diagnostic test in which we recomputed voting-power shares using only non-author votes, while also controlling for author-choice positioning and approving-choice status. For this purpose, we derived a new sample from \textit{Main}, keeping only proposals in which the author voted, at least one accepted non-author vote had positive voting power, and total non-author voting power was positive.
We report three specifications on this sample: (i) Pooled, which includes all selected proposal choices; (ii) Author $\times$ Position, which splits the sample into `author choice in first position' vs. `author choice elsewhere'; and (iii) Author $\times$ Stance, which splits it into `author choice with Approve stance' vs. `author choice with any other stance'.

\begin{table}[!tbp]
\centering
\caption{Author-specific diagnostics for author-choice AMEs}
\label{tab:robustness-author-choice-diagnostics}
\footnotesize
\setlength{\tabcolsep}{2.5pt}
\renewcommand{\arraystretch}{1.10}
\begin{tabularx}{\linewidth}{@{}>{\RaggedRight\arraybackslash}X>{\RaggedRight\arraybackslash}p{0.18\linewidth}r|rr|rr}
\toprule
 & & & \multicolumn{2}{c|}{Author-Position Split} & \multicolumn{2}{c}{Author-Stance Split} \\
Author's VP & N & Pooled & First pos. & Non-first & Approve & Non-Approve \\
\midrule
 Included & 118,483 & 45.62 & 57.52 & 33.99 & 55.54 & 43.89 \\
 Removed & 118,483 & 29.68 & 42.41 & 17.65 & 45.41 & 27.60 \\
\bottomrule
\end{tabularx}
\parbox{\linewidth}{\footnotesize \textit{Notes}: the sample includes proposals from \textit{Main} in which the author voted, with at least one accepted positive-power non-author vote and positive total non-author voting power. Both rows include all choices from the same proposals. Removed shares are recomputed directly from non-author votes. Cells report response-scale author-choice AMEs in percentage points, averaged equally across Basic and Single-choice voting systems.}
\end{table}

Table \ref{tab:robustness-author-choice-diagnostics} reports the results. In the top row, we report the AME values when the author's voting power is included; in the bottom row, we report the AME values when the author's voting power is excluded.
On the pooled sample, the observed author-choice AME is 45.62 percentage points and falls to 29.68 percentage points after subtracting the author's own support, which suggests that the author-choice pattern is not mechanically reducible to authors voting for their own choices.
The position split shows that the association is present for both first-position and non-first author choices, albeit weaker. The approve split likewise shows positive author-choice AMEs for both approving and non-approving author choices, with the latter showing a weaker association as expected.

One concern is that proposal authors may select choices that have already attracted substantial support. To investigate this possibility, we examine the timing of authors' votes relative to those of other participants. Among proposals with at least ten voters (N = 24,191), authors cast the first recorded vote in 54.2\% of cases and vote within the first decile of the voting sequence in 64.9\% of cases. In 83.5\% of proposals, authors vote before half of the other participants have cast their votes; at the time authors vote, the median share of voting power already cast by other voters is zero (more details in Appendix \ref{asec:author_vote_timing}). These findings suggest that the author-choice advantage cannot be explained simply by authors following earlier votes. However, they do not rule out the possibility that authors already know which choices others prefer, for example through discussions or coordination before voting begins.

Taken together, the regression evidence refines the descriptive pattern from Section~\ref{sec:desc_analysis}. This evidence remains a descriptive author-choice association, and not a causal claim about persuasion or manipulation. Yet, across all specifications, author-selected choices show a large and robust conditional association with voting-power shares, with the association remaining visible even after separating author participation from author choice, and after removing author voting power. 

\section{Discussion} %
\label{sec:discussion}

Our results show that voting power in DAO governance is systematically associated with three features of a ballot choice: its position in the list, its stance, and whether the proposal author backs it.
In this section, we interpret these associations and discuss implications for the implementation of DAOs and limits to our work.

\subsection{Interpreting the Biases}

Of the three conditional associations, the author advantage is the largest and most robust. 
The mechanical support that authors provide by casting their vote for their own choice explains part, but not most of the pattern: removing the author's voting power lowers the author advantage from 45.6 to 29.7 percentage points in the sample in which authors voted and other voters contributed positive voting power. The remaining two thirds come from other voters. The advantage remains nearly unchanged when the outcome is the share of votes rather than of voting power (57.6 percentage points), so it is not driven only by a few large holders.
What remains is consistent with other voters following the author's lead, for instance because the author's visible vote steers later voters toward the same option~\citep{muchnik2013social}, %
or because less-informed voters treat the author's choice as an informative cue about a proposal they have not examined in detail~\citep{feddersen1996swing}.  
In the latter case, the author advantage should be larger where voters have weaker incentives to acquire information. This is what we observe: the advantage is roughly halved in the largest DeFi protocols, where stakes are high and participation is broader, and it drops to 39.7 percentage points once proposals with very low participation are excluded. This pattern parallels the finding that ballot-order effects are larger in low-information races \citep{koppell2004effects}. %

The approval advantage behaves differently. It falls from 27.1 to 7.7 percentage points once we condition on the author having voted, indicating that much of what appears as a general preference for approving choices moves together with the author's own support.
At the same time, it is the largest of the three associations in the top-TVL protocols (45--52 percentage points). Both facts are consistent with a selection process that takes place before the vote: in large protocols, proposals are typically discussed in forums and tested informally before being posted on Snapshot, so those that reach a formal vote are mostly those expected to pass~\citep{faqir2021comparative,messias2024understanding}. %

Position, finally, plays a smaller and mostly indirect role.
On its own, being listed first is associated with an advantage of 7.7 percentage points, and of 4.3 percentage points in proposals where the author voted.
Position matters more when it interacts with author selection: after removing the author's own voting power, the author advantage is more than twice as large when the author's choice is listed first (42.4 percentage points) as when it is not (17.6 percentage points). Because proposers decide which options appear on the ballot and in which order, the ordering of choices is not neutral: it is one of the ways in which the author's preferred option is made more prominent to voters.
Placement therefore seems to matter mainly in combination with the author's endorsement, consistent with agenda setters steering outcomes through the design of the choice set \citep{romer1978political}, rather than through a mechanical ballot-order effect of the kind estimated by \citet{ho2008estimating}.

\subsection{Implications for Governance Design}

The findings point to a practical transparency issue in DAO governance. Since author-selected choices systematically receive more voting power, voters and observers may need clearer information about who created a proposal, which option the author selected, and how much of the final score comes from the author's own voting power.
A related concern is that large governance token holders and other vested contributors can shape outcomes in ways that ordinary voters may not easily observe~\citep{buterin2021coinvoting}; similarly, opposing positions may be under-represented and near-unanimous outcomes may reflect limited deliberation.
This does not imply that author influence is improper. Proposal authors often have relevant expertise and may legitimately signal which option implements the proposal as intended. Still, the concentration of voting power around author-selected choices matters for accountability, especially when authors are pseudonymous, affiliated with organizations, or connected to multiple governance spaces.

The results also speak to a recurring question in the economics of organizations: how much of a voting outcome reflects voters' independent preferences, and how much the influence of whoever controls the agenda. In corporations, this concern motivated disclosure rules for proxy voting and scrutiny of proxy advisors \citep{malenko2016role}. DAOs face an analogous problem. Because a substantial part of the support for an option is associated with the author's choice, the author's influence is larger than their token holdings imply, and lopsided outcomes should not automatically be read as broad consensus.

Moreover, these findings show that interface choices are analytically important. %
Snapshot ballots present options in an ordered list and expose voting activity and running tallies under many configurations, so voters can see that support is highly uneven before they cast their own vote.
For governance design, that means choice order, author-choice visibility, and author-vote visibility should be documented and, where possible, made auditable rather than treated as neutral interface details. For empirical interpretation, highly skewed voting outcomes should not automatically be read as unambiguous preference aggregation: they may also reflect proposal screening, prominent author signals, list order, and the public accumulation of voting power during the vote.

Three low-cost design responses can be drawn from this: reporting the author's own voting power explicitly in the final tally and additional information on the author itself; randomizing the order of non-default choices; and, where herding is a concern, hiding interim tallies until the vote closes. %

\subsection{Limitations}
The analysis is observational and therefore quantifies systematic associations rather than causal effects.
Consistent with the positive-outcome skew discussed in Section~\ref{sec:background}, high approval rates may partly reflect proposal screening before formal voting: proposals that fail informal ``temperature checks'' or forum discussions 
may be abandoned or revised before reaching Snapshot.
Our dataset therefore observes only proposals that survive this filtering, so lopsided approval rates may partly reflect a survival process rather than how voters treat proposals once balloted.
This boundary cannot be removed by simply collecting more pre-vote material, because the voting power shares, as well as choices, choice order, and author selection are defined only for formal Snapshot ballots.
The residual author-choice association after removing author voting power may reflect information, coordination, ballot design, social influence, or unmeasured proposal quality.
Social-influence interpretations also depend on vote and tally visibility during the voting period: privacy-preserving mechanisms such as the Shutter plugin~\citep{nelson2022snapshot}, which encrypts votes on Snapshot while the poll is open and decrypts them only after it closes, may follow different dynamics, but current evidence on such plugins is not yet strong enough to identify that channel in our sample. Future work could combine Snapshot data with forum discussions, proposal histories, author vote timing, and on-chain records to separate these mechanisms more directly, while treating proposals that never reach a vote as outside the ballot-level regression framework used. %

\section{Conclusion}
\label{sec:Conclusion}

This paper examines how voting power is distributed across choices in governance proposals executed on Snapshot, the leading off-chain voting platform for DAOs, and the extent to which the observed high approval rates can be explained by three biases: \textit{position}, capturing the importance of the list ordering, \textit{approval}, capturing the systematic impact of a choice's stance, and \textit{author}, capturing reputation or social influence signals.

Across the observed data, three regularities stand out. 
First, choices listed in the first position receive only slightly more voting power than later choices, with an advantage of 7.7 percentage points. Position matters more in combination with the author's choice: net of the author's own votes, the author advantage is more than twice as large when the author's choice is listed first.
Second, approving choices are also favored, with an estimated advantage of 27.1 percentage points. However, the estimate falls to 7.7 percentage points once we condition on the author having voted, so much of it moves together with the proposer's support.
Third, author-selected choices show the largest conditional association with voting-power shares, with an average marginal effect of 58.8 percentage points in our main specification.
In proposals where the author voted and other voters contributed positive voting power, the author's own tokens account for about one third of the estimate: removing their voting power lowers the advantage from 45.6 to 29.7 percentage points; the remaining two thirds come from other voters.  The advantage is smaller where stakes and participation are higher: it is roughly halved in the largest DeFi protocols and falls to 39.7 percentage points once proposals with very low participation are excluded.
The associations remain large across multiple samples and specifications.

These results suggest that the regularities of corporate voting, where the party that sets the agenda tends to win, reappear in DAOs even without a formal hierarchy. %
The author bias pattern is consistent with more than one mechanism. Author-selected choices receive mechanical support because authors cast voting power for their own preferred options; the residual author advantage is consistent with less-informed voters deferring to the proposer, with later voters following a visible early vote, and with authors selecting choices that are already more likely to attract support. 
Moreover, because we only observe proposals that reach a formal vote, part of the approval advantage may reflect selection before the ballot rather than how voters treat proposals once they are listed.
The present design distinguishes these associations more cleanly than raw averages, but it does not isolate a single causal channel.

Overall, the paper provides evidence that DAO voting outcomes are shaped not only by formal token-weighted aggregation.
They are also associated with interface position, stance structure, and especially the choices selected by proposal authors. Those patterns do not by themselves establish misconduct or manipulation, but they show that descriptive features of the ballot and of the proposal author are central to understanding how voting power accumulates in decentralized governance.

\clearpage

\clearpage

\bibliography{literature.bib}

\clearpage

\appendix

\section*{Supplementary Material}

\section{Decentralized Autonomous Organizations}

\begin{figure}[h]
	\centering
	\begin{tikzpicture}[
		box/.style={draw, rounded corners, minimum height=1cm, minimum width=4cm, align=center},
		arrow/.style={->, thick},
		timeline/.style={thick},
		vote/.style={draw=blue!70, thick},
		votetext/.style={blue!70},
		scale=0.9, 
		transform shape
		]
		
		\node[box, minimum width=3.5cm] (create) at (-0.5,1.5) {Create a Proposal};
		\node[box, draw=blue!70, text=blue!70, minimum width=2.5cm] (vote) at (4,1.5) {Cast Votes};
		\node[box] (execute) at (9,1.5) {Execute the Proposal};
		
		\draw[->, arrow] (create) -- (vote);
		\draw[->, arrow] (vote) -- (execute);
		
		\draw[timeline, ->] (-2,0) -- (11,0);
		
		\foreach \x/\label in {-2/-2, 0/0, 2/2, 4/4, 6/6, 8/8, 10/10} {
			\draw (\x,0.15) -- (\x,-0.15);
			\node[below] at (\x,-0.2) {\label};
		}
		
		\draw[vote] (1,0) -- (7,0);
		\filldraw[vote, fill=blue!70] (1,0) circle (3pt);
		\filldraw[vote, fill=blue!70] (7,0) circle (3pt);
		
		\node[votetext, below=1cm] at (4,0) {3. Voting Period};
		
		\node[below=0.8cm, align=center] at (-1,0) {2. Determine\\Voting Power};
		\node[below=0.8cm, align=center] at (1,0) {1. Create\\Proposal};
		\node[below=0.8cm, align=center] at (9,0) {4. Execute\\Proposal};
		
	\end{tikzpicture}
	\caption{\textbf{Timeline of a DAO governance process}. The decision-making process can be divided into three time windows: the pre-voting period, the voting period, and the post-voting period.}
	\label{fig:dao_timeline}
\end{figure}
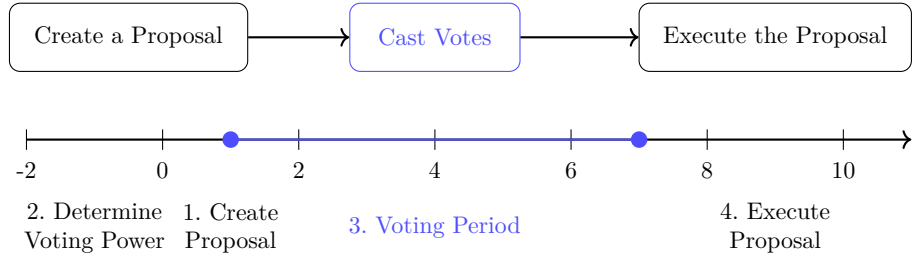

Figure~\ref{fig:dao_timeline} summarizes the governance sequence used throughout the empirical analysis. A proposal is created, voting power is determined by the rules of the relevant DAO space, token holders cast votes during the voting window, and the winning choice can then be executed or used as a governance signal.

\section{Data Processing}
\label{sec:app_data}

The original dataset, obtained from Snapshot.org, contains \DaoBiasRawVoteN{} individual votes spread across \DaoBiasRawProposalN{} governance proposals from \DaoBiasRawSpaceN{} DAO spaces, for a total of \DaoBiasRawChoiceN{} choices, over the period Oct 2020 to Nov 2023. Below, we discuss the various data cleaning procedures that we applied.

\subsection{Initial Processing}

We conducted a series of sanity checks and cleaning procedures in two stages.
At the proposal level, following \citet{kitzler2024governance}, we removed proposals labeled as `pending' or `invalid' and retained those marked as `final'. We also eliminated proposals with nonsensical titles, template text in the body, or template text in the title, such as ``Lorem ipsum dolor sit amet, consectetur adipiscing elit [...]'', ``new title'', ``new proposal'', or ``Title of your proposal goes here''. We further excluded proposals with less than two choices and proposals marked by Snapshot's ``flagged'' field.

At the vote level, we identified invalid votes, i.e., votes not conforming to the choices admitted by the proposal. These include plain malformed votes, such as votes containing an address rather than an integer, as well as votes containing an integer pointing to a non-existing choice, such as index 4 when only three choices are available. Snapshot.org marks these votes as invalid,\footnote{See for instance: \url{https://snapshot.org/\#/testsnap.eth/proposal/0x1b1c6e19edb5358110aa20e6e261d92b22bc2422925d759b5968e190f33ecbf5}.} while the API still exposes them. We excluded malformed votes and votes pointing to unavailable choices, and recorded duplicate-ballot and invalid-vote indicators as diagnostics. The cleaning code records all cleaning decisions for reproduction.

After filtering to final proposals, the vote-merge validation starts from \DaoBiasStageThreeLoadedFinalProposalN{} proposals. The resulting analytic dataset contains \DaoBiasFinalSpaceN{} spaces, with \DaoBiasFinalProposalN{} proposals for a total of \DaoBiasFinalChoiceN{} choices and \DaoBiasFinalVoteMassN{} favorite-mode ballot mass. This amounts to an average of \DaoBiasFinalChoicePerProposalMean{} choices per proposal ($\pm$ \DaoBiasFinalChoicePerProposalStd{}), with a median of \DaoBiasFinalChoicePerProposalMedian{}; spaces on average have \DaoBiasFinalProposalPerSpaceMean{} proposals ($\pm$ \DaoBiasFinalProposalPerSpaceStd{}), with a median of \DaoBiasFinalProposalPerSpaceMedian{}. The final dataset was checked to be free from invalid or duplicated votes and from all proposals that were flagged with issues by the Snapshot team.

We additionally excluded \DaoBiasExcludedSnapshotFlaggedProposalN{} proposals marked by Snapshot's ``flagged'' field, corresponding to \DaoBiasExcludedSnapshotFlaggedProposalPct\% of the \DaoBiasRawSnapshotFlaggedProposalN{} raw proposals with this flag. We further classify spaces as `mature' when they fulfill all of the following three criteria from \cite{kitzler2024governance}: (i) at least five followers, (ii) more than two voters in any proposal, and (iii) more than one proposal. We evaluate the quality of these procedures by looking at the retention of mature spaces. Encouragingly, we find that only 5\% of the eliminated spaces were classified as mature.

\subsection{Processing Votes}\label{appendix_processing_votes}

The primary voting-power outcome in the paper is choice-level voting-power share. Most proposals (\DaoBiasSinglePreferenceProposalPct\%) follow a single-preference choice model (see Table~\ref{tab_voting_systems_catalog} in main text), which makes the computation of the voting power share straightforward (i.e., choice's voting power divided by the total voting power in a proposal).

However, in some voting systems (as also discussed in Section~\ref{sec:background}), users can express preferences for multiple choices with a single vote. 
For example, in weighted choice models, users can select multiple options and assign weights reflecting their relative preference.
In these cases, we need to uniformly map users' raw votes to a set of expressed choices for vote-count diagnostics. 
To do so, we split multiple-preference ballots into one or multiple choice-level rows while conserving each ballot's unit mass. 

We test two diagnostic mappings for assigning ballot mass to choices in the presence of multiple preferences: (i) favorite-choice, and (ii) support-fraction. The former assigns each ballot's unit mass to the favorite choice or choices; approval ballots split mass equally across the distinct approved choices, and weighted or quadratic ballots keep the maximum positive-weight choices and split ties equally. The latter divides ballot mass across supported choices; for weighted and quadratic ballots, positive weights are normalized across all selected choices.
Table \ref{tab_voting_systems_proc} reports how users' raw votes are mapped to a set of expressed choices in different multiple-preference voting systems.

Both mappings may lead to more choice-level rows than ballot rows in multiple-choice voting systems, but neither duplicates a ballot's mass as a full vote for every selected choice. The vote-expansion procedure processed \DaoBiasVoteExpansionFinishedProposalN{} proposals and produced \DaoBiasVoteExpansionFavoriteRowN{} favorite-choice rows, with \DaoBiasVoteExpansionFavoriteErrorRowN{} favorite-choice parse-error rows retained as diagnostics.

In Table~\ref{tab:robustness-author-choice-diagnostics}, we ask how much support the author's chosen option receives from other voters. We exclude the author's vote, add up the voting power that other voters assign to each choice, and divide each total by the voting power cast by all other voters in that proposal. We include only proposals where at least one other voter casts positive voting power and the total voting power remaining after excluding the author is positive. This comparison uses Basic and Single-choice proposals.

Practically, between the two mapping systems the differences are small and they occur only for weighted and quadratic proposals, single-choice, basic, ranked-choice, and approval have zero mapping differences. Quantitatively, we register differences across the mappings in 2,721 weighted proposals and 917 quadratic proposals for a total of 30,929 proposal-choice pairs. Only 8,263.57 out of 54,978,057 valid ballot mass, about 0.015\% overall, changes across the two systems, with a median redistributed proposal mass of 3.47 percentage points, 95th percentile is 40.0 percentage points, max is 87.13 percentage points.

Across all the variations in Table \ref{tab:robustness-variation-key}, author choice is defined with the favorite-choice mapping. For variations 1-11, this either is the only relevant mapping or is practically identical to support-fraction under the included voting systems. In variation 12 (Full), this is a substantive choice: weighted and quadratic ballots are included, but the author-choice regressor remains favorite-choice based.

We continue to identify the author's chosen option from their recorded vote, even when we exclude their voting power from the outcome. In Basic and Single-choice proposals, the favorite-choice and support-fraction mappings identify the same option. Table~\ref{tab:robustness-author-choice-diagnostics} therefore compares support for the same author-selected choices with and without the author's own voting power.

\begin{table}
	\centering
	\footnotesize
	\caption{Multi-preference vote processing}
	\label{tab_voting_systems_proc}
    \begin{tabular*}{\linewidth}{@{\extracolsep{\fill}}llll}
		\toprule
		\makecell{\textbf{Voting}\\\textbf{system}} &
		\makecell{\textbf{Raw ballot}} &
		\makecell{\textbf{Favorite-choice rows}\\\textbf{choice: mass}} &
		\makecell{\textbf{Support-fraction rows}\\\textbf{choice: mass}} \\
		\midrule
		Basic         & \makecell[l]{1 \\ {\scriptsize\emph{rows: 1, mass: 1}}} &
		\makecell[l]{1: 1 \\ {\scriptsize\emph{rows: 1, mass: 1}}} &
		\makecell[l]{1: 1 \\ {\scriptsize\emph{rows: 1, mass: 1}}} \\
		\specialrule{0.001pt}{2pt}{4pt}
		Single-choice & \makecell[l]{2 \\ {\scriptsize\emph{rows: 1, mass: 1}}} &
		\makecell[l]{2: 1 \\ {\scriptsize\emph{rows: 1, mass: 1}}} &
		\makecell[l]{2: 1 \\ {\scriptsize\emph{rows: 1, mass: 1}}} \\
		\specialrule{0.001pt}{2pt}{4pt}
		Weighted      & \makecell[l]{\{1: 50, 2: 50\} \\ {\scriptsize\emph{rows: 1, mass: 1}}} &
		\makecell[l]{1: 1/2 \\ 2: 1/2 \\ {\scriptsize\emph{rows: 2, mass: 1}}} &
		\makecell[l]{1: 1/2 \\ 2: 1/2 \\ {\scriptsize\emph{rows: 2, mass: 1}}} \\
		\specialrule{0.001pt}{2pt}{4pt}
		Quadratic     & \makecell[l]{\{1: 3, 2: 1\} \\ {\scriptsize\emph{rows: 1, mass: 1}}} &
		\makecell[l]{1: 1 \\ {\scriptsize\emph{rows: 1, mass: 1}}} &
		\makecell[l]{1: 3/4 \\ 2: 1/4 \\ {\scriptsize\emph{rows: 2, mass: 1}}} \\
		\specialrule{0.001pt}{2pt}{4pt}
		Ranked-choice & \makecell[l]{[4, 2, 3, 1] \\ {\scriptsize\emph{rows: 1, mass: 1}}} &
		\makecell[l]{4: 1 \\ {\scriptsize\emph{rows: 1, mass: 1}}} &
		\makecell[l]{4: 1 \\ {\scriptsize\emph{rows: 1, mass: 1}}} \\
		\specialrule{0.001pt}{2pt}{4pt}
		Approval      & \makecell[l]{[1, 2, 4] \\ {\scriptsize\emph{rows: 1, mass: 1}}} &
		\makecell[l]{1: 1/3 \\ 2: 1/3 \\ 4: 1/3 \\ {\scriptsize\emph{rows: 3, mass: 1}}} &
		\makecell[l]{1: 1/3 \\ 2: 1/3 \\ 4: 1/3 \\ {\scriptsize\emph{rows: 3, mass: 1}}} \\
		\bottomrule
	\end{tabular*}
	\parbox{\linewidth}{\footnotesize \textit{Notes}: initially, a raw ballot is a single row in the dataset carrying unit mass ({\scriptsize\emph{rows: 1, mass: 1}}); the mapping expands it into one or more rows that always sum to the same unit mass, so no ballot is double-counted even when the number of rows grows. The favorite-choice mapping assigns mass to the maximum-weight choice(s) (ties split equally) and equally splits approval ballots across all approved choices. The support-fraction mapping normalizes positive weights across all selected choices for weighted and quadratic ballots, splits approval ballots equally, and keeps the top-ranked choice only for ranked-choice ballots.} 
\end{table}

\section{Choice Stance Detection}\label{appendix_choice_stance_detection}

The choices associated with a proposal can be expressed with simple binary responses, numbers, strings of text, or other arbitrary content, and may include indications to approve or reject its content, or more complex options.
We thus try to classify each choice of all proposals to a simplified \textit{stance}, defined as the intended meaning or position expressed by a voting choice within a proposal.
To classify the stance of a choice, we need to first make an important distinction between proposals having a \textit{direct effect} on the governance of its organization. 
We assign a stance of ``Reject'' or ``Approve'' only to choices of proposals that are directly rejecting or approving governance changes; often in such proposals there exists a third choice to abstain from making a decision, which we label ``Abstain'' (such a choice helps reach the quorum, without expressing a preference).

Finally, we assign the label ``Other'' to all the choices of proposals that do not have a direct governance effect, and more broadly to all the choices that cannot be directly classified as either ``Reject'', ``Approve'', or ``Abstain.''

Importantly, a proposal must offer at least one approval and one refusal choice, so that any choice could be labeled as either ``Approve'' or ``Reject''; otherwise, all choices are labeled as ``Other.'' For instance, simply selecting an alternative between a number of choices does not follow an approve/reject pattern, because there is no possibility to object to the proposal altogether; in this case, all choices will be labeled as ``Other.'' However, if there is one choice rejecting all the alternatives, we mark this choice as ``Reject'', and the other as ``Approve'' (more ``Reject'' and ``Abstain'' choices in the same proposals are also allowed). Before this recoding, we registered 4,274 reject-only and 1,945 approve-only proposals that we normalized to ``Other'' (see Table~\ref{tab:robustness-stance-sensitivity}).

Having stated our classification goal, we implemented it using rule-based heuristics and validated via large language models and human evaluation, as described below.

\subsection{Heuristics}\label{appendix_heuristics}

The heuristic procedure consists of two substantive classification steps followed by a normalization and exception-handling step. First, a registry-backed rule classifier assigns an initial stance to each choice using exact, prefix, suffix, high-confidence-contains, and low-confidence-contains patterns. Second, we apply targeted contextual checks for cases that cannot be reliably resolved from isolated choice text, including negated action labels, stance-bearing residual options, and manually inspected emoji choices. Finally, we reconcile known exceptions and enforce the final proposal-level consistency criteria, including space- and proposal-level overrides, per-proposal stance-sequence overrides,
and the normalization of undocumented one-sided Approve/Reject proposals to \textit{Other}.

\paragraph{Registry-backed rule classifier.} The general keyword rules are defined in the stance-pattern registry used by the classifier. The registry starts from conservative stance-specific exact terms and expands them into prefix, suffix, and contains tiers, with short or ambiguous tokens kept exact-only. Tables~\ref{tab:choice-stance-heuristics-reject}--\ref{tab:choice-stance-heuristics-abstain} summarize the registry tiers and their resolution rules by stance.

\begingroup
\small
\setlength{\LTleft}{0pt}
\setlength{\LTright}{0pt}
\setlength{\LTcapwidth}{0.92\textwidth}
\setlength{\tabcolsep}{3pt}
\renewcommand{\arraystretch}{1.12}
\begin{longtable}{@{}>{\RaggedRight\arraybackslash}p{0.13\textwidth}>{\RaggedRight\arraybackslash}p{0.31\textwidth}>{\RaggedRight\arraybackslash}p{0.50\textwidth}@{}}
\caption{Reject stance heuristic registry}\label{tab:choice-stance-heuristics-reject}\\
\toprule
Tier & Rule & Terms \\
\midrule
\endfirsthead
\caption[]{Reject stance heuristic registry (continued)}\\
\toprule
Tier & Rule & Terms \\
\midrule
\endhead
\bottomrule
\endfoot
Exact & Trimmed lower-case choice equals the canonical term. & English: againist, against, aganist, decline, declined, denied, deny, disagree, disagreed, disapporve, disapprove, disapproved, dissaprove, dissent, do not, do not accept, do not approve, do not proceed \\
 &  & English: do not support, do nothing, don't, don't accept, don't agree, don't approve, don't support, dont agree, dont support, dont't proceed, i deny, i disagree, i dissent, i do not accept, i do not agree, i do not approve, i do not support, i don't accept \\
 &  & English: i don't agree, i don't approve, i don't support, i dont agree, i dont support, i refuse, i reject, ino, n, nae, nah, nay, negative, negative vote, no, no support, nope, not accept \\
 &  & English: not agree, not approve, not approved, not in favor, not support, oppose, opposed, refuse, refused, reject, rejected \\
 &  & Non-English: contra, no apruebo, non, non approvo, {\CJKFont não}, {\CJKFont нет}, {\CJKFont 不会}, {\CJKFont 不可以}, {\CJKFont 不同意}, {\CJKFont 不支持}, {\CJKFont 不是}, {\CJKFont 不能}, {\CJKFont 反对}, {\CJKFont 反対}, {\CJKFont 否}, {\CJKFont 没有} \\
Prefix & Left-trimmed choice starts with the term followed by space or punctuation. & English: againist, against, aganist, decline, declined, denied, deny, disagree, disagreed, disapporve, disapprove, disapproved, dissaprove, dissent, do not, do not accept, do not approve, do not proceed \\
 &  & English: do not support, do nothing, don't, don't accept, don't agree, don't approve, don't support, dont agree, dont support, dont't proceed, i deny, i disagree, i dissent, i do not accept, i do not agree, i do not approve, i do not support, i don't accept \\
 &  & English: i don't agree, i don't approve, i don't support, i dont agree, i dont support, i refuse, i reject, ino, nae, nah, nay, negative, negative vote, no, no support, nope, not accept, not agree \\
 &  & English: not approve, not approved, not in favor, not support, oppose, opposed, refuse, refused, reject, rejected \\
 &  & Non-English: contra, no apruebo, non, non approvo, {\CJKFont não}, {\CJKFont нет}, {\CJKFont 不会}, {\CJKFont 不可以}, {\CJKFont 不同意}, {\CJKFont 不支持}, {\CJKFont 不是}, {\CJKFont 不能}, {\CJKFont 反对}, {\CJKFont 反対}, {\CJKFont 否}, {\CJKFont 没有} \\
Suffix & Right-trimmed choice ends with a space plus the term. & English: againist, against, aganist, decline, declined, denied, deny, disagree, disagreed, disapporve, disapprove, disapproved, dissaprove, dissent, do not, do not accept, do not approve, do not proceed \\
 &  & English: do not support, do nothing, don't, don't accept, don't agree, don't approve, don't support, dont agree, dont support, dont't proceed, i deny, i disagree, i dissent, i do not accept, i do not agree, i do not approve, i do not support, i don't accept \\
 &  & English: i don't agree, i don't approve, i don't support, i dont agree, i dont support, i refuse, i reject, ino, nae, nah, nay, negative, negative vote, no, no support, nope, not accept, not agree \\
 &  & English: not approve, not approved, not in favor, not support, oppose, opposed, refuse, refused, reject, rejected \\
 &  & Non-English: contra, no apruebo, non, non approvo, {\CJKFont não}, {\CJKFont нет}, {\CJKFont 不会}, {\CJKFont 不可以}, {\CJKFont 不同意}, {\CJKFont 不支持}, {\CJKFont 不是}, {\CJKFont 不能}, {\CJKFont 反对}, {\CJKFont 反対}, {\CJKFont 否}, {\CJKFont 没有} \\
High-confidence contains & Choice contains a space-delimited or punctuation-delimited registry term. Confident matches resolve to the stance unless a higher-priority tier already matched. & English: againist, against, aganist, decline, declined, denied, deny, disagree, disagreed, disapporve, disapprove, disapproved, dissaprove, dissent, do not approve, do not support, do nothing, don't accept \\
 &  & English: don't agree, don't approve, don't support, dont agree, dont support, dont't proceed, i deny, i disagree, i dissent, i do not agree, i do not approve, i do not support, i don't agree, i don't approve, i don't support, i refuse, i reject, ino \\
 &  & English: nae, nah, nay, negative vote, no support, nope, not accept, not agree, not approve, not in favor, not proceed, not support, oppose, opposed, refuse, refused, reject, rejected \\
 &  & Non-English: contra, non approvo, {\CJKFont não}, {\CJKFont нет}, {\CJKFont 不会}, {\CJKFont 不可以}, {\CJKFont 不同意}, {\CJKFont 不支持}, {\CJKFont 不是}, {\CJKFont 不能}, {\CJKFont 反对}, {\CJKFont 反対}, {\CJKFont 否}, {\CJKFont 没有} \\
Low-confidence contains & Low-confidence-only matches resolve to Other; conflicts remain visible in the audit trail. & None \\
\end{longtable}
\endgroup

\begingroup
\small
\setlength{\LTleft}{0pt}
\setlength{\LTright}{0pt}
\setlength{\LTcapwidth}{0.92\textwidth}
\setlength{\tabcolsep}{3pt}
\renewcommand{\arraystretch}{1.12}
\begin{longtable}{@{}>{\RaggedRight\arraybackslash}p{0.13\textwidth}>{\RaggedRight\arraybackslash}p{0.31\textwidth}>{\RaggedRight\arraybackslash}p{0.50\textwidth}@{}}
\caption{Approve stance heuristic registry}\label{tab:choice-stance-heuristics-approve}\\
\toprule
Tier & Rule & Terms \\
\midrule
\endfirsthead
\caption[]{Approve stance heuristic registry (continued)}\\
\toprule
Tier & Rule & Terms \\
\midrule
\endhead
\bottomrule
\endfoot
Exact & Trimmed lower-case choice equals the canonical term. & English: accept, affirmative, agree, agreed, approve, approve!, approved, aye, for, i agree, i support, in favor, ok, pass, positive, positive vote, proceed, s \\
 &  & English: support, syes, y, yae, yay, yes, yes! \\
 &  & Non-English: a favor, approvo, aprovar, apruebo, oui, si, sim, {\CJKFont sí}, {\CJKFont да}, {\CJKFont 会}, {\CJKFont 可以}, {\CJKFont 同意}, {\CJKFont 支持}, {\CJKFont 是}, {\CJKFont 有}, {\CJKFont 能}, {\CJKFont 賛成}, {\CJKFont 赞成} \\
Prefix & Left-trimmed choice starts with the term followed by space or punctuation. & English: accept, affirmative, agree, agreed, approve, approve!, approved, aye, for, i agree, i support, in favor, pass, positive, positive vote, proceed, support, syes \\
 &  & English: yae, yay, yes, yes! \\
 &  & Non-English: a favor, approvo, aprovar, apruebo, oui, si, sim, {\CJKFont sí}, {\CJKFont да}, {\CJKFont 可以}, {\CJKFont 同意}, {\CJKFont 支持}, {\CJKFont 賛成}, {\CJKFont 赞成} \\
Suffix & Right-trimmed choice ends with a space plus the term. & English: affirmative, agreed, approve!, approved, aye, for, i agree, i support, in favor, pass, positive, positive vote, proceed, syes, yae, yay, yes, yes! \\
 &  & Non-English: a favor, approvo, aprovar, apruebo, oui, si, sim, {\CJKFont sí}, {\CJKFont да}, {\CJKFont 同意}, {\CJKFont 支持}, {\CJKFont 賛成}, {\CJKFont 赞成} \\
High-confidence contains & Choice contains a space-delimited or punctuation-delimited registry term. Confident matches resolve to the stance unless a higher-priority tier already matched. & English: affirmative, agreed, approve!, approved, aye, in favor, positive vote, syes, yae, yay, yes! \\
 &  & Non-English: a favor, approvo, aprovar, apruebo, oui, sim, {\CJKFont sí}, {\CJKFont да}, {\CJKFont 同意}, {\CJKFont 支持}, {\CJKFont 賛成}, {\CJKFont 赞成} \\
Low-confidence contains & Low-confidence-only matches resolve to Other; conflicts remain visible in the audit trail. & English: accept, agree, approve, support, yes \\
\end{longtable}
\endgroup

\begingroup
\small
\setlength{\LTleft}{0pt}
\setlength{\LTright}{0pt}
\setlength{\LTcapwidth}{0.92\textwidth}
\setlength{\tabcolsep}{3pt}
\renewcommand{\arraystretch}{1.12}
\begin{longtable}{@{}>{\RaggedRight\arraybackslash}p{0.13\textwidth}>{\RaggedRight\arraybackslash}p{0.31\textwidth}>{\RaggedRight\arraybackslash}p{0.50\textwidth}@{}}
\caption{Abstain stance heuristic registry}\label{tab:choice-stance-heuristics-abstain}\\
\toprule
Tier & Rule & Terms \\
\midrule
\endfirsthead
\caption[]{Abstain stance heuristic registry (continued)}\\
\toprule
Tier & Rule & Terms \\
\midrule
\endhead
\bottomrule
\endfoot
Exact & Trimmed lower-case choice equals the canonical term. & English: abstain, i abstain, maybe, not sure \\
 &  & Non-English: {\CJKFont abstenção}, abster, je m'abstiens, {\CJKFont 弃权}, {\CJKFont 棄権} \\
Prefix & Left-trimmed choice starts with the term followed by space or punctuation. & English: abstain, i abstain, maybe, not sure \\
 &  & Non-English: {\CJKFont abstenção}, abster, je m'abstiens, {\CJKFont 弃权}, {\CJKFont 棄権} \\
Suffix & Right-trimmed choice ends with a space plus the term. & English: abstain, i abstain, maybe, not sure \\
 &  & Non-English: {\CJKFont abstenção}, abster, je m'abstiens, {\CJKFont 弃权}, {\CJKFont 棄権} \\
High-confidence contains & Choice contains a space-delimited or punctuation-delimited registry term. Confident matches resolve to the stance unless a higher-priority tier already matched. & English: abstain \\
 &  & Non-English: {\CJKFont abstenção}, abster, je m'abstiens, {\CJKFont 弃权}, {\CJKFont 棄権} \\
Low-confidence contains & Low-confidence-only matches resolve to Other; conflicts remain visible in the audit trail. & None \\
\end{longtable}
\endgroup

The raw classifier stores the resolved stance, match source, matched pattern, canonical pattern, all match sources, and a conflict indicator. Confident exact, prefix, suffix, and high-confidence contains matches determine the raw stance. Low-confidence contains terms are still stored in the audit trail, but by themselves resolve to ``Other''; when they occur alongside a confident match, the confident match determines the stance. Cases containing both Approve and Reject evidence are not silently discarded at this stage: the raw classifier resolves by tier order and records an approve/reject conflict flag for later review.

\paragraph{Targeted contextual checks and final audit.} After raw registry classification, the notebook writes raw audit CSVs for choice-level matches and proposal-level stance states. It then applies the notebook-level targeted checks and writes a second pair of audit CSVs for the post-notebook-rule checkpoint. Finally, documented metadata rules apply whole-space exclusions, whole-proposal overrides, per-proposal stance-sequence overrides, and one-sided normalization. The final checkpoint writes choice- and proposal-level audit CSVs after overrides. Targeted checks included choices beginning with English negation tokens ``dont'', ``do not'', and ``don't''; ``not'' was evaluated but excluded because it was too inclusive. These tokens proxy for proposal-rejecting choices that are often paired with approval choices that pattern matching alone cannot identify (e.g., Buy vs Don't buy or Transfer vs Don't transfer). Stance-specific targeted terms included, for Reject, ``neither'' and ``none of the above''; for Abstain, ``Invalid question/options'', ``revisit'', ``rework'', ``revision'', and ``discuss''. Terms evaluated for Approve, including ``option'', ``price'', ``\$'', and ``will'', were not sufficiently associated with approval choices under our definition. Manual inspection also covered stance-bearing emojis, namely:

\begin{itemize}
    \item[APPROVE]: 
{\EmojiFont\char"2705},
{\EmojiFont\char"2714},
{\EmojiFont\char"270C},
{\EmojiFont\char"1F525},
{\EmojiFont\char"1F642},
{\EmojiFont\char"1F44D}
   \item[REJECT]:
   {\EmojiFont\char"274C},
{\EmojiFont\char"1F44E},
{\EmojiFont\char"1F4A9},
{\EmojiFont\char"1F480}
\item[ABSTAIN]: 
{\EmojiFont\char"1F914},
{\EmojiFont\char"1F937},
{\EmojiFont\char"1F943}
\end{itemize}

\begin{table}[!tbp]
\centering
\caption{Audit of one-sided stance recoding for the approval-stance sensitivity analysis}
\label{tab:robustness-stance-sensitivity}
\small
\setlength{\tabcolsep}{3pt}
\renewcommand{\arraystretch}{1.10}
\begin{tabularx}{\linewidth}{@{}>{\RaggedRight\arraybackslash}p{0.20\linewidth}>{\RaggedRight\arraybackslash}Xrrr@{}}
\toprule
When & Proposal pattern & Proposals & Choices & Spaces \\
\midrule
Before recoding & Reject, no Approve & 4,274 & 14,295 & 1,313 \\
Before recoding & Approve, no Reject & 1,945 & 7,708 & 892 \\
After recoding & No Approve or Reject labels & 75,408 & 273,825 & 15,370 \\
After recoding & Approve and Reject both present & 51,873 & 124,393 & 7,378 \\
\bottomrule
\end{tabularx}
\parbox{\linewidth}{\footnotesize \textit{Notes}: ``Before recoding'' counts undocumented one-sided proposals that were later normalized to Other. ``After recoding'' counts the final post-recoding buckets used by the stance sensitivity audit.}
\end{table}

\subsection{Large Language Models (LLMs) Evaluations}\label{appendix_llm_eval}

To assess the robustness of our choice stance classification, we compare the heuristics with external LLM evaluations of both open source and commercial models, namely Llama 3.1 and 3.3 70 Billion Instruct, OpenAI GPT 4o API (OAI), and ChatGPT 4o.
Because LLMs may themselves be sensitive to presentation order, we do not use them as behavioral subjects or as direct evidence of voter bias; instead, we use them only as auxiliary stance classifiers and validate their labels against rule-based heuristics, multiple model outputs, and human annotations. The prompts for the LLMs are reported in Appendix~\ref{sec_llm_prompts}. Llama 3.3 was evaluated only on non-Basic proposals, so its evaluation universe differs from those of the other models.

\paragraph{Stance detection.} We instructed each LLM to assign a stance (as in `Abstain', `Approve', `Other', or `Reject') to each choice in a proposal, and to assign `Other` to all choices of proposals not being true governance proposals, to follow our heuristics classification goals closely. The use of LLMs also allows us to examine whether the body of a proposal implicitly or explicitly contains a cue pointing readers toward a voting direction. We treat this \textit{suggested stance} as an LLM-coded textual cue in the proposal text, not as verified author intent or evidence that voters were persuaded. We thus obtain two variables for each LLM that respectively capture whether and in which direction the text suggests a vote, and classify each choice as `Abstain', `Approve', `Other', or `Reject'.

We note that some proposals could not be parsed correctly by the LLMs, and in a few instances, a different number of choices was interpolated by the LLM compared to the original proposal. These errors are typically associated with proposals having a large number of choices ($> 10$). We discarded the proposals where we observed this behavior. Within the final analytic sample of 398,218 choices, usable stance labels cover 207,842 choices for Llama 3.1 (52.19\%), 249,466 for Llama 3.3 (62.65\%), 225,112 for OAI (56.53\%), and 216,721 for ChatGPT (54.42\%).

\begin{figure}
    \centering
    \includegraphics[width=1\linewidth]{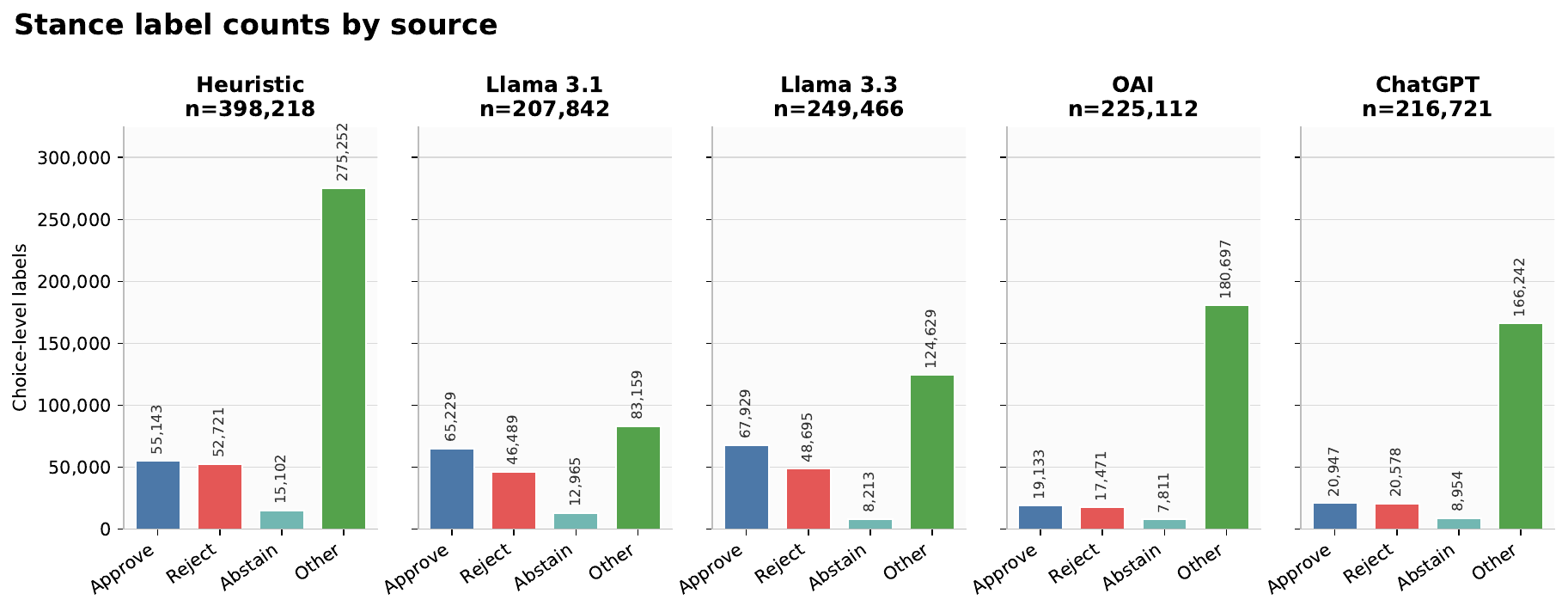}
    \caption{\textbf{Agreement across stance classification methods}. Bars count source-specific non-missing choice-level stance labels for the heuristic, Llama 3.1, Llama 3.3, OAI, and ChatGPT sources. Labels are counted within each source-specific denominator and are limited to Approve, Reject, Abstain, and Other.}
    \label{fig:llms_agree}
\end{figure}

To measure the agreement across our classification methods, we compute Cohen's Kappa between our heuristics and each LLM, using model-specific denominators so Llama 3.3's sparser coverage is handled consistently. The values are \DaoBiasLlmKappaLlama{} for Llama 3.1, \DaoBiasLlmKappaLlamaThreePointThree{} for Llama 3.3, \DaoBiasLlmKappaOai{} for OAI, and \DaoBiasLlmKappaChatgpt{} for ChatGPT, indicating \DaoBiasLlmKappaLlamaAgreement{}, \DaoBiasLlmKappaLlamaThreePointThreeAgreement{}, \DaoBiasLlmKappaOaiAgreement{}, and \DaoBiasLlmKappaChatgptAgreement{} agreement, respectively. The agreement across our classification methods is also displayed in Fig. \ref{fig:llms_agree}, highlighting two patterns: (i) \DaoBiasLlmKappaHighestSource{} has the highest agreement with the heuristics, and (ii) the proposals we removed due to classification problems from LLMs are mostly from the `Other' stance, which is often associated with proposals with a large number of choices in the heuristics classification (see Fig. \ref{fig:distr_choice_stance}).

\paragraph{Suggested stance detection.} We also asked whether the proposal text itself appeared to steer voters toward a particular outcome. For each proposal, an LLM would identify a suggested voting direction, such as Approve, Reject, Abstain, or Other, or it would say that the text did not suggest any voting direction. We treated the latter as a reviewed but neutral proposal, not as missing data. A proposal was considered unreviewed and discarded only when a given source produced no usable label after parsing and coverage filters.

\begin{figure}
    \centering
    \includegraphics[width=1\linewidth]{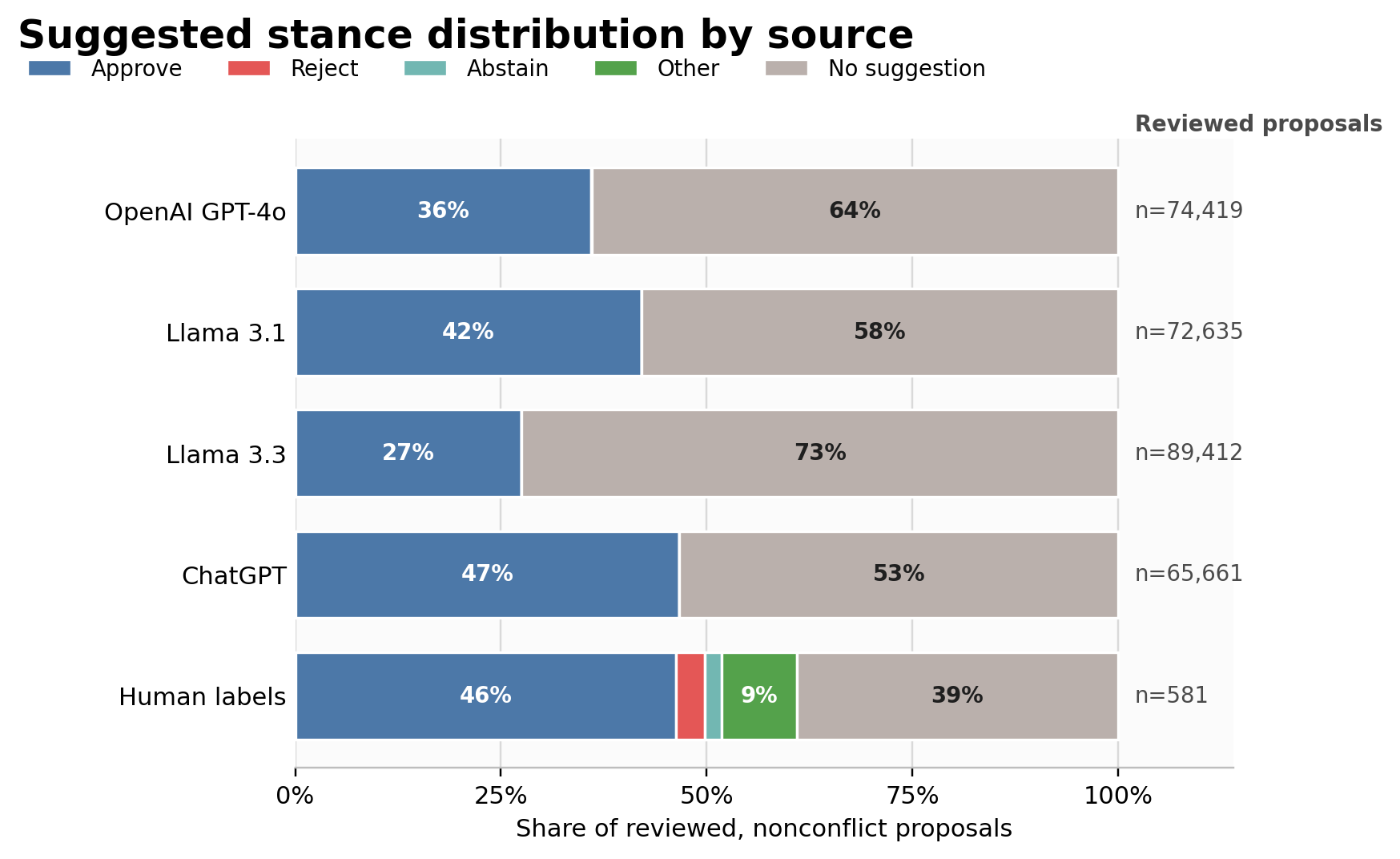}
    \caption{\textbf{Suggested-stance labels by source.} Bars show the share of proposals for which the source found no suggested voting direction or found a textual cue suggesting Approve, Reject, Abstain, or Other. Details about human labels are available in Sec. \ref{appendix:crowdwork_eval}, but we anticipate here the relevant result.}
    \label{fig:llms_stance_suggested}
\end{figure}

The share of proposals with no detected voting suggestion ranges from \DaoBiasSuggestedNoSuggestionPctMin\% for \DaoBiasSuggestedNoSuggestionPctMinSource{} to \DaoBiasSuggestedNoSuggestionPctMax\% for \DaoBiasSuggestedNoSuggestionPctMaxSource{}. Specifically, no-suggestion labels account for \DaoBiasSuggestedNoSuggestionPctLlama\% of reviewed, nonconflict proposal-level outputs for Llama 3.1, \DaoBiasSuggestedNoSuggestionPctLlamaThreePointThree\% for Llama 3.3, \DaoBiasSuggestedNoSuggestionPctOai\% for OAI, and \DaoBiasSuggestedNoSuggestionPctChatgpt\% for ChatGPT. These shares are source-specific: each model is measured only on proposals for which it produced a usable, nonconflicting label. This distinction is especially important for Llama 3.3, which covers a smaller sample of proposals: \DaoBiasSuggestedLlamaThreePointThreeReviewedProposalN{} reviewed proposals, \DaoBiasSuggestedLlamaThreePointThreeNotReviewedProposalN{} not-reviewed proposals, \DaoBiasSuggestedLlamaThreePointThreeBasicExclusionProposalN{} Basic-proposal coverage exclusions, \DaoBiasSuggestedLlamaThreePointThreeConflictProposalN{} conflicts, \DaoBiasSuggestedLlamaThreePointThreeNoSuggestionProposalN{} no-suggestion proposals, and \DaoBiasSuggestedLlamaThreePointThreePositiveProposalN{} positive suggested-stance proposals. Thus, a missing Llama 3.3 label should not be read as a neutral proposal or as a no-suggestion decision. When an automated model does detect a positive suggested stance, the stance is almost always Approve; Reject, Abstain, and Other cues are rare (see Fig. \ref{fig:llms_stance_suggested}).

Finally, we ask whether a detected suggestion points to the stance that later received the most voting power. In Table \ref{tab:stance_suggest_recall_precision_f1} we summarize how model-detected textual cues align with realized voting outcomes, using each model's own reviewed sample. Approve is the only positive suggested-stance category with broad coverage, accounting for at least \DaoBiasSuggestedApprovePositivePctMin\% of positive automated-model predictions. Across automated models, Approve precision ranges from \DaoBiasSuggestedApprovePrecisionMin{} to \DaoBiasSuggestedApprovePrecisionMax{}, while recall ranges from \DaoBiasSuggestedApproveRecallMin{} to \DaoBiasSuggestedApproveRecallMax{}, and F1 ranges from \DaoBiasSuggestedApproveFOneMin{} to \DaoBiasSuggestedApproveFOneMax{}. Reject, Abstain, and Other positive suggested-stance categories appear in too few cases for stable interpretation, so their metric cells should be read as sparse diagnostics rather than as source-level performance estimates.

\begin{table}[ht]
\centering
\caption{Detected suggested stance compared with the winning stance}
\label{tab:stance_suggest_recall_precision_f1}
\begingroup
\small
\setlength{\tabcolsep}{3pt}
\begin{tabularx}{\linewidth}{@{}>{\RaggedRight\arraybackslash}Xlrrrrrr@{}}
\toprule
Source & Stance & Pred. & Winner & TP & Precision & Recall & F1 \\
\midrule
OpenAI GPT-4o & APPROVE & 26737 & 40070 & 20509 & 0.767064 & 0.511829 & 0.613978 \\
OpenAI GPT-4o & REJECT & 31 & 5376 & 11 & 0.354839 & 0.002046 & 0.004069 \\
OpenAI GPT-4o & ABSTAIN & 1 & 749 & 0 & \multicolumn{3}{c}{Sparse ($n<30$)} \\
OpenAI GPT-4o & OTHER & 10 & 28017 & 7 & \multicolumn{3}{c}{Sparse ($n<30$)} \\
Llama 3.1 & APPROVE & 30540 & 39422 & 22676 & 0.742502 & 0.575212 & 0.648238 \\
Llama 3.1 & REJECT & 3 & 5285 & 1 & \multicolumn{3}{c}{Sparse ($n<30$)} \\
Llama 3.1 & ABSTAIN & 0 & 740 & 0 & \multicolumn{3}{c}{Sparse ($n<30$)} \\
Llama 3.1 & OTHER & 0 & 26986 & 0 & \multicolumn{3}{c}{Sparse ($n<30$)} \\
Llama 3.3 & APPROVE & 24523 & 34870 & 16016 & 0.653101 & 0.459306 & 0.539323 \\
Llama 3.3 & REJECT & 23 & 4884 & 6 & \multicolumn{3}{c}{Sparse ($n<30$)} \\
Llama 3.3 & ABSTAIN & 0 & 466 & 0 & \multicolumn{3}{c}{Sparse ($n<30$)} \\
Llama 3.3 & OTHER & 1 & 49000 & 1 & \multicolumn{3}{c}{Sparse ($n<30$)} \\
ChatGPT & APPROVE & 30599 & 34640 & 19477 & 0.636524 & 0.562269 & 0.597097 \\
ChatGPT & REJECT & 0 & 4558 & 0 & \multicolumn{3}{c}{Sparse ($n<30$)} \\
ChatGPT & ABSTAIN & 0 & 701 & 0 & \multicolumn{3}{c}{Sparse ($n<30$)} \\
ChatGPT & OTHER & 0 & 25578 & 0 & \multicolumn{3}{c}{Sparse ($n<30$)} \\
\bottomrule
\end{tabularx}
\endgroup

\parbox{\linewidth}{\footnotesize \textit{Notes}: For each source, we treat the detected suggested stance as a prediction (column Pred.) and we count how many times the prediction was actually true (column TP) over all the times a given stance won (column Winner). With this framework, we can compute precision, recall, and F1 to capture alignment between textual suggestions and realized voting outcomes.}
\end{table}

\clearpage

\subsubsection{Prompts for LLMs}\label{sec_llm_prompts}

\paragraph{Llama models}

{\scriptsize
\begin{verbatim}

Reading the title, body, and choices of following proposal, you have to assign one of
the following category to each choice in the proposal:

- APPROVE: supports implementing the proposal's main action or change. This includes 
adopting new policies, updating code, adding members, transferring money, establishing 
partnerships, organizing events, or any action that moves the proposal forward (including
reversals or cancellations if that's what the proposal is asking for).
- REJECT: opposes implementing the proposal's main action - it wants to maintain the 
current state and prevent the proposed change from happening.
- ABSTAIN: leaves the decision to other voters.
- OTHER: if the choice cannot be categorized as either APPROVE, REJECT, or ABSTAIN

A proposal can have multiple choices with the same category.

You must label ALL choices of a proposal as OTHER when:
    
- There is only one choice in a proposal.
- The choices are opinions on issues unrelated to the DAO (e.g., opinion polls about politics,
weather, life-style, or coins).
- The set of choices of a proposal contains one or more APPROVE choices, but no REJECT choice 
(and vice versa).

Please do **not** repeat the question in your answer, and include _only_ one valid JSON 
object containing the following properties:

*   effect: Briefly describe the primary consequence(s) of the proposal if it is approved. 
Use N/A if the proposal would not result in any direct action or immediate consequence.
*   stances: The stance of all the choices in a JSON array; i.e., if there are 3 choices, the 
output should look like [x,y,z], where x, y, z are the stance of the first, second and 
third choice.
*   suggested\_stance: If the way the proposal is written (considering title, body, and 
choices) in a way that suggests or invites voters to vote for choices with a particular 
stance, please write it here, else write N/A.


Do not add any other text to your answer.

Here are the title, body, and choices (separated by '{sep}'), of the proposal:

---

### Proposal:

**Title:**
{title}

**Body:**
{body}

**Choices:**
{choices}

\end{verbatim}
}

\clearpage

\paragraph{OpenAI}

{\scriptsize

\begin{verbatim}

Your task is to categorize the stance of the choices of proposals from 
Decentralised Autonomous Organisation (DAO) on Snapshot.org.

It is crucial to be precise.

Every choice can have one of the following stances:

- APPROVE: selecting this choice brings about a change in the DAO, for instance a change in the
code or in the members of the board or in the governance more broadly, or it involves
transferring money, or establishing a new partnership, or organizing or actively participating
in an event.
- REJECT: selecting this choice cancels the changes caused by choices with APPROVE stance.
- ABSTAIN: selecting this choice leaves the decision to other voters.
- OTHER: if the choice cannot be categorized as either APPROVE, REJECT, or ABSTAIN

You must labels _all_ choices of a proposal as OTHER when:

- There is only one choice in a proposal.
- The choices are opinions on issues unrelated to the DAO (e.g., opinion polls about politics, 
weather, life-style, or coins).
- The set of choices of a proposal contains one or more APPROVE choices, but no REJECT choice 
(and vice versa).

Please do **not** repeat the question in your answer, and include _only_ one valid JSON object containing
the following properties:

*   effect: Briefly describe the primary consequence(s) of the proposal if it is approved. Use N/A
if the proposal would not result in any direct action or immediate consequence.
*   stances: The stance of all the choices in a JSON array; i.e., if there are 3 choices, the output
should look like [x,y,z], where x, y, z are the stance of the first, second and third choice.
*   suggested\_stance: If the way the proposal is written (considering title, body, and choices) in a way
that suggests or invites voters to vote for choices with a particular stance, please write it here, 
else write N/A.

Do not add any other text to your answer.


\end{verbatim}
}

\clearpage

\subsection{Human Evaluations}\label{appendix_human_eval}

We perform an additional robustness check on the heuristic stance classification. Our aim is to measure the extent to which our heuristics agree with human evaluations, thereby assessing their external validity. We first create a stratified multi-language sample of random proposals and then assign these to crowdworkers for evaluation; the evaluation was implemented with NodeGame~\citep{balietti_nodegame_2017}. 

\subsubsection{Ethical statement}

The study followed standard ethical procedures for incentivized experiments involving human participants. All participants were at least 18 years old and provided informed consent before participating. Participants were informed about the study procedures and their voluntary participation and could withdraw at any time without penalty. The study involved minimal foreseeable risks, and no personally identifiable information was collected. In accordance with § 6(1) of the Statute of the Ethics Committee of the University of Mannheim, ethical review was not required, as the study did not involve circumstances warranting consultation with the ethics committee.

\subsubsection{Sample construction}

We select a set of representative proposals as follows. First, we computed the embeddings of the content of each proposal (title, body, and choices) using a multilingual sentence transformer model (intfloat/multilingual-e5-large); second, we reduced the dimensionality using UMAP; third, we computed 50 coherent clusters using K-Means; finally, we picked 20 random proposals within each cluster for a total of 1,000 proposals to further sample in the next step.

Next, we remove all proposals with less than two choices and more than \DaoBiasHumanMaxChoiceN{} choices (more than \DaoBiasHumanMaxChoiceN{} choices would be tiresome for humans to review); then, to ensure multi-language support, the evaluated sample includes proposals in the most common non-English languages in the dataset: Chinese (\DaoBiasHumanChineseProposalN{}), Japanese (\DaoBiasHumanJapaneseProposalN{}), and Spanish (\DaoBiasHumanSpanishProposalN{}). The remainder (\DaoBiasHumanEnglishProposalN{}) was filled exclusively from English-language entries\footnote{This is to avoid assigning proposals of different languages to evaluators.}. To ensure diversity across different content clusters, we sampled roughly equally from each cluster.\footnote{Korean proposals were selected but eventually not evaluated due to an assignment bug, so the total number of evaluated proposals is \DaoBiasHumanEvaluatedProposalN{}.}

\subsubsection{Crowd evaluation}

Using the online labor market for research Prolific, we assign a group of evaluators (N = \DaoBiasHumanEvaluatorN{}) the task of manually labeling five proposals, so that each proposal is reviewed at least \DaoBiasHumanProposalReviewMinN{} times. We used platform filters to select crowdworkers fluent in English and, in addition, having Japanese, Chinese, or Spanish as their primary language for the non-English sets. Because crowdworkers are unlikely to have background knowledge of DAOs, we implemented a two-step quiz procedure. At step 1, we explained what DAOs are and how voting takes place in DAOs (one multiple-choice quiz question). At step 2 we explained the rules for labeling the choices of governance proposals (five mandatory multiple-choice quiz questions). Participants who failed twice at any step were required to leave the task; this ensured an informed base for the decisions of all crowdworkers.

For each proposal, crowdworkers had three subtasks: (i) to label the stance of each choice (`Abstain', `Approve', `Other', or `Reject'), (ii) to mark if the proposal was real or spam, and (iii) to indicate whether the text of the proposal is written in a way that suggests a given action (e.g., `Approve').

At the end of the task, crowdworkers were asked to express how clear the task and proposals were. 86.1\% of them completely or somewhat agreed that the task was clear, while 74.9\% of them completely or somewhat agreed that the proposals were clear.
Participants who completed the entire process received a fixed payment of US\$2 for the task, and it took on average \DaoBiasHumanAverageTaskMinutes{} minutes to complete it. Further information on crowdworkers' demographics is reported in Table~\ref{tab:demographics}.

\begin{table}
\centering
\caption{Demographic breakdown of evaluators}
\label{tab:demographics}
\begin{tabular}{lc@{\hspace{1cm}}c@{\hspace{1cm}}c}
\toprule
Category & All evaluators & Fast evaluators & Normal evaluators \\
& (N =354) & (N = 6 ) & (N = 348)\\
\midrule
Gender  &  &  & \\
\midrule
Male & 210 & 2 & 208 \\
Female & 142 & 4 & 138 \\
Non-binary & 2 & 0 & 2 \\
\midrule
Age Group  &  &  & \\
\midrule
18-20 & 11 & 0 & 11 \\
21-25 & 43 & 1 & 42 \\
26-30 & 66 & 3 & 63 \\
31-35 & 48 & 2 & 46 \\
36-40 & 59 & 0 & 59 \\
41-45 & 35 & 0 & 35 \\
46-50 & 23 & 0 & 23 \\
51-55 & 22 & 0 & 22 \\
56-60 & 18 & 0 & 18 \\
61-65 & 14 & 0 & 14 \\
66-70 & 7 & 0 & 7 \\
71-75 & 8 & 0 & 8 \\
\midrule
Education   &  &  & \\
\midrule
High-School & 65 & 0 & 65 \\
College & 147 & 3 & 144 \\
Grad School & 142 & 3 & 139 \\
\midrule 
Prior knowledge of DAOs  &  &  & \\
\midrule
No & 286 & 6 & 280 \\
Yes & 68 & 0 & 68 \\
\midrule 
Time to conduct tasks  &  &  & \\
\midrule
Avg. Time & 14.09 & 3.98 & 14.27 \\
\midrule 
Task clarity  &  &  & \\
\midrule
Completely agree & 165 & 1 & 164 \\
Somewhat agree & 140 & 4 & 136 \\
Neither agree nor disagree & 22 & 1 & 21 \\
Somewhat disagree & 26 & 0 & 26 \\
Completely disagree & 1 & 0 & 1 \\
\midrule 
Proposal clarity  &  &  & \\
\midrule
Completely agree & 118 & 0 & 118 \\
Somewhat agree & 147 & 4 & 143 \\
Neither agree nor disagree & 39 & 0 & 39 \\
Somewhat disagree & 45 & 2 & 43 \\
Completely disagree & 5 & 0 & 5 \\
\bottomrule
\end{tabular}
\parbox{\linewidth}{\footnotesize \textit{Notes}: the evaluators are N = 375. The demographics are computed on the restricted set of users who completed the last task (i.e., providing user information and feedback).}
\end{table}

\subsubsection{Comparison of heuristics and human evaluations}\label{appendix:crowdwork_eval}

In the final step, we aggregate the evaluations provided by different crowdworkers and assign each stance a unique label (`Abstain', `Approve', `Other', or `Reject').

To do so, we adopt a majority rule approach and assign a label to a stance only when it is possible to identify one single most occurring label; if two or more evaluations are tied as the most occurring, we leave the label unassigned. For instance, if the evaluations for one stance are [Reject, Reject, Other, Other], that stance is left unassigned.
Next, we fill in unassigned stances when the label can be inferred from the remaining choices of the same proposal. Specifically, proposals that include one `Approve' typically also include a `Reject' option (and vice-versa). We thus identify cases where a proposal has exactly one `Approve' (`Reject') choice, one unassigned choice, and all remaining choices labeled as `Other' or `Abstain'. If at least one annotator assigned the missing label to the `Reject' (`Approve') option, then we label the non-assigned choice accordingly. 
As an illustrative example, a proposal with choices labeled as \{N/A, Reject, Abstain\} after the first step will be updated to \{Approve, Reject, Abstain\} if at least one crowdworker evaluated the missing label as `Approve'.

We flag evaluations that are potentially carried out with low effort in two complementary ways: first, we identify those where users spent unusually little time completing their task (i.e., less than five minutes, considering that the task requires approximately ten minutes to be conducted). Second, we group proposals per language, normalize their body lengths, and identify those with short evaluation time (less than 20 seconds) despite being relatively long proposals (above the 33rd percentile of all proposals).

Lastly, we measure to what extent the heuristics are consistent with the crowdworkers' evaluations. The Cohen's Kappa of the heuristics and the human evaluators is \DaoBiasHumanKappa{}, which indicates \DaoBiasHumanKappaAgreement{} agreement. This comparison covers \DaoBiasHumanLabeledChoiceN{} choices, and the result does not change when removing the \DaoBiasHumanLowEffortEvaluatorN{} evaluators flagged as low effort.

\clearpage

\section{Additional Material and Info}

In this section, we report additional regression and statistical tables to support the claims of the paper, divided by the section of the main paper in which they are most closely referenced.

\subsection{Position Bias}

Table \ref{table_stats_vp_n} and Table \ref{table_stats_3biases_bytype_firstchoice} show descriptive statistics about voting power, participation, and the three biases under study.

\begin{table}
\centering
\caption{Choice position, voting-power share, and individual-vote share}
\label{table_stats_vp_n}
\begin{tblr}{
colspec={
Q[l]
Q[r, colsep=6pt]
Q[r, colsep=6pt]
Q[r, colsep=20pt]
Q[r, colsep=6pt]
Q[r, colsep=10pt]
Q[r]
},
hline{1}={1-7}{solid, black, 0.1em},
hline{2}={2-3}{solid, black, 0.05em},
hline{2}={4-5}{solid, black, 0.05em},
hline{3}={1-7}{solid, black, 0.05em},
hline{14}={1-7}{solid, black, 0.1em},
cell{1}{2}={c=2}{halign=c},
cell{1}{4}={c=2}{halign=c},
}
& Voting Power &  & Number of votes &  & Percent & N \\
Choice Pos. & Mean & SD & Mean & SD &  &  \\
1 & \num{0.70} & \num{0.39} & \num{0.71} & \num{0.37} & \num{31.96} & 127,281 \\
2 & \num{0.22} & \num{0.35} & \num{0.22} & \num{0.32} & \num{31.96} & 127,281 \\
3 & \num{0.14} & \num{0.26} & \num{0.13} & \num{0.24} & \num{9.87} & 39,303 \\
4 & \num{0.14} & \num{0.24} & \num{0.13} & \num{0.22} & \num{3.48} & 13,839 \\
5 & \num{0.11} & \num{0.22} & \num{0.11} & \num{0.19} & \num{2.04} & 8,130 \\
6 & \num{0.09} & \num{0.18} & \num{0.08} & \num{0.15} & \num{1.28} & 5,110 \\
7 & \num{0.07} & \num{0.15} & \num{0.07} & \num{0.13} & \num{0.97} & 3,853 \\
8 & \num{0.06} & \num{0.14} & \num{0.06} & \num{0.13} & \num{0.81} & 3,209 \\
9 & \num{0.05} & \num{0.13} & \num{0.06} & \num{0.12} & \num{0.69} & 2,766 \\
10 & \num{0.05} & \num{0.12} & \num{0.05} & \num{0.10} & \num{0.62} & 2,482 \\
10+ & \num{0.01} & \num{0.06} & \num{0.01} & \num{0.05} & \num{16.31} & 64,964 \\
\end{tblr}
\parbox{\linewidth}{\footnotesize Notes:  The sample is the final cleaned dataset with \DaoBiasFinalChoiceN{} choices. Outcomes are mean choice-level voting-power share and mean favorite-choice vote share by choice position; N indicates the number of choices with a given position index; positions above 10 grouped as 10+.}
\end{table}

\begin{table}
\centering
\caption{Voting power, frequency of approval and author choices of the first choice by voting type}
\label{table_stats_3biases_bytype_firstchoice}
\begin{tblr}[         %
]                     %
{                     %
colspec={Q[]Q[]Q[]Q[]Q[]Q[]Q[]Q[]Q[]},
hline{2}={9}{solid, black, 0.05em},
hline{3}={1-9}{solid, black, 0.05em},
hline{2}={2,4,6,8}{solid, black, 0.05em, l=-0.5},
hline{2}={3,5,7}{solid, black, 0.05em, r=-0.5},
hline{1}={1-9}{solid, black, 0.1em},
hline{9}={1-9}{solid, black, 0.1em},
cell{1}{1}={}{halign=c},
cell{1}{2}={c=2}{halign=c},
cell{1}{3}={}{halign=c},
cell{1}{4}={c=2}{halign=c},
cell{1}{5}={}{halign=c},
cell{1}{6}={c=2}{halign=c},
cell{1}{7}={}{halign=c},
cell{1}{8}={c=2}{halign=c},
cell{1}{9}={}{halign=c},
cell{2-8}{1}={}{halign=l},
cell{2-8}{2}={}{halign=r},
cell{2-8}{3}={}{halign=r},
cell{2-8}{4}={}{halign=r},
cell{2-8}{5}={}{halign=r},
cell{2-8}{6}={}{halign=r},
cell{2-8}{7}={}{halign=r},
cell{2-8}{8}={}{halign=r},
cell{2-8}{9}={}{halign=r},
}                     %
& Voting Power &  & Approve &  & Author Choice &  & All &  \\
type & Mean & SD & Mean & SD & Mean & SD & Percent & N \\
Basic & \num{0.79} & \num{0.34} & \num{0.77} & \num{0.42} & \num{0.84} & \num{0.37} & \num{7.05} & 8,975 \\
Single-choice & \num{0.71} & \num{0.39} & \num{0.38} & \num{0.49} & \num{0.73} & \num{0.44} & \num{85.95} & 109,398 \\
\hline
Approval & \num{0.42} & \num{0.37} & \num{0.20} & \num{0.40} & \num{0.67} & \num{0.47} & \num{1.21} & 1,537 \\
Quadratic & \num{0.50} & \num{0.42} & \num{0.30} & \num{0.46} & \num{0.69} & \num{0.46} & \num{1.41} & 1,793 \\
Ranked-choice & \num{0.40} & \num{0.38} & \num{0.19} & \num{0.39} & \num{0.51} & \num{0.50} & \num{0.71} & 907 \\
Weighted & \num{0.47} & \num{0.41} & \num{0.36} & \num{0.48} & \num{0.64} & \num{0.48} & \num{3.67} & 4,671 \\
\end{tblr}
\parbox{\linewidth}{\footnotesize \textit{Notes}: the sample is the final cleaned dataset with \DaoBiasFinalChoiceN{} choices.}
\end{table}

\clearpage
\subsection{Approval Bias}

\begin{figure}
    \centering
    \includegraphics[width=\linewidth]{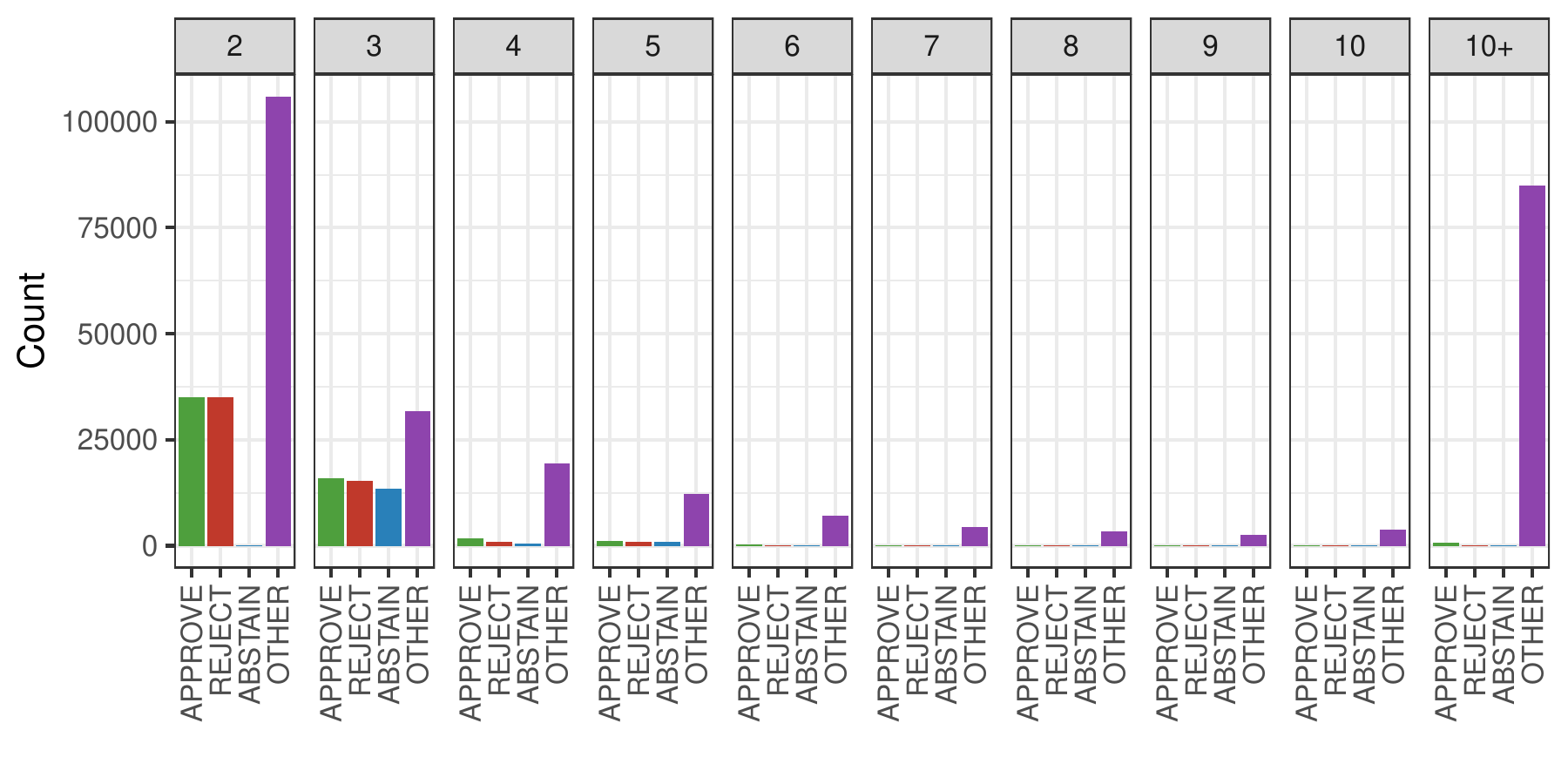}
    \caption{\textbf{Choice stance counts by number of choices in a proposal (full sample).}}
    \label{fig:distr_choice_stance}
\end{figure}

Table \ref{table_stats_stance_choice_idx} reports the stance (Approve, Reject, Abstain, Other) shares for every position across all proposals. Fig. \ref{fig:distr_choice_stance} reports the stance counts by the number of choices in the proposal. Together, they show that the stance Other is especially common in proposals with many choices, while two- and three-choice proposals more often follow an Approve/Reject/Abstain structure. 

\begin{figure}
    \centering
    \includegraphics[width=\linewidth]{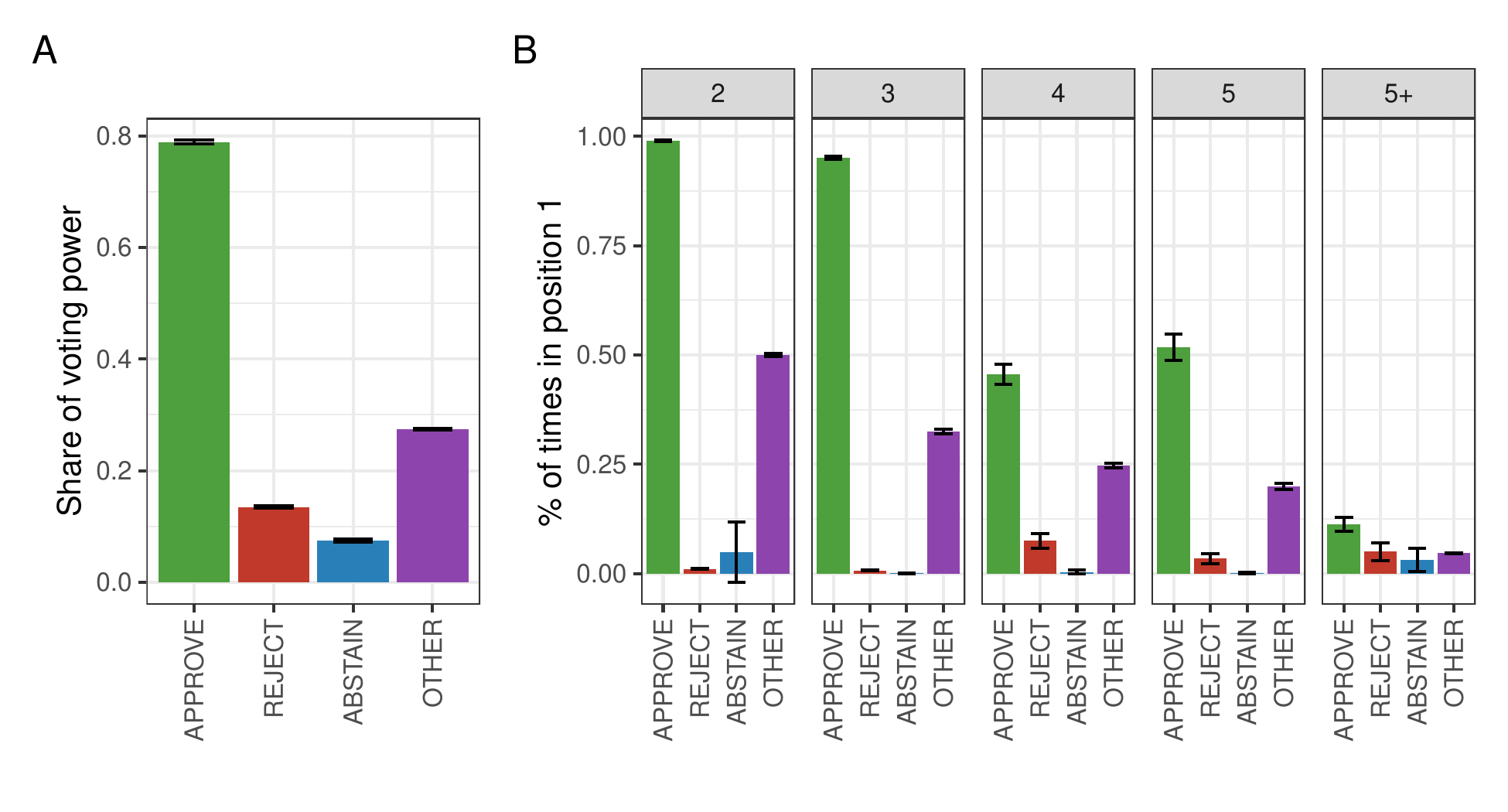}
    \caption{\textbf{Voting power and position by stance (full sample).} \textbf{A.} Mean choice-level voting-power share by stance. \textbf{B.} Share of choices with a given stance that are in position 1, by number of choices in the proposal. These shares are conditional on stance and therefore do not describe the overall distribution of stances. Confidence intervals are 95\% confidence intervals of the means.}
    \label{fig:approve_bias_app}
\end{figure}

Fig. \ref{fig:approve_bias_app} tackles a different question. Panel A compares the average voting-power share of choices with each stance. Panel B reports, within each stance and choice-count group, the share of choices that appear in the first ballot position. Thus, Fig. \ref{fig:distr_choice_stance} describes how common each stance is, whereas Fig. \ref{fig:approve_bias_app} describes voting power and ballot placement conditional on stance. From Panel B, we learn that when an Approve choice is found in a proposal, it is almost always placed in position 1 for proposals with two or three choices; overall Approve is the most common stance across all sub-panels. 

\begin{figure}
    \centering
    \includegraphics[width=1\linewidth]{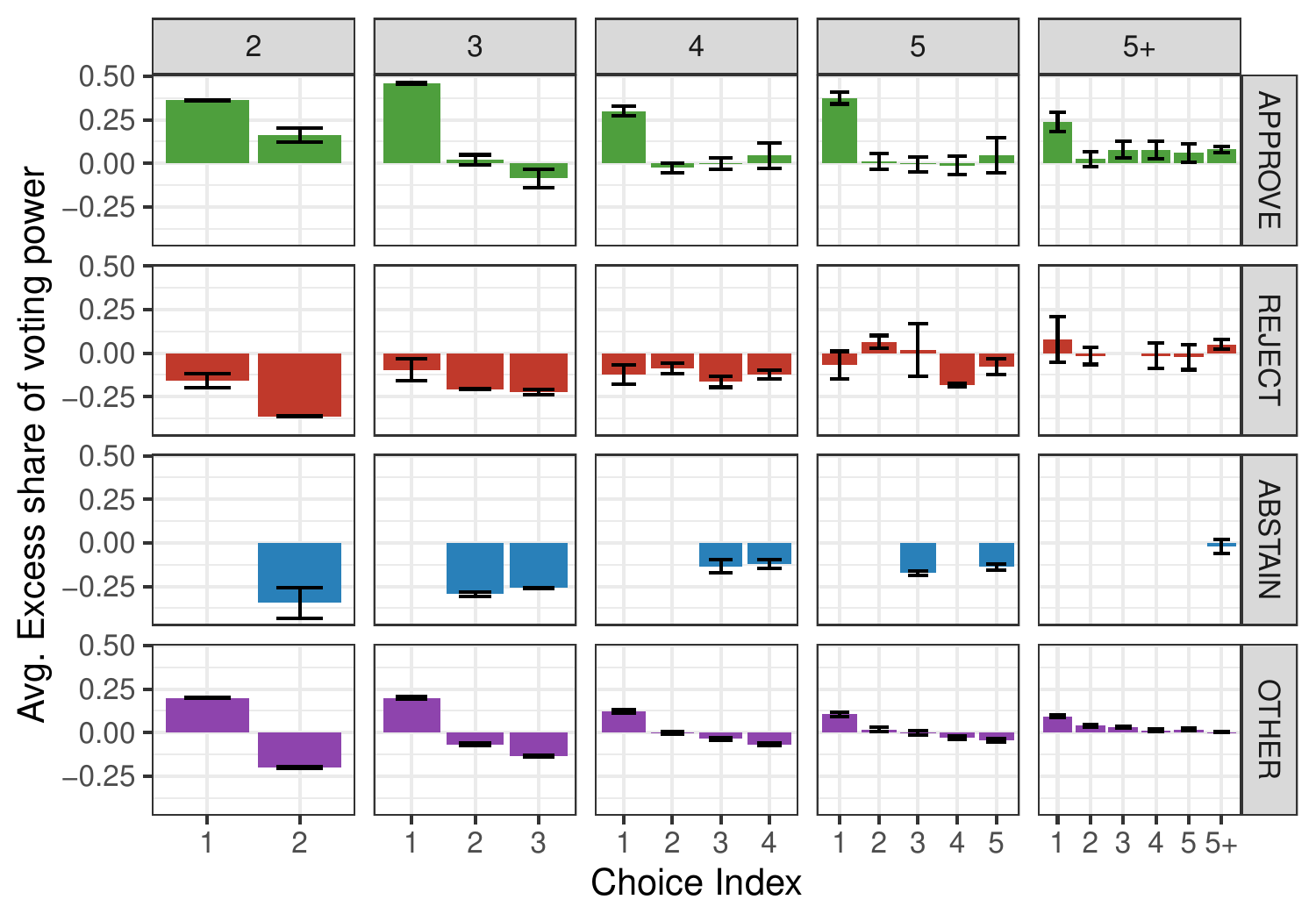}
    \caption{\textbf{Excess voting power by stance, choice position, and number of choices (full sample).} Only stance-position-num-choice combinations with more than 10 observations are kept in the plot. Error bars are 95\% confidence intervals of the means.}
    \label{fig:avg_excess_vp_choice_stance_by_num_choices}
\end{figure}

Finally, Fig. \ref{fig:avg_excess_vp_choice_stance_by_num_choices} extends Fig. \ref{fig:avg_excess_vp_choice_stance} in the main text, disaggregating the excess/deficit of voting power by the number of choices in a proposal. The qualitative findings remain the same. 

\begin{table}
\centering
\caption{Stance distribution by choice position}
\label{table_stats_stance_choice_idx}
\begin{tblr}{
colspec={Q[]Q[]Q[]Q[]Q[]Q[]Q[]Q[]Q[]Q[]Q[]},
hline{2}={11}{solid, black, 0.05em},
hline{3}={1-11}{solid, black, 0.05em},
hline{1}={1-11}{solid, black, 0.1em},
hline{14}={1-11}{solid, black, 0.1em},
cell{1}{2}={c=2}{halign=c},
cell{1}{4}={c=2}{halign=c},
cell{1}{6}={c=2}{halign=c},
cell{1}{8}={c=2}{halign=c},
cell{1}{10}={c=2}{halign=c},
cell{2-13}{1}={}{halign=l},
cell{2-13}{2-11}={}{halign=r},
}
& Approve &  & Reject &  & Abstain &  & Other &  & All &  \\
Choice Position & Mean & SD & Mean & SD & Mean & SD & Mean & SD & Percent & N \\
1 & \num{0.40} & \num{0.49} & \num{0.00} & \num{0.07} & \num{0.00} & \num{0.01} & \num{0.59} & \num{0.49} & \num{31.96} & 127,281 \\
2 & \num{0.01} & \num{0.12} & \num{0.39} & \num{0.49} & \num{0.00} & \num{0.05} & \num{0.59} & \num{0.49} & \num{31.96} & 127,281 \\
3 & \num{0.02} & \num{0.15} & \num{0.03} & \num{0.18} & \num{0.35} & \num{0.48} & \num{0.60} & \num{0.49} & \num{9.87} & 39,303 \\
4 & \num{0.03} & \num{0.16} & \num{0.06} & \num{0.24} & \num{0.03} & \num{0.17} & \num{0.89} & \num{0.32} & \num{3.48} & 13,839 \\
5 & \num{0.02} & \num{0.14} & \num{0.02} & \num{0.13} & \num{0.06} & \num{0.24} & \num{0.90} & \num{0.30} & \num{2.04} & 8,130 \\
6 & \num{0.01} & \num{0.11} & \num{0.01} & \num{0.12} & \num{0.01} & \num{0.11} & \num{0.96} & \num{0.19} & \num{1.28} & 5,110 \\
7 & \num{0.02} & \num{0.13} & \num{0.01} & \num{0.08} & \num{0.00} & \num{0.07} & \num{0.97} & \num{0.16} & \num{0.97} & 3,853 \\
8 & \num{0.01} & \num{0.10} & \num{0.01} & \num{0.11} & \num{0.00} & \num{0.06} & \num{0.98} & \num{0.16} & \num{0.81} & 3,209 \\
9 & \num{0.01} & \num{0.11} & \num{0.00} & \num{0.06} & \num{0.00} & \num{0.04} & \num{0.98} & \num{0.13} & \num{0.69} & 2,766 \\
10 & \num{0.01} & \num{0.10} & \num{0.01} & \num{0.08} & \num{0.01} & \num{0.11} & \num{0.97} & \num{0.16} & \num{0.62} & 2,482 \\
10+ & \num{0.01} & \num{0.09} & \num{0.00} & \num{0.04} & \num{0.00} & \num{0.02} & \num{0.99} & \num{0.10} & \num{16.31} & 64,964 \\
\end{tblr}
\end{table}

\clearpage

\subsection{Author Bias}

\begin{figure}[hb]
    \centering
    \includegraphics[width=1\linewidth]{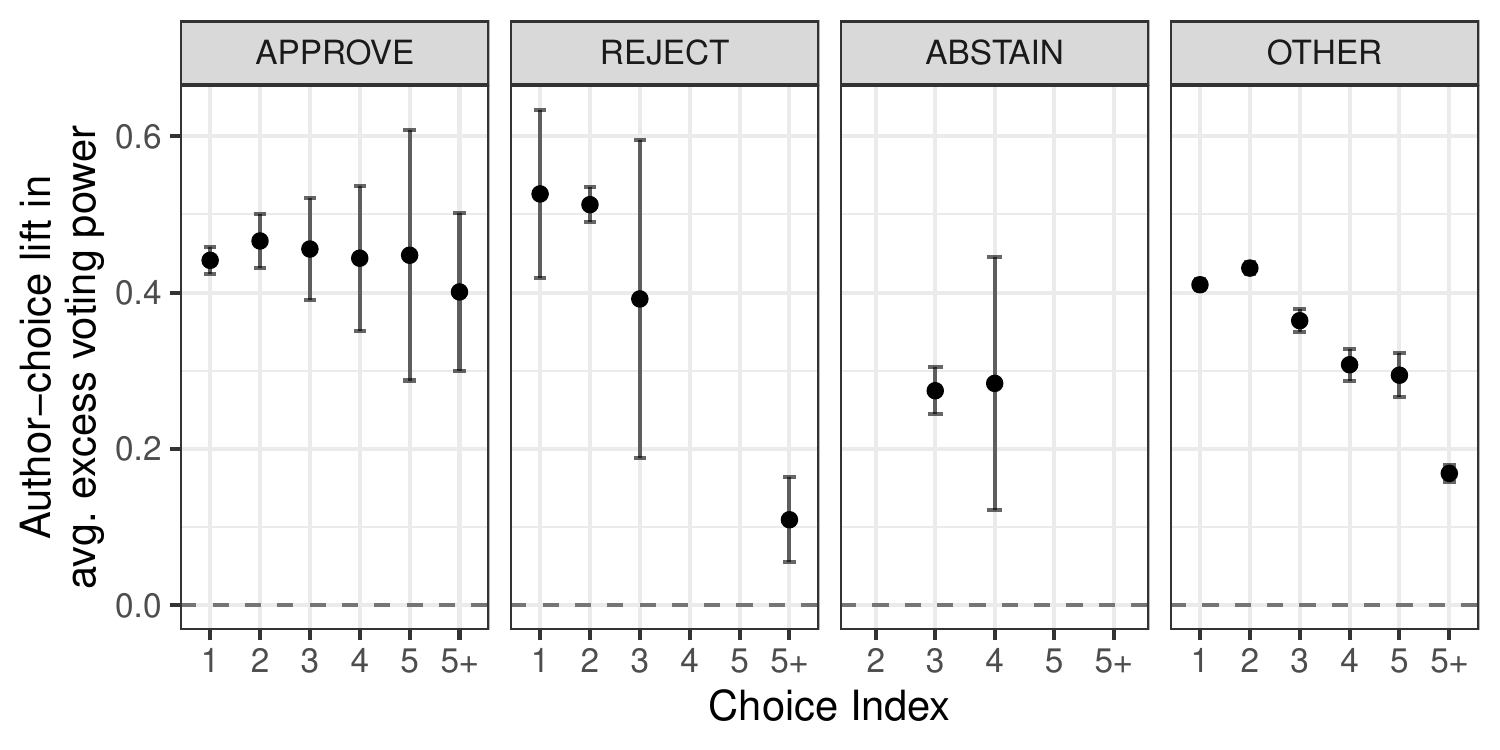}
    \caption{\textbf{Author boost by stance and choice position.} Each dot in the plot is the difference between the point estimate of the average excess voting power of a choice of an author vs other choices. Error bars are 95\% confidence intervals of the means.}
    \label{fig:author_boost_by_stance_and_pos}
\end{figure}

Fig. \ref{fig:author_boost_by_stance_and_pos} shows the boost in excess voting power when an author chooses a choice at a given position with a given stance.

\subsubsection{Timing of Proposal Authors' Votes}
\label{asec:author_vote_timing}

One concern with the author-choice association is that proposal authors
may select alternatives that have already attracted substantial support.
To assess this possibility, we examine the timing of authors' votes
relative to other participants. Using timestamps of individual recorded
ballots, we reconstruct the voting sequence within each proposal and
identify the author's position in that sequence.\footnote{Same-second ballots cannot be ordered
    unambiguously, so only strictly earlier ballots count as preceding
    the author's vote.} We also calculate the
share of non-author voters and the share of eventual non-author voting
power recorded before the author casts a vote. The analysis retains only
proposals with a valid author ballot and consistent voting records.

\begin{figure}[htbp]
    \centering%
        \includegraphics[width=0.85\textwidth]{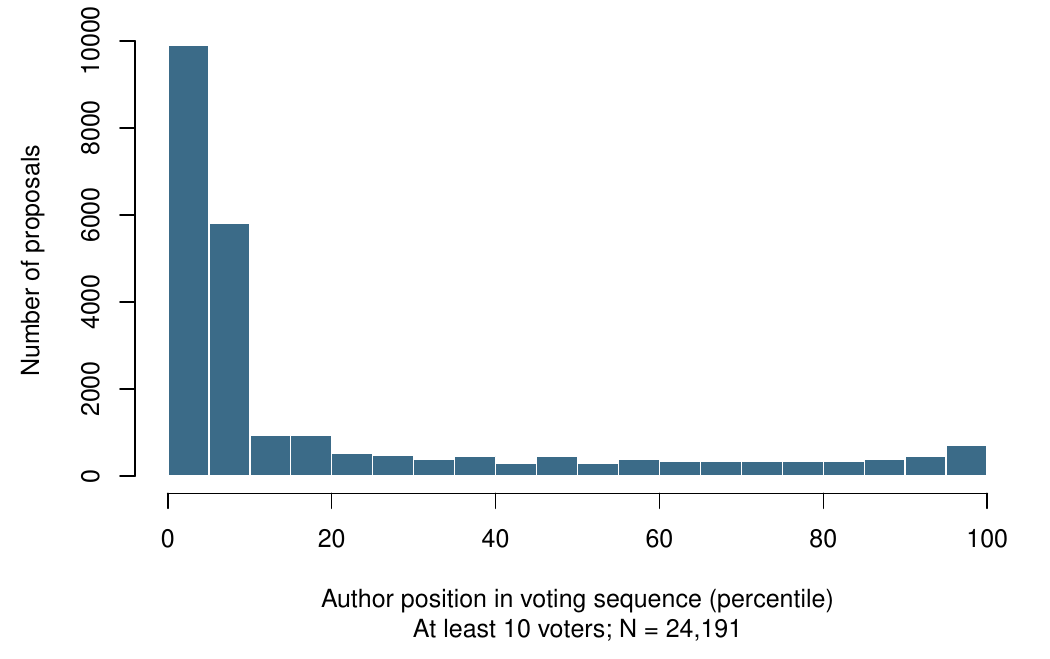}%
    \caption{\textbf{Timing of proposal authors' votes.}
    Distribution of the author's position in the recorded voting sequence
    among proposals with at least ten voters ($N=24{,}191$). Position is
    measured by chronological rank divided by the number of voters,
    expressed as a percentage. Smaller values indicate earlier votes;
    bins span five percentage points.}
    \label{fig:author_vote_timing}
\end{figure}

Figure~\ref{fig:author_vote_timing} plots the distribution of authors'
positions in the voting sequence for proposals with at least ten voters.
Authors tend to vote early: in 54.2\% of these proposals, no non-author
ballot was recorded before the author's vote, and in 64.9\% the author
voted within the first decile of the voting sequence.

Table~\ref{tab:author_vote_timing} reports the timing statistics under alternative minimum-participation thresholds. The concentration of author votes early in the sequence persists in more heavily participated proposals. Among proposals with at least ten voters, authors vote before half of the other participants in 83.5\% of cases. At the time of the author's vote, the median share of eventual non-author voting power already recorded is zero (mean: 16.3\%).

\begin{table}[htbp]
    \centering
    \caption{\textbf{Timing of proposal authors' votes by participation level}}
    \label{tab:author_vote_timing}
    \small
    \begin{tabular}{rrrrrr}
        \toprule
        Minimum & Num. & Author & First & Before half & Prior non-author VP \\
        voters  & proposals & first (\%) & decile (\%) & of others (\%) & mean (\%) \\
        \midrule
        5  & 33,126 & 59.6 & 47.4 & 84.5 & 15.5 \\
        10 & 24,191 & 54.2 & 64.9 & 83.5 & 16.3 \\
        20 & 16,412& 51.1 & 66.8 & 83.7 & 15.7 \\
        \bottomrule
    \end{tabular}

    \vspace{0.3em}
    \begin{minipage}{0.96\textwidth}
    \footnotesize
    \textit{Notes:} ``Author first'' means that no non-author
    ballot was recorded strictly before the author's ballot. ``First
    decile'' indicates that the author's rank divided by the number of
    distinct voters does not exceed $0.10$. ``Before half of others''
    indicates that fewer than half of the other voters had cast their
    ballots when the author voted. ``Prior non-author VP'' measures the share of voting power cast by other voters before the author voted, relative to the total voting power cast by other voters. Proposals in which no other voter contributed voting power are excluded from this calculation.
    \end{minipage}
\end{table}

Taken together, these descriptive patterns limit explanations based
solely on authors responding to votes that had already been recorded:
authors frequently participate before most other voters. However,
early voting does not demonstrate that authors influence subsequent
voters. Authors may anticipate community preferences or coordinate
with other participants before the formal vote.

\clearpage

\subsection{Regression Robustness Specifications and Sensitivity Tables}

Table~\ref{tab:robustness-three-bias-summary} reports the shared robustness grid for the three focal AMEs. Cells show response-scale AMEs with Bonferroni/FWER-adjusted DAO-space clustered-bootstrap confidence intervals, using the same 36-AME family as Figure~\ref{fig:three-bias-forest}. This replaces separate one-off inference tables for the primary, author-voted-only, and participation-control rows, so every shared robustness specification is reported with the same three-bias contrast set and uncertainty convention.
We include \textit{Vote-count} as a shared outcome robustness row, keeping the vote-count-share sensitivity inside this common grid rather than in a standalone table. 
The first-vs-second position AME is a counterfactual prediction contrast, not a first-vs-rest effect and not the coefficient on choice rank. For each fitted model, we predict outcomes on the estimation sample with choice rank set to 1 and then to 2, recomputing the squared-rank term in each counterfactual, and average the response-scale difference.

\begin{table}[!tbp]
\centering
\caption{Three-bias AMEs across shared robustness specifications}
\label{tab:robustness-three-bias-summary}
\small
\setlength{\tabcolsep}{2pt}
\renewcommand{\arraystretch}{1.10}
\begin{tabularx}{\linewidth}{@{}>{\RaggedRight\arraybackslash}X>{\RaggedLeft\arraybackslash}p{0.19\linewidth}>{\RaggedLeft\arraybackslash}p{0.19\linewidth}>{\RaggedLeft\arraybackslash}p{0.19\linewidth}@{}}
\toprule
Specification & \makecell[r]{Author\\choice} & \makecell[r]{First vs\\second} & \makecell[r]{Approve\\stance} \\
\midrule
Main & \makecell[r]{58.78\\{\scriptsize [53.82, 63.74]}} & \makecell[r]{7.75\\{\scriptsize [5.41, 10.08]}} & \makecell[r]{27.14\\{\scriptsize [20.52, 33.76]}} \\
Vote-count & \makecell[r]{57.64\\{\scriptsize [53.46, 61.83]}} & \makecell[r]{8.25\\{\scriptsize [5.71, 10.79]}} & \makecell[r]{29.72\\{\scriptsize [24.25, 35.18]}} \\
Auth-voted & \makecell[r]{65.88\\{\scriptsize [61.34, 70.42]}} & \makecell[r]{4.29\\{\scriptsize [2.37, 6.21]}} & \makecell[r]{7.72\\{\scriptsize [4.66, 10.78]}} \\
Expanded & \makecell[r]{48.02\\{\scriptsize [43.34, 52.70]}} & \makecell[r]{6.29\\{\scriptsize [4.24, 8.35]}} & \makecell[r]{24.64\\{\scriptsize [18.65, 30.62]}} \\
Full & \makecell[r]{44.56\\{\scriptsize [39.84, 49.27]}} & \makecell[r]{1.16\\{\scriptsize [-0.27, 2.59]}} & \makecell[r]{29.07\\{\scriptsize [22.82, 35.33]}} \\
Vote-control & \makecell[r]{58.78\\{\scriptsize [53.81, 63.75]}} & \makecell[r]{7.74\\{\scriptsize [5.41, 10.08]}} & \makecell[r]{27.14\\{\scriptsize [20.52, 33.76]}} \\
TVL-25 & \makecell[r]{27.12\\{\scriptsize [7.59, 46.65]}} & \makecell[r]{7.65\\{\scriptsize [-9.97, 25.28]}} & \makecell[r]{51.72\\{\scriptsize [30.48, 72.96]}} \\
TVL-50 & \makecell[r]{32.61\\{\scriptsize [14.46, 50.77]}} & \makecell[r]{10.70\\{\scriptsize [-5.18, 26.57]}} & \makecell[r]{46.41\\{\scriptsize [28.25, 64.56]}} \\
TVL-100 & \makecell[r]{32.92\\{\scriptsize [16.12, 49.73]}} & \makecell[r]{11.17\\{\scriptsize [-4.12, 26.46]}} & \makecell[r]{45.13\\{\scriptsize [27.65, 62.62]}} \\
Proposal FE & \makecell[r]{56.63\\{\scriptsize [51.09, 62.17]}} & \makecell[r]{5.94\\{\scriptsize [3.75, 8.12]}} & \makecell[r]{35.69\\{\scriptsize [27.25, 44.13]}} \\
Low-part. & \makecell[r]{39.73\\{\scriptsize [34.70, 44.76]}} & \makecell[r]{8.97\\{\scriptsize [6.05, 11.89]}} & \makecell[r]{32.51\\{\scriptsize [24.59, 40.43]}} \\
No top-5 & \makecell[r]{59.31\\{\scriptsize [55.62, 63.01]}} & \makecell[r]{6.91\\{\scriptsize [5.31, 8.50]}} & \makecell[r]{30.08\\{\scriptsize [25.83, 34.32]}} \\
\bottomrule
\end{tabularx}
\vspace{0.25em}
\parbox{\linewidth}{\footnotesize Notes: Cells report response-scale average marginal effects (AMEs) in percentage points with Bonferroni/FWER-adjusted 95\% DAO-space clustered-bootstrap confidence intervals in brackets. Intervals are adjusted across the 36 shared AMEs shown in this table. For each specification, the model is fit on the rows selected by that specification. For the author-choice column, the AME is then averaged over the subset of that model's fitted rows
where \texttt{author\_voted}=1. In the \textit{Auth-voted} row, this subset is the whole fitted sample; in the other rows, it is usually only part of the fitted sample. The author-choice
  column compares author-selected with non-author-selected choices among rows from proposals where the author voted. The first-vs-second column reports the predicted difference between
  choice rank 1 and choice rank 2 using the rank-plus-rank-squared position specification. The approve-stance column reports the response-scale AME for the approval stance, comparing choices coded as approving with choices coded reject, abstain, or other.}
\end{table}

\subsubsection{First-choice position sensitivity}

Tables~\ref{tab:robustness-first-choice-specifications} and~\ref{tab:robustness-first-choice-sensitivity} report the sensitivity check that replaces the rank-plus-rank-squared position controls with a binary first-choice indicator. The coefficient ladder shows the main single-preference specification, while the AME grid mirrors the shared robustness specifications.

\begin{table}[!tbp]
\centering
\caption{First-choice-position sensitivity specifications}
\label{tab:robustness-first-choice-specifications}
\small
\setlength{\tabcolsep}{4pt}
\renewcommand{\arraystretch}{1.10}
\begin{tabularx}{\linewidth}{@{}>{\RaggedRight\arraybackslash}Xrrrrrr@{}}
\toprule
& (1) & (2) & (3) & (4) & (5) & (6) \\
\midrule
Author choice & \makecell[r]{3.250***\\(0.088)} & \makecell[r]{3.250***\\(0.088)} & \makecell[r]{3.237***\\(0.088)} & \makecell[r]{3.237***\\(0.088)} & \makecell[r]{3.242***\\(0.087)} & \makecell[r]{1.937***\\(0.092)} \\
Approving choice & \makecell[r]{1.301***\\(0.134)} & \makecell[r]{1.301***\\(0.134)} & \makecell[r]{1.277***\\(0.135)} & \makecell[r]{1.277***\\(0.135)} & \makecell[r]{1.273***\\(0.135)} & \makecell[r]{1.287***\\(0.135)} \\
First choice & \makecell[r]{1.286***\\(0.052)} & \makecell[r]{1.286***\\(0.052)} & \makecell[r]{1.259***\\(0.051)} & \makecell[r]{1.259***\\(0.051)} & \makecell[r]{1.251***\\(0.051)} & \makecell[r]{1.257***\\(0.051)} \\
\hline
Author voted & \makecell[r]{-1.499***\\(0.039)} & \makecell[r]{-1.499***\\(0.039)} & \makecell[r]{-1.494***\\(0.040)} & \makecell[r]{-1.472***\\(0.038)} & \makecell[r]{-1.474***\\(0.038)} & \makecell[r]{-0.883***\\(0.040)} \\
Relative proposal date &  & \makecell[r]{-0.000\\(0.000)} & \makecell[r]{-0.000\\(0.000)} & \makecell[r]{-0.000\\(0.000)} & \makecell[r]{-0.000\\(0.000)} & \makecell[r]{-0.000\\(0.000)} \\
Number of choices &  &  & \makecell[r]{-0.104***\\(0.016)} & \makecell[r]{-0.104***\\(0.016)} & \makecell[r]{-0.105***\\(0.016)} & \makecell[r]{-0.106***\\(0.016)} \\
Author choice in first six &  &  &  & \makecell[r]{-0.026*\\(0.011)} & \makecell[r]{-0.026*\\(0.011)} & \makecell[r]{-0.683***\\(0.044)} \\
Author choice x first-six visibility &  &  &  &  &  & \makecell[r]{1.443***\\(0.092)} \\
Relative choice text length &  &  &  &  & \makecell[r]{-0.120+\\(0.064)} & \makecell[r]{-0.123+\\(0.063)} \\
\hline
Voting-type controls & Yes & Yes & Yes & Yes & Yes & Yes \\
N & 293816 & 293816 & 293816 & 293816 & 293816 & 293816 \\
\bottomrule
\end{tabularx}
\vspace{0.25em}
\parbox{\linewidth}{\footnotesize Notes: Coefficient estimates are on the logit scale with DAO-space clustered standard errors in parentheses. Significance markers on coefficients: + p below 0.1, * p below 0.05, ** p below 0.01, *** p below 0.001.}
\end{table}

\begin{table}[!tbp]
\centering
\caption{First-choice position-control three-bias AMEs across shared robustness specifications}
\label{tab:robustness-first-choice-sensitivity}
\small
\setlength{\tabcolsep}{4pt}
\renewcommand{\arraystretch}{1.10}
\begin{tabularx}{\linewidth}{@{}>{\RaggedRight\arraybackslash}Xrrr@{}}
\toprule
Specification & \makecell[r]{Author\\choice} & \makecell[r]{First vs\\non-first} & \makecell[r]{Approve\\stance} \\
\midrule
Main & 56.20 & 19.66 & 18.57 \\
Vote-count & 55.05 & 20.38 & 20.95 \\
Auth-voted & 64.62 & 9.10 & 5.52 \\
Expanded & 46.85 & 19.44 & 16.67 \\
Full & 40.92 & 17.07 & 16.03 \\
Vote-control & 56.20 & 19.66 & 18.57 \\
TVL-25 & 18.01 & 43.17 & 22.27 \\
TVL-50 & 21.24 & 45.64 & 20.03 \\
TVL-100 & 21.45 & 45.94 & 19.32 \\
Proposal FE & 54.88 & 16.37 & 25.66 \\
Low-part. & 34.67 & 25.68 & 21.19 \\
No top-5 & 56.83 & 18.14 & 21.63 \\
\bottomrule
\end{tabularx}
\vspace{0.25em}
\parbox{\linewidth}{\footnotesize Notes: Cells report response-scale average marginal effects (AMEs) in percentage points. This grid mirrors the main three-bias summary in Table \ref{tab:robustness-three-bias-summary} but replaces the rank-plus-rank-squared position controls with a binary first-choice indicator. The first-vs-non-first column reports the predicted response-scale difference between first-choice and non-first-choice rows.}
\end{table}

Table~\ref{tab:robustness-top-space-exclusion} documents the DAO spaces excluded in the \textit{No top-5} row of the shared robustness table.

\begin{table}[!tbp]
\centering
\caption{Top spaces excluded in robustness model}
\label{tab:robustness-top-space-exclusion}
\small
\setlength{\tabcolsep}{6pt}
\renewcommand{\arraystretch}{1.08}
\begin{tabular}{@{}rl@{}}
\toprule
Rank & Space \\
\midrule
1 & cakevote.eth \\
2 & snapshot.dcl.eth \\
3 & pancake \\
4 & index-coop.eth \\
5 & polls.lenster.xyz \\
\bottomrule
\end{tabular}
\end{table}

\clearpage
\subsection{Matching of Top-100 DeFi Protocols}\label{sec:app:defillama}

\paragraph{Data collection.}  We downloaded the complete set of approximately 7,500 protocols returned by DefiLlama APIs. Protocols classified under centralized-exchange categories were excluded prior to analysis, to focus on decentralized finance protocols only, the kind of protocols one would expect to have a DAO on Snapshot.

For each eligible protocol, we retrieved the historical TVL records and filtered to retain only the daily total value locked observations associated with the Ethereum chain. The study period was defined as 1 October 2020 through 30 November 2023, inclusive. Observations outside this window were discarded. Because multiple observations may occur for the same UTC calendar date, records were deduplicated at the UTC-date level, retaining the last available observation for each protocol-date pair.

To ensure sufficient temporal coverage, protocols were required to have Ethereum TVL observations for at least 80\% of the study period, corresponding to a minimum of 925 observed days. Protocols below this coverage threshold were excluded. Missing dates were not imputed: no zero-filling, forward-filling, or other interpolation procedure was applied.

Eligible protocols were ranked according to the median of their observed daily Ethereum TVL values during the study period. The median, rather than the mean, was used as the primary ranking statistic to reduce sensitivity to short-lived TVL spikes or extreme observations. Protocols were ordered in descending median Ethereum TVL, with ties resolved alphabetically.

\paragraph{Matching procedures.} We constrained matching to the top 100 DefiLlama's protocols ranked as specified above. The fields \texttt{governanceID} and \texttt{governanceIDs} sometimes contain Snapshot identifiers encoded as \texttt{snapshot:<space-id>}; all such identifiers are extracted and stored. Because DefiLlama organizes many protocols into parent--child hierarchies, governance identifiers are often attached only to the parent record rather than to each child protocol. We therefore query the website-backing \texttt{/lite/protocols2} endpoint to retrieve parent-level governance identifiers, and each child protocol inherits its parent's Snapshot identifiers in a separate parent field, so that the provenance of direct and inherited matches remains distinct. A protocol--space pair is accepted automatically when the Snapshot space appears in either the direct or inherited DefiLlama fields.

The automatic matches are then audited against local Snapshot data. Accepted pairs remain eligible only if the Snapshot space exists in our final cleaned dataset. Going beyond exact matching, we then construct plausible Snapshot identifiers from each protocol's slug and name by normalizing text, stripping common version or category suffixes such as \texttt{-v2}, and appending common Snapshot suffixes such as \texttt{.eth}, \texttt{dao.eth}, and \texttt{gov.eth}. These fuzzy candidates are used for audit and manual review, but they are not accepted automatically. Protocols that remain unresolved are reviewed manually, and the final results are shown in Tables \ref{tab:robustness-top100-tvl-dao-spaces-part1} and \ref{tab:robustness-top100-tvl-dao-spaces-part2}.

\begin{table}[!tbp]
\centering
\caption{Top 100 matched DAO Snapshot-space ranks used in the Top-TVL robustness subset (ranks 1-25)}
\label{tab:robustness-top100-tvl-dao-spaces-part1}
\scriptsize
\setlength{\tabcolsep}{2pt}
\renewcommand{\arraystretch}{1.05}
\begin{tabularx}{\linewidth}{@{}rr>{\RaggedRight\arraybackslash}p{0.16\linewidth}>{\RaggedRight\arraybackslash}X>{\RaggedRight\arraybackslash}p{0.11\linewidth}rrr@{}}
\toprule
\makecell[r]{\#} & \makecell[r]{DefiLlama\\rank} & \makecell[l]{DefiLlama\\protocol} & \makecell[l]{Snapshot\\space} & Category & \makecell[r]{TVL} & \makecell[r]{Prop.} & \makecell[r]{Choices} \\
\midrule
1 & 2 & \makecell[l]{Lido} & {\urlstyle{same}\url{lido-snapshot.eth}} & \makecell[l]{Liquid\\Staking} & \$6,249,613,456 & 156 & 329 \\
2 & 3 & \makecell[l]{Curve DEX} & {\urlstyle{same}\url{curve.eth}} & Dexs & \$4,584,909,044 & 124 & 287 \\
3 & 4 & \makecell[l]{Convex Finance} & {\urlstyle{same}\url{cvx.eth}} & Yield & \$4,079,001,323 & 744 & 10,238 \\
4 & 5 & \makecell[l]{Aave V2} & {\urlstyle{same}\url{aave.eth}} & Lending & \$4,027,204,550 & 457 & 1,392 \\
4 & 35 & \makecell[l]{Aave V1} & {\urlstyle{same}\url{aave.eth}} & Lending & \$80,000,026 & 457 & 1,392 \\
5 & 6 & \makecell[l]{Compound V2} & {\urlstyle{same}\url{comp-vote.eth}} & Lending & \$2,697,179,105 & 5 & 17 \\
6 & 7 & \makecell[l]{Uniswap V3} & {\urlstyle{same}\url{uniswap}} & Dexs & \$2,532,881,366 & 105 & 282 \\
6 & 8 & \makecell[l]{Uniswap V2} & {\urlstyle{same}\url{uniswap}} & Dexs & \$1,740,993,807 & 105 & 282 \\
6 & 77 & \makecell[l]{Uniswap V1} & {\urlstyle{same}\url{uniswap}} & Dexs & \$8,548,008 & 105 & 282 \\
7 & 10 & \makecell[l]{Balancer V2} & {\urlstyle{same}\url{balancer.eth}} & Dexs & \$758,661,665 & 625 & 1,561 \\
7 & 16 & \makecell[l]{Balancer V1} & {\urlstyle{same}\url{balancer.eth}} & Dexs & \$279,586,764 & 625 & 1,561 \\
8 & 12 & \makecell[l]{Yearn Finance} & {\urlstyle{same}\url{veyfi.eth}} & \makecell[l]{Yield\\Aggregator} & \$649,626,401 & 7 & 23 \\
9 & 13 & \makecell[l]{SushiSwap} & {\urlstyle{same}\url{sushigov.eth}} & Dexs & \$615,839,109 & 201 & 585 \\
10 & 14 & \makecell[l]{Synthetix v1+v2} & {\urlstyle{same}\url{synthetix-stakers-poll.eth}} & Synthetics & \$592,131,590 & 2 & 4 \\
11 & 15 & \makecell[l]{dYdX V3} & {\urlstyle{same}\url{dydxgov.eth}} & Derivatives & \$379,380,032 & 35 & 74 \\
12 & 18 & \makecell[l]{Nexus Mutual} & {\urlstyle{same}\url{community.nexusmutual.eth}} & Insurance & \$267,782,615 & 56 & 137 \\
13 & 20 & \makecell[l]{RenVM} & {\urlstyle{same}\url{ren-project.eth}} & Bridge & \$244,749,194 & 31 & 139 \\
14 & 21 & \makecell[l]{Frax} & {\urlstyle{same}\url{frax.eth}} & \makecell[l]{Algo\\Stables} & \$205,967,755 & 348 & 712 \\
15 & 23 & \makecell[l]{Harvest Finance} & {\urlstyle{same}\url{harvestfi.eth}} & \makecell[l]{Yield\\Aggregator} & \$194,147,321 & 7 & 15 \\
16 & 25 & \makecell[l]{Bancor V2.1} & {\urlstyle{same}\url{bancornetwork.eth}} & Dexs & \$173,900,182 & 542 & 1,392 \\
17 & 26 & \makecell[l]{xDAI Stake Bridge} & {\urlstyle{same}\url{xdaistake.eth}} & \makecell[l]{Canonical\\Bridge} & \$168,526,652 & 214 & 471 \\
18 & 27 & \makecell[l]{StakeWise V2} & {\urlstyle{same}\url{stakewise.eth}} & \makecell[l]{Liquid\\Staking} & \$135,455,186 & 73 & 167 \\
19 & 28 & \makecell[l]{Badger DAO} & {\urlstyle{same}\url{badgerdao.eth}} & \makecell[l]{Yield\\Aggregator} & \$132,500,580 & 104 & 280 \\
20 & 29 & \makecell[l]{Loopring} & {\urlstyle{same}\url{loopringdao.eth}} & Dexs & \$123,843,786 & 14 & 188 \\
21 & 31 & \makecell[l]{Index Coop} & {\urlstyle{same}\url{index-coop.eth}} & Indexes & \$101,927,589 & 977 & 2,424 \\
21 & 32 & \makecell[l]{Set Protocol} & {\urlstyle{same}\url{index-coop.eth}} & Indexes & \$94,881,084 & 977 & 2,424 \\
22 & 36 & \makecell[l]{Stake DAO} & {\urlstyle{same}\url{stakedao.eth}} & Yield & \$74,601,704 & 88 & 259 \\
23 & 37 & \makecell[l]{Idle} & {\urlstyle{same}\url{idlefinance.eth}} & \makecell[l]{Yield\\Aggregator} & \$70,569,685 & 110 & 430 \\
24 & 37 & \makecell[l]{Idle} & {\urlstyle{same}\url{staking.idlefinance.eth}} & \makecell[l]{Yield\\Aggregator} & \$70,569,685 & 101 & 352 \\
25 & 39 & \makecell[l]{KEEP Network} & {\urlstyle{same}\url{keepstakers.eth}} & \makecell[l]{Cross\\Chain\\Bridge} & \$70,020,929 & 10 & 67 \\
\bottomrule
\end{tabularx}
\vspace{0.25em}
\parbox{\linewidth}{\footnotesize Notes: This part reports Top-TVL sample ranks 1-25. Not all protocols found in DefiLLama can be matched in our Snapshot database; multiple DefiLlama protocols can be matched to the same Snapshot space.}
\end{table}

\begin{table}[!tbp]
\centering
\caption{Top 100 matched DAO Snapshot-space ranks used in the Top-TVL robustness subset, continued (ranks 26-100)}
\label{tab:robustness-top100-tvl-dao-spaces-part2}
\scriptsize
\setlength{\tabcolsep}{2pt}
\renewcommand{\arraystretch}{1.05}
\begin{tabularx}{\linewidth}{@{}rr>{\RaggedRight\arraybackslash}p{0.16\linewidth}>{\RaggedRight\arraybackslash}X>{\RaggedRight\arraybackslash}p{0.11\linewidth}rrr@{}}
\toprule
\makecell[r]{\#} & \makecell[r]{DefiLlama\\rank} & \makecell[l]{DefiLlama\\protocol} & \makecell[l]{Snapshot\\space} & Category & \makecell[r]{TVL} & \makecell[r]{Prop.} & \makecell[r]{Choices} \\
\midrule
26 & 39 & \makecell[l]{KEEP Network} & {\urlstyle{same}\url{threshold.eth}} & \makecell[l]{Cross\\Chain\\Bridge} & \$70,020,929 & 67 & 214 \\
27 & 40 & \makecell[l]{Ribbon} & {\urlstyle{same}\url{gauge.rbn.eth}} & \makecell[l]{Options\\Vault} & \$61,640,461 & 35 & 232 \\
28 & 40 & \makecell[l]{Ribbon} & {\urlstyle{same}\url{rbn.eth}} & \makecell[l]{Options\\Vault} & \$61,640,461 & 34 & 88 \\
29 & 41 & \makecell[l]{Fei Protocol} & {\urlstyle{same}\url{fei.eth}} & \makecell[l]{Algo\\Stables} & \$61,030,545 & 136 & 423 \\
30 & 42 & \makecell[l]{Homora V2} & {\urlstyle{same}\url{alpha-finance-lab.eth}} & \makecell[l]{Leveraged\\Farming} & \$58,282,004 & 1 & 2 \\
31 & 43 & \makecell[l]{CREAM Lending} & {\urlstyle{same}\url{cream-finance.eth}} & Lending & \$52,907,346 & 58 & 127 \\
31 & 85 & \makecell[l]{CreamSwap} & {\urlstyle{same}\url{cream-finance.eth}} & Dexs & \$4,952,526 & 58 & 127 \\
32 & 46 & \makecell[l]{Notional V2} & {\urlstyle{same}\url{notional.eth}} & Lending & \$43,636,953 & 30 & 69 \\
33 & 47 & \makecell[l]{Origin Dollar} & {\urlstyle{same}\url{origingov.eth}} & \makecell[l]{Yield\\Aggregator} & \$40,954,614 & 65 & 312 \\
34 & 47 & \makecell[l]{Origin Dollar} & {\urlstyle{same}\url{ousdgov.eth}} & \makecell[l]{Yield\\Aggregator} & \$40,954,614 & 97 & 861 \\
35 & 48 & \makecell[l]{mStable CDP} & {\urlstyle{same}\url{gov.dhedge.eth}} & CDP & \$40,248,021 & 70 & 153 \\
35 & 88 & \makecell[l]{Chamber Vaults} & {\urlstyle{same}\url{gov.dhedge.eth}} & Indexes & \$4,386,828 & 70 & 153 \\
36 & 48 & \makecell[l]{mStable CDP} & {\urlstyle{same}\url{mstablegovernance.eth}} & CDP & \$40,248,021 & 142 & 411 \\
36 & 88 & \makecell[l]{Chamber Vaults} & {\urlstyle{same}\url{mstablegovernance.eth}} & Indexes & \$4,386,828 & 142 & 411 \\
37 & 49 & \makecell[l]{Saddle Finance} & {\urlstyle{same}\url{saddlefinance.eth}} & Dexs & \$38,370,975 & 88 & 593 \\
38 & 50 & \makecell[l]{Vesper} & {\urlstyle{same}\url{vsp.eth}} & \makecell[l]{Yield\\Aggregator} & \$38,096,025 & 25 & 67 \\
39 & 51 & \makecell[l]{dForce Lending} & {\urlstyle{same}\url{dforcenet.eth}} & Lending & \$35,203,566 & 65 & 131 \\
40 & 52 & \makecell[l]{Rari Capital} & {\urlstyle{same}\url{fuse.eth}} & \makecell[l]{Yield\\Aggregator} & \$33,884,090 & 144 & 354 \\
41 & 53 & \makecell[l]{Unslashed} & {\urlstyle{same}\url{unslashed.eth}} & Insurance & \$33,712,560 & 30 & 79 \\
42 & 58 & \makecell[l]{SharedStake} & {\urlstyle{same}\url{sharedstake.eth}} & \makecell[l]{Liquid\\Staking} & \$29,779,030 & 50 & 165 \\
43 & 59 & \makecell[l]{Rook} & {\urlstyle{same}\url{rook.eth}} & Dexs & \$28,495,996 & 43 & 96 \\
44 & 60 & \makecell[l]{Ooki} & {\urlstyle{same}\url{ooki.eth}} & Lending & \$26,325,261 & 8 & 16 \\
45 & 62 & \makecell[l]{Metronome V1} & {\urlstyle{same}\url{metronome.eth}} & Yield & \$23,430,580 & 19 & 57 \\
46 & 64 & \makecell[l]{Rhino.fi} & {\urlstyle{same}\url{rhinofi.vote}} & Bridge & \$19,079,766 & 19 & 62 \\
47 & 65 & \makecell[l]{Parallel Protocol V2} & {\urlstyle{same}\url{mimo.eth}} & CDP & \$18,920,030 & 114 & 347 \\
48 & 66 & \makecell[l]{TrueFi} & {\urlstyle{same}\url{truefigov.eth}} & \makecell[l]{Uncoll.\\Lending} & \$17,146,372 & 77 & 167 \\
49 & 67 & \makecell[l]{Bella Protocol} & {\urlstyle{same}\url{bella}} & Yield & \$15,051,240 & 1 & 5 \\
50 & 69 & \makecell[l]{Pickle} & {\urlstyle{same}\url{pickle.eth}} & \makecell[l]{Yield\\Aggregator} & \$13,971,060 & 51 & 150 \\
51 & 71 & \makecell[l]{DFX V2} & {\urlstyle{same}\url{dfx.eth}} & Dexs & \$13,327,258 & 32 & 69 \\
52 & 72 & \makecell[l]{PoolTogether V3} & {\urlstyle{same}\url{poolpool.pooltogether.eth}} & \makecell[l]{Yield\\Lottery} & \$11,460,793 & 65 & 151 \\
53 & 72 & \makecell[l]{PoolTogether V3} & {\urlstyle{same}\url{pooltogether.eth}} & \makecell[l]{Yield\\Lottery} & \$11,460,793 & 24 & 64 \\
54 & 74 & \makecell[l]{Gnosis Protocol v1} & {\urlstyle{same}\url{gnosis.eth}} & \makecell[l]{Prediction\\Market} & \$9,516,819 & 82 & 227 \\
55 & 75 & \makecell[l]{88mph} & {\urlstyle{same}\url{88mph.eth}} & Lending & \$9,282,355 & 11 & 22 \\
56 & 76 & \makecell[l]{1inch} & {\urlstyle{same}\url{1inch.eth}} & \makecell[l]{DEX\\Aggregator} & \$8,854,452 & 47 & 124 \\
57 & 78 & \makecell[l]{OnX Finance} & {\urlstyle{same}\url{onx-finance.eth}} & \makecell[l]{Yield\\Aggregator} & \$8,466,727 & 12 & 82 \\
58 & 79 & \makecell[l]{BarnBridge} & {\urlstyle{same}\url{barnbridge.eth}} & \makecell[l]{Yield\\Aggregator} & \$7,990,989 & 32 & 75 \\
59 & 84 & \makecell[l]{Hakka Finance} & {\urlstyle{same}\url{hakka.eth}} & Derivatives & \$4,957,516 & 15 & 50 \\
60 & 91 & \makecell[l]{PieDAO} & {\urlstyle{same}\url{piedao.eth}} & Indexes & \$3,850,940 & 111 & 273 \\
61 & 93 & \makecell[l]{Perlin} & {\urlstyle{same}\url{perlinx.eth}} & Dexs & \$3,814,817 & 3 & 6 \\
62 & 94 & \makecell[l]{Dego Finance} & {\urlstyle{same}\url{dego}} & Services & \$3,670,970 & 8 & 37 \\
63 & 99 & \makecell[l]{xToken} & {\urlstyle{same}\url{xxtka.eth}} & \makecell[l]{Liquidity\\Manager} & \$2,821,323 & 7 & 24 \\
\bottomrule
\end{tabularx}
\vspace{0.25em}
\parbox{\linewidth}{\footnotesize Notes: This part reports Top-TVL sample ranks 26-100. Not all protocols found in DefiLLama can be matched in our Snapshot database; multiple DefiLlama protocols can be matched to the same Snapshot space.}
\end{table}

\end{document}